 \documentclass[aps,prd, 12pt, notitlepage,a4paper,floatfix,showpacs,nofootinbib,superscriptaddress,tikz]{revtex4-2}
\pdfoutput=1
\usepackage[utf8]{inputenc}
\usepackage[usenames,dvipsnames]{color}  
\usepackage{graphicx}
\RequirePackage[colorlinks=true,urlcolor=red,anchorcolor=red,citecolor=red,filecolor=red,linkcolor=red,menucolor=red,pagecolor=red,linktocpage=true,pdfproducer=medialab,pdfa=true]{hyperref}
\usepackage{setspace}
\usepackage{upgreek}
\usepackage{braket}
\usepackage{xcolor}
\usepackage{microtype}
\usepackage{lipsum}
\usepackage{placeins}
\usepackage{mathrsfs}
\usepackage[bbgreekl]{mathbbol}

\usepackage[font=small]{caption}
\usepackage[font=small]{subcaption}
\usepackage{stackengine}
\usepackage{amsmath}
\usepackage{amssymb}
\usepackage[normalem]{ulem}
\usepackage{graphicx}
\makeatletter
\def\p@subsection{}
\makeatother
\usepackage{float}
\usepackage{IEEEtrantools}
\usepackage{tikz-feynman}
\usepackage{caption}
\usepackage{subcaption}
\usepackage{adjustbox}
\usepackage{ragged2e}
\usepackage{mathrsfs}
\tikzfeynmanset{compat=1.0.0} 

\makeatletter
\def\p@subsection{}
\makeatother

\definecolor{darkred}{rgb}{0.6,0,0}

\definecolor{linkcolor}{rgb}{0,0,0.5}

\def\gsim{\raise0.3ex\hbox{$\;>$\kern-0.75em\raise-1.1ex\hbox{$\sim\;$}}}
\def\lsim{\raise0.3ex\hbox{$\;<$\kern-0.75em\raise-1.1ex\hbox{$\sim\;$}}}

\def\beqn#1{\begin{equation}\label{#1}}
\def\eeqn{\end{equation}}

\def\beqa#1{\begin{eqnarray}\label{#1}}
\def\eeqa{\end{eqnarray}}

\newcommand {\ignore}[1]{}

\def\A4{$A_4$}
\def\Z4{$\mathcal{Z}_4$ }
\def\Z2{$\mathcal{Z}_2$}
\def\Z3{$\mathcal{Z}_3$}

\def\321{$\mathrm{SU(3) \otimes SU(2) \otimes U(1)}$ }

\def\red{\color{red}{}}

 \newcommand{\AddrIISERB}{Department of Physics,
 Indian Institute of Science Education and Research - Bhopal \\
 Bhopal Bypass Road, Bhauri, Bhopal, India}

 \newcommand{\AddrIITBHU}{ Department of Physics, Indian Institute of Technology (BHU), Varanasi 221005, India}
\begin{document}

\title{\color{BrickRed} Dirac Scoto Inverse-Seesaw from $A_4$ Flavor Symmetry }

\author{Ranjeet Kumar}\email{ranjeet20@iiserb.ac.in}
\affiliation{\AddrIISERB}
\author{Newton Nath}\email{nnath.phy@iitbhu.ac.in} \affiliation{\AddrIITBHU} 
\author{Rahul Srivastava}\email{rahul@iiserb.ac.in}
\affiliation{\AddrIISERB}
\author{Sushant Yadav}\email{sushant20@iiserb.ac.in}
\affiliation{\AddrIISERB}

\begin{abstract}
  \vspace{1cm} 
  \noindent
We present a Dirac scotogenic-like one loop radiative model where the stability of dark matter is intricately linked to the breaking of $A_4$ flavor symmetry. This breaking induces a $\mathcal{Z}_2$ dark symmetry, stabilizing the dark matter candidate. The breaking of $A_4 \to \mathcal{Z}_2$ leads to cutting the loop and facilitating a ``scoto inverse-seesaw" mass mechanism responsible for neutrino mass generation. This elucidates the explicit explanation of two mass-squared differences, $\Delta m^2_{\rm{atm}}$ and $\Delta m^2_{\rm{sol}}$ observed in neutrino oscillations. Our model accounts for normal and inverted ordering of neutrino masses, revealing sharp correlations between $\sum m_i$ and $\langle m_{\beta} \rangle$. It also shows strong compatibility with current data in the $\delta_{CP}$–$\theta_{23}$ plane. Moreover, stringent constraints on scalar masses narrow down the viable dark matter mass regions, accommodating $SU(2)_L$  singlet and doublet scalar dark matter as well as  fermionic dark matter. Additionally, our model presents a viable avenue for addressing lepton flavor violating decays while remaining consistent with current experimental constraints.
\end{abstract}

\maketitle

\section{Introduction}
\label{sec:intro}

The discovery of non-zero neutrino masses and mixing in neutrino oscillation experiments~\cite{SNO:2001kpb,Super-Kamiokande:1998kpq}, along with compelling evidence that approximately 85\% of the Universe's total matter exists as dark matter (DM)~\cite{Planck:2018vyg}, has drawn significant interest from particle physicists. These observations point to clear shortcomings of the otherwise successful Standard Model (SM), which cannot account for neutrino masses or the nature of DM. Additionally, the question of whether neutrinos are Dirac or Majorana particles in nature has remained a long-standing puzzle, continuing to attract considerable attention in the field. These interesting aspects have driven efforts to explore physics beyond the Standard Model (BSM) for a deeper understanding of such phenomena. From a theoretical viewpoint, various models have been proposed in the literature to explain the tiny masses of neutrinos and the origin of DM within a unified framework. One widely acclaimed model is the ``scotogenic" mechanism~\cite{ma2006verifiable}, a minimal extension of the SM that links DM stability to neutrino mass generation and treats neutrinos as Majorana particles.
Although these models offer insight into the smallness of neutrino masses and DM stability, they are often criticized for their difficulty in explaining the flavor structure of the leptonic sector, and the two different mass scales observed in neutrino oscillation experiments~\cite{Capozzi:2021fjo, deSalas:2020pgw, Esteban:2020cvm} namely, atmospheric, $\Delta m^2_{\rm {atm}}$, and the solar, $\Delta m^2_{\rm {sol}}$.
The scoto-seesaw model, recently addressed in \cite{Rojas:2018wym}, is one possible way to explain the origin of the two different mass scales of neutrino oscillation in addition to DM. Within this framework, $\Delta m^2_{\rm {atm}}$ arises from scoto-loop, while $\Delta m^2_{\rm {sol}}$ appears at the tree level seesaw~\cite{Rojas:2018wym,Barreiros:2020gxu,Mandal:2021yph,Barreiros:2022aqu, Ganguly:2023jml,Ganguly:2022qxj,VanDong:2023xbd,Leite:2023gzl,Kumar:2023moh,VanDong:2024lry}. However, scoto-seesaw models do not shed any light on the mixing and flavor structure of the leptonic sector like other neutrino mass mechanisms. 
 
Experimental evidence for the Majorana nature of neutrinos remains inconclusive, as indicated by the results from searches for neutrinoless double beta decay \cite{Schechter:1981bd}. This leaves open the possibility that neutrinos may instead be Dirac particles, thereby motivating further investigation into their Dirac nature. While the canonical scotogenic model \cite{ma2006verifiable} was originally formulated for Majorana neutrinos, the scotogenic framework is sufficiently general to accommodate Dirac neutrinos as well~\cite{Gu:2007ug,Farzan:2012sa}. Indeed, in recent years, several attempts have been made to explore Dirac scotogenic models, as proposed in~\cite{Ma:2016mwh,Bonilla:2018ynb,Dasgupta:2019rmf,Guo:2020qin,CentellesChulia:2024iom}. This model extends the SM by incorporating radiative neutrino mass generation, while ensuring that neutrinos remain Dirac particles.

To address the challenges related to the origin of neutrino masses, their flavor structure, the presence of two distinct mass scales, and the stability of DM, a recent approach based on flavor symmetries has been developed~\cite{Kumar:2024zfb,Borah:2024gql,Kumar:2024jot}. Within this framework, it has been demonstrated that employing the $A_4$ flavor symmetry~\cite{Ma:2001dn,Babu:2002dz,Altarelli:2005yx} allows for predictive solutions to the leptonic flavor structure. In addition,  the spontaneous breaking of $A_4$ to its subgroup $\mathcal{Z}_2$~\cite{Hirsch:2010ru,Boucenna:2011tj,DeLaVega:2018bkp} ensures DM stability while also leading to the scoto-seesaw mass mechanism. Refs.~\cite{Kumar:2024zfb}, explore the Majorana scotogenic loop, whereas Ref.~\cite{Borah:2024gql} investigates the Dirac scotogenic loop. Additionally, the authors of~\cite{Kumar:2024zfb} emphasize that the mass of DM is tightly constrained by the breaking of $A_4$ symmetry. They further provide conclusive evidence for the scalar DM scenario by ruling out the possibility of fermionic DM. 

In this work, we propose a framework that explains the Dirac nature of neutrinos within the leptonic flavor structure while simultaneously addressing both scalar and fermionic DM scenarios. Overall, our flavor-symmetric approach provides a highly predictive framework for both neutrino sector phenomenology and DM studies, offering compelling theoretical insights into BSM physics.
\begin{figure}[h!]
\centering
\includegraphics[width=0.5\textwidth]{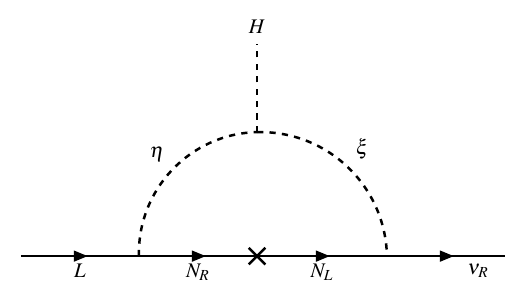}
\caption{\centering Dirac scotogenic model for neutrino mass generation.}
\label{fig:UVloop}
\end{figure}
To serve our purpose, we employ $A_4$ flavor symmetry on a Dirac scotogenic-like radiative loop as shown in Fig.~\ref{fig:UVloop}. The spontaneous symmetry breaking (SSB) of $A_4$ to its $\mathcal{Z}_2$ subgroup stabilizes the DM candidate. This breaking also leads to the cutting of the loop, inducing a hybrid ``scoto inverse-seesaw" mass mechanism. Note that the inverse-seesaw mass mechanism proposed in ~\cite{Mohapatra:1986bd} requires a small mass parameter, usually called $\mu$-term. This naturally emerges in our formalism due to the presence of an $SU(2)_L$ singlet scalar acquiring a vacuum expectation value (VEV), which can be very small. Recently, authors of \cite{CentellesChulia:2020dfh} have addressed this mechanism for both Majorana and Dirac neutrinos. In our formalism, the Dirac neutrino masses are generated at the tree level via the inverse-seesaw mechanism and at the loop level through the scoto-loop. 

Here, we also consider an accidental $U(1)_{B-L}$ symmetry, inherently present in the SM, and its breaking into a $\mathcal{Z}_3$ subgroup to prevent tree-level couplings before SSB and to forbid the Majorana neutrino mass term.
The chiral anomaly-free $B-L$ symmetry requires the ($-4,-4,5$) charge assignment for the right-handed neutrinos $\nu_{R_i}$~\cite{Ma:2014qra,Ma:2015raa,Ma:2015mjd}. Unlike the Majorana version~\cite{Kumar:2024zfb}, our model features both normal ordering (NO) and inverted ordering (IO) of neutrino masses. Furthermore, the $A_4$ symmetric approach imposes an upper bound on the scalar masses based on the two mass-squared differences of neutrino, thereby constraining the allowed parameter space for DM candidates. Also, we study the different charged lepton flavor violation (CLFV) processes and check their viability under different current and future proposed experiments. 

This paper is organized as follows. In Sec. \ref{sec:model}, we outline the model setup based on $A_4$ flavor symmetry and discuss the scalar mass spectrum as well as the neutrino sector. We discuss the emergence of the inverse-seesaw mechanism at tree level and scoto-loop due to the breaking of $A_4$ symmetry. In Sec.~\ref{sec:nusecres}, we show the numerical results and correlation of neutrino sector observables for both NO and IO of neutrino masses.  We discuss our dark sector results in Sec.~\ref{sec:dm} and show that the mass of DM candidates is very constrained. In Sec.~\ref{sec:lfv}, we perform the numerical analysis for CLFV processes, and finally, concluding remarks are made in Sec.~\ref{sec:conc}.

\section{Model SET-UP} \label{sec:model}
We utilize a flavor-symmetric Dirac scotogenic model based on the $A_4$ symmetry group to account for neutrino oscillation parameters and DM phenomenology. In addition to the SM particles, we introduce BSM particles: scalars, $\eta$ and $\xi$, which are $SU(2)_L$ doublet and singlet, respectively, and both transform as triplets under the \A4 symmetry. In this formalism, singlet fermions $N_L$ and $N_R$ are also introduced and transformed as triplets under the \A4 symmetry. 
We also add three right-handed neutrinos $\nu_{R}$, which transform as singlets $(1, 1'', 1')$ under the \A4 symmetry. The SM Higgs doublet $\Phi$ transforms as trivial singlet $(1)$, the charged lepton doublets $L_i$ and singlets $e_{R_i} \, (i = 1,2,3)$ transform as $(1, 1', 1'')$ under the \A4 symmetry, respectively. For more details on the \A4 flavor symmetry and its multiplication rules, see App.~\ref{app:A4}. The particle content and their transformation rule under the  $ SU(2)_L \otimes U(1)_Y$ are summarized in the second column of Tab.~\ref{tab:FieldCharge}. The last two columns present the transformation properties of different fields under the \A4 symmetry and the $U(1)_{B-L}$ as well as their residual subgroup $\mathcal{Z}_2$ and $\mathcal{Z}_3$, respectively. 
\begin{center}
\begin{table}[!h] 
\begin{tabular}{ |c||c||c||c|c| } 
 \hline
 Fields & $SU(2)_L\otimes U(1)_Y$ & {\red \A4}  ${\color{Mahogany} \mathbf{\to}}$ $\mathcal{Z}_2$  & $U(1)_{B-L}$ ${\color{Mahogany} \mathbf{\to}}$ ${\color{Mulberry}\boldsymbol{{\mathcal{Z}}_3}}$  \\
 \hline
$L_i$ & (2, -1) &  {\red(1, $1'$, $1''$)} ${\color{Mahogany} \mathbf{\to}}$ (${+}$, $+$, $+$)  & -1 ${\color{Mahogany} \mathbf{\to}}$ ${\color{Mulberry}\boldsymbol{\omega^2}}$  \\
$e_{R_i}$ & (1, -2) &  {\red(1, $1'$, $1''$)} ${\color{Mahogany} \mathbf{\to}}$ ($+$, $+$, $+$) & -1 ${\color{Mahogany} \mathbf{\to}}$ ${\color{Mulberry}\boldsymbol{\omega^2}}$ \\
$H$ & (2, 1) &  {\red 1} ${\color{Mahogany} \mathbf{\to}}$ $+$ & 0 ${\color{Mahogany} \mathbf{\to}}$ ${\color{Mulberry}\boldsymbol{ 1}}$\\
\hline \hline
$\nu_{R_i}$ & (1, 0) &  {\red(1, $1''$, $1'$)} ${\color{Mahogany} \mathbf{\to}}$ ($+$, $+$, $+$)  & (-4, -4, 5) ${\color{Mahogany} \mathbf{\to}}$ ${\color{Mulberry}\boldsymbol{\omega^2}}$\\
 $N_L$ & (1, 0) &  {\red 3} ${\color{Mahogany} \mathbf{\to}}$ ($+$, $-$, $-$)  & -4 ${\color{Mahogany} \mathbf{\to}}$ ${\color{Mulberry}\boldsymbol{\omega^2}}$\\
 $N_R$ & (1, 0) &  {\red 3} ${\color{Mahogany} \mathbf{\to}}$ ($+$, $-$, $-$)  & -4 ${\color{Mahogany} \mathbf{\to}}$ ${\color{Mulberry}\boldsymbol{\omega^2}}$ \\
 $\eta$ & (2, 1) &  {\red 3} ${\color{Mahogany} \mathbf{\to}}$ ($+$, $-$, $-$) & 3 ${\color{Mahogany} \mathbf{\to}}$ ${\color{Mulberry}\boldsymbol{1}}$\\
$\xi$ & (1, 0) &  {\red 3} ${\color{Mahogany} \mathbf{\to}}$ ($+$, $-$, $-$)  & 0 ${\color{Mahogany} \mathbf{\to}}$ ${\color{Mulberry}\boldsymbol{ 1}}$\\
 \hline
\end{tabular}
\caption{Field content and transformation properties of different fields under the  $ SU(2)_L\otimes U(1)_Y$, $A_4$, and  $U(1)_{B-L}$ symmetry, where $i = 1, 2, 3$ represents generation indices.}
\label{tab:FieldCharge}
\end{table}
\end{center}
The charge assignments in Tab.~\ref{tab:FieldCharge} ensure that, following the $A_4$ symmetry breaking by the VEVs of $\eta$ and $\xi$, all the SM particles and $\nu_{R_i}$'s remain even under the residual $\mathcal{Z}_2$ symmetry. In contrast, the triplet BSM particles may acquire odd charges under the residual $\mathcal{Z}_2$ symmetry. The components of \A4 triplets $\eta, \ \xi$ and $N$ split into two, the first components transforming as even while the other components have odd charges under $\mathcal{Z}_2$, see App.~\ref{app:A4} for more details. The odd particles will eventually become dark sector particles as discussed in Sec.~\ref{sec:intro}. This discussion will focus on the scenario, where $U(1)_{B-L}$ breaks down to the residual \Z3 subgroup\footnote{The explicit breaking of the \( U(1)_{B-L} \) symmetry arises from the cubic interaction term \( H^\dagger \eta\, \xi \) in the scalar potential. Here, \( H \) and \( \xi \) carry zero \( B-L \) charge, while \( \eta \) has a charge of three. Consequently, this interaction alters \( B-L \) in steps of three, leading to the breaking of \( U(1)_{B-L} \) in units of three. As a result, the residual symmetry is the discrete subgroup \( \mathcal{Z}_3 \).  
Since \( B-L \) is violated in multiples of three, the transformations that preserve the term \( H^\dagger \eta\, \xi \) must correspond to those where \( B-L \) charges differ by multiples of three. This implies that the \( U(1)_{B-L} \) phase factor, \( e^{i\alpha(B-L)} \), is constrained such that  
$e^{i\alpha(B-L)} = e^{i\alpha(B-L+3n)}$ for any integer \( n \). This periodicity signifies that the continuous \( U(1)_{B-L} \) symmetry is broken down to the discrete subgroup \( \mathcal{Z}_3 \).}, a natural outcome when the $B-L$ symmetry is broken in a unit of three. The presence of the residual \Z3 subgroup ensures the Dirac nature of neutrinos. 

Some of the salient features of the model are as follows:
\begin{itemize}
    \item The dimension-4 Dirac neutrino mass term of the form,  $\bar{L}_i\tilde{H}\nu_{R_j}$ and the bare Majorana mass term, $ \bar{\nu}_{R_i}^c \nu_{R_j}$, are prohibited by both the \A4 and  $U(1)_{B-L}$ symmetries, except for $i=j=1$. Specifically, although the terms $\bar{L}_1\tilde{H}\nu_{R_1}$ and $ \bar{\nu}_{R_1}^c \nu_{R_1}$ are allowed by $A_4$, they are forbidden due to the $U(1)_{B-L}$ charge assignments.
    \item Similarly, the effective dimension-5  Weinberg operator $\bar{L}_i^c H H L_j$, which generates Majorana neutrino masses, is forbidden by both the \A4 and $U(1)_{B-L}$ symmetries, except for $i=j=1$. The term corresponding to $i=1$ is allowed by the \A4 but remains forbidden by $U(1)_{B-L}$ symmetry.
    \item The effective dimension-5 operator $\bar{L}_i H\nu_{R_j}\xi$, which generates Dirac neutrino masses, is forbidden by both the \A4 and  $U(1)_{B-L}$ symmetry.
    \item  The effective dimension-5 operator $\bar{L_i}\,\eta\,\, \nu_{R_j} \,  \xi$, where   $i, j = 1, 2$ is allowed by the symmetries and requires UV-completion. This term emerges after the breaking of the $A_4$ symmetry, as shown in the top-right panel of Fig.~\ref{fig:breakA4}.
    %
        
                \item In this minimal setup, one neutrino remains massless because the Yukawa interaction $Y^{\prime} \bar{N}_{L} \xi \nu_{R}$ is forbidden for the third generation right-handed neutrino due to $B-L$ charge assignments. However, the massless neutrino can be made massive by adding a $SU(2)_L$ singlet but $A_4$ triplet scalar $\zeta_9$ with $B-L$ charge $-9$, allowing the Yukawa term $y_9 \bar{N}_L \zeta_9 \nu_{R_3}$. After SSB, $\zeta_9$ acquires a VEV  $(u_9, 0, 0)$ preserving the $A_4 \to \mathcal{Z}_2$ breaking. Then through the scoto-seesaw mechanism, we can generate a light Dirac neutrino mass for the $(\nu_{L_3}, \nu_{R_3})$ pair. Note that, since one massless neutrino is allowed by the current oscillation data, we do not develop this analysis further.
\end{itemize}
The spontaneous breaking of the $A_{4}$ symmetry to its residual $\mathcal{Z}_2$ subgroup results in a hybrid scoto-seesaw mechanism, wherein an interplay between the tree level type-I seesaw and the one-loop level scotogenic mechanism facilitates the generation of neutrino masses. Moreover, the residual $\mathcal{Z}_2$ symmetry assumes the role of the scotogenic dark symmetry. Consequently, the lightest $\mathcal{Z}_2$ odd particle is inherently stable and emerges as a potential DM candidate.

\subsection{Scalar sector} 
\label{sec:scsec}
In this section, we discuss the scalar sector of our model and the emergence of $\mathcal{Z}_2$ dark symmetry from $A_4$ flavor symmetry. The scalar potential of our model is provided in App.~\ref{app:pot}. The $A_4$ symmetry is broken spontaneously by the VEVs of the scalars $\eta$, $\xi$. We aim to obtain the unbroken $\mathcal{Z}_2$ subgroup as a residual symmetry. This can be accomplished by the following VEV alignments of $\eta$ and $\xi$ components
\begin{align} \label{eq:vevalig}
  \langle \eta_1 \rangle &=  \frac{v_2}{\sqrt{2}}, \ \langle \eta_2 \rangle = 0 = \langle \eta_3 \rangle \, , \nonumber \\
    \langle \xi_1 \rangle &=  u , \ \quad \langle \xi_2 \rangle = 0 = \langle \xi_3 \rangle \, .
\end{align}
Under the residual $\mathcal{Z}_2$ symmetry, these fields transform as
\begin{align}
\eta_1 \to +\eta_1 \, , \quad \eta_{2} \to -\eta_{2}, \quad \eta_{3} \to -\eta_{3}\,,  \nonumber \\
\xi_1 \to +\xi_1 \, , \quad \xi_{2} \to -\xi_{2}, \quad \xi_{3} \to -\xi_{3}\,.
\end{align} 
The SM $H$, being a singlet of $A_4$ transforms as $H \to + H$ under the residual $\mathcal{Z}_2$, and hence $\langle H \rangle =  \frac{v_1}{\sqrt{2}}$. 
After the SSB, the $SU(2)_L$ doublet scalars H, $\eta_{i}$ and singlet scalar $\xi_{i}$ can be expressed as 
\begin{align}
&H =  \begin{pmatrix}                   
 H^+ \\
 \frac{v_1+\phi_1+ i \sigma_1}{\sqrt{2}}
           \end{pmatrix}  ,\,\, \quad 
 \eta_1 =  \begin{pmatrix}                   
 \eta_1^+ \\
 \frac{v_2+\phi_2+ i \sigma_2}{\sqrt{2}}
           \end{pmatrix}  ,\,\,  \quad 
 \eta_2 =  \begin{pmatrix}                   
 \eta_2^+ \\
 \frac{\eta_2^R+ i \eta_2^I}{\sqrt{2}}
           \end{pmatrix}  ,\,\,  \quad 
 \eta_3 =  \begin{pmatrix}                   
 \eta_3^+ \\
 \frac{\eta_3^R+ i \eta_3^I}{\sqrt{2}}
           \end{pmatrix}  ,\,\, \nonumber \\  &\xi_1=u+\phi_3, \quad \xi_2 \equiv \xi^R_2, \quad \xi_3 \equiv \xi^R_3  \,,
\end{align} 
where $\phi_1,\phi_2,\phi_3,\eta_2^R,\eta_3^R,\xi_2^R, \xi_3^R$ are CP even particles,  whereas CP odd particles are $\sigma_1,\sigma_2,\eta_2^I,\eta_3^I$. The SM VEV is given by $\sqrt{v_1^2 + v_2^2}= v_{SM} \approx 246 $ GeV. 

One can compute the scalar mass spectrum from the scalar potential,
\begin{align}\label{eq:scalarpot1} 
V&=  V_H +V_{\eta} +V_{\xi} +V_{H \eta} + V_{H\xi} + V_{\eta \xi} + V_{H\eta \xi} + h.c. \ .
\end{align}
See Eq.~\eqref{eq:pot1} in App.~\ref{app:pot} for the complete description of the potential. We want to mention that, unlike the scenario in Ref.~\cite{Kumar:2024zfb}, the fields $\eta_2$ and $\eta_3$ do not mix in the present framework. This distinction arises because, in this model, $\eta$ is charged under $U(1)_{B-L}$, which forbids terms such as $\eta^3 H$, ensuring that no mixing occurs between $\eta_2$ and $\eta_3$. The real component of the scalar $\eta$ mixes with the real scalar $\xi$, with $\eta^R_2$ mixing with $\xi^R_2$ and $\eta^R_3$ with $\xi^R_3$.

The mass matrices are given as follows:
\begin{align} 
&\mathcal{M}^2_{2R}= \begin{bmatrix}
\Lambda_{\eta_{2}} v_2^2-\alpha_{2} u^2 & \alpha_1 v_2 u \\
\alpha_1 v_2 u &\Lambda_{\xi_{2}} u^2-\alpha_{3} v_2^2 \\
\end{bmatrix}, \, \quad
\mathcal{M}^2_{3R}= \begin{bmatrix}
\Lambda_{\eta_{2}} v_2^2-\alpha_{3} u^2 & \alpha_1 v_2 u \\
\alpha_1 v_2 u &\Lambda_{\xi_{2}} u^2-\alpha_{2} v_2^2 \\
\end{bmatrix} \;.
\end{align}
The details about these mass matrices and the mass spectrum of other scalars of the model have been presented in the App.~\ref{app:pot}. The $\eta^R_{2}$ mixes with $\xi^R_{2}$ and after the mixing the mass eigenstates $\chi_1$ and $\chi_2$ are given as follows
\begin{align} \label{eq:eigenstates}
    \chi_1  &= \cos{\theta_2} \eta^R_2 + \sin{\theta_2} \xi^R_2, \quad 
    \chi_2  = -\sin{\theta_2} \eta^R_2 + \cos{\theta_2} \xi^R_2 \ .
\end{align}
 The mixing angle is given by 
\begin{align} \label{eq:mixangle}
&\tan {2 \theta_{2}} =\frac{2 \alpha_1  v_2u}{\Lambda_{\xi_{2}} u^2-\Lambda_{\eta_{2}} v_2^2+\alpha_{2} u^2-\alpha_{3}v_2^2} \ .
\end{align} 
The mass eigenvalues corresponding to the mass eigenstates can be written as follows
\begin{align} \label{eq:scmasses}
    m^2_{1R}&= \left( \Lambda_{\eta_{2}} v_2^2-\alpha_{2} u^2 \right) \cos^2{\theta_2} + \left( \Lambda_{\xi_{2}} u^2-\alpha_{3} v_2^2 \right) \sin^2{\theta_2}- 2 \alpha_1 v_2 u  \sin{\theta_2}  \cos{\theta_2} \,,\nonumber \\
    m^2_{2R}&= \left( \Lambda_{\eta_{2}} v_2^2-\alpha_{2} u^2 \right) \sin^2{\theta_2} + \left( \Lambda_{\xi_{2}} u^2-\alpha_{3} v_2^2 \right) \cos^2{\theta_2}+2 \alpha_1 v_2 u  \sin{\theta_2}  \cos{\theta_2} \,.
\end{align}
Similarly, $\eta^R_{3}$ mixes with $\xi^R_{3}$ and after the mixing the mass eigenstates $\chi_3$ and $\chi_4$ are given as follows
\begin{align} \label{eq:eigenstates1}
    \chi_3  &= \cos{\theta_3} \eta^R_3 + \sin{\theta_3} \xi^R_3, \quad
   \chi_4  = -\sin{\theta_3} \eta^R_3 + \cos{\theta_3} \xi^R_3 \ .
\end{align}
The mixing angle in this case is given by
\begin{align} \label{eq:mixangle}
&\tan {2 \theta_{3}} =\frac{2 \alpha_1 v_2u}{\Lambda_{\xi_{2}} u^2-\Lambda_{\eta_{2}} v_2^2+\alpha_{3} u^2-\alpha_{2}v_2^2} \ .
\end{align} 
The mass eigenvalues corresponding to the mass eigenstates can be written as follows
\begin{align} \label{eq:scmasses}
    m^2_{3R}&= \left( \Lambda_{\eta_{2}} v_2^2-\alpha_{3} u^2 \right) \cos^2{\theta_3} + \left( \Lambda_{\xi_{2}} u^2-\alpha_{2} v_2^2 \right) \sin^2{\theta_3}- 2 \alpha_1 v_2 u  \sin{\theta_3}  \cos{\theta_3} \,,\nonumber \\
    m^2_{4R}&= \left( \Lambda_{\eta_{2}} v_2^2-\alpha_{3} u^2 \right) \sin^2{\theta_3} + \left( \Lambda_{\xi_{2}} u^2-\alpha_{2} v_2^2 \right) \cos^2{\theta_3}+ 2 \alpha_1 v_2 u  \sin{\theta_3}  \cos{\theta_3} \ .
\end{align}
\begin{figure}[t!]
    \centering
    \includegraphics[height=7cm]{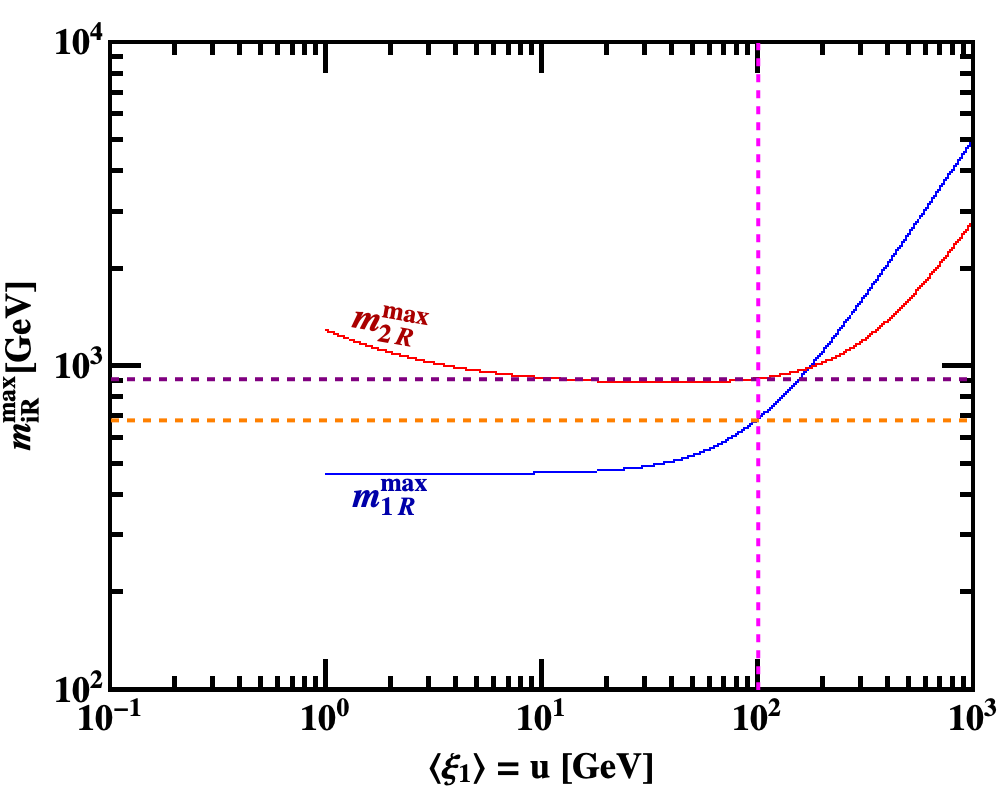}
    \caption{\centering Correlation between the upper bound of dark sector scalar masses and singlet scalar VEV.}
    \label{fig:masslim}
\end{figure}
For a small mixing between the dark scalar particles, $\chi_1$ and $\chi_3$ will primarily behave as doublet scalars, while $\chi_2$ and $\chi_4$ will predominantly exhibit singlet nature. This follows from Eqs.~\eqref{eq:eigenstates} and \eqref{eq:eigenstates1}, which reduce to $\chi_{1,3} \sim \eta_{2,3}^R$ and $\chi_{2,4} \sim \xi_{2,3}^R$ in the limit of small mixing. Hence, in the scalar DM case, we will have two different types of DM candidates that will provide different natures of the DM particle (see Sec.~\ref{sec:dm}). For the sake of definiteness, we have taken $\chi_1 \ (\chi_2$)  as doublet (singlet) DM candidates, making them the lightest particle in the dark sector. Note that taking $\chi_3$ and $\chi_4$ as a DM candidate will not change the results of our analysis.

In Fig.~\ref{fig:masslim}, we have shown a correlation of the maximum value of the mass of dark sector particle, $m^{\rm max}_{iR}$, with the singlet scalar VEV, $\langle \xi_1 \rangle=u$. The numerical values of other parameters are chosen such that $m_{iR}$ is the maximum for a given value of $u$. Since, the doublets VEV follows $\sqrt{v_1^2+v_2^2}=v_{SM}$ and $\lambda$ has perturbativity limit, the maximum value of scalars masses $(m_{iR}; \ i=1,2,3,4)$ is governed by $u$. While performing the numerical analysis, we observed that there is an upper limit on $u$ ($\sim 100$ GeV), beyond which the neutrino oscillation data can no longer be explained, as discussed in Sec.~\ref{sec:nusecres}. The limit on $u$ will restrict the dark scalar masses $m_{1R}$ and $m_{2R}$ to be $\approx 700 \ \rm{GeV}$ \footnote{Note that, as pointed out in~\cite{Kumar:2024zfb}, scalar masses also have an upper bound, but this arises solely from the perturbativity constraint. In contrast, in the present model, the upper bound is determined by both the perturbativity limit and the restriction on the value of $u$ imposed by neutrino observables.} and $\approx 900 \ \rm{GeV}$, respectively. A Similar behavior is observed for $m_{3R}$ and $m_{4R}$, as their analytical expressions closely resemble those of $m_{1R}$ and $m_{2R}$. Hence, the DM candidate in our model has an upper bound $\leq 700\rm {GeV}$ \footnote{It is important to point out that even if the DM candidate is a fermion, this mass limit still applies as the DM candidate has to be the lightest in the dark sector particles.}. This specific feature of our model arises directly from the \A4 $\to \mathcal{Z}_2$ breaking, distinguishing it from conventional Dirac scotogenic and scoto-seesaw models.
\subsection{Yukawa sector}
Referring to the charge assignments specified in Tab.~\ref{tab:FieldCharge}, the Yukawa Lagrangian governing the leptonic sector can be expressed as follows
\begin{align}\label{eq:YukawaLag}
- \mathcal{L}_y & = y_{11}(\overline{L}_1)_1 H (e_{R_1})_1 +  y_{22}(\overline{L}_2)_{1^{\prime\prime}} H (e_{R_2})_{1^{\prime}} + y_{33}(\overline{L}_3)_{1^{\prime}} H (e_{R_3})_{1^{\prime\prime}} + y_1(\overline{L}_1)_1(\tilde{\eta}  N_R)_1 \nonumber \\ 
&  + y_2(\overline{L}_2)_{1^{\prime\prime}} (\tilde{\eta}  N_R)_{1^{\prime}} + y_3(\overline{L}_3)_{1^{\prime}}(\tilde{\eta} N_R)_{1^{\prime\prime}} + y'_1 (\overline{N}_L \xi)_1 (\nu_{R_1})_1 + y'_2 ( \overline{N}_L \xi)_{1^{\prime}} (\nu_{R_2})_{1^{\prime\prime}} \nonumber \\ 
&+ M(\overline{N}_L N_R)_1 +  h.c.\;
\end{align}
Exploiting the $A_4$ multiplication rule (see App.~\ref{app:A4}), we can simplify Eq.~\eqref{eq:YukawaLag} as
\begin{align}\label{eq:YukawaLag2}
- \mathcal{L}_y & = y_{11}\overline{L}_1 H e_{R_1} +  y_{22}\overline{L}_2 H e_{R_2} + y_{33}\overline{L}_3 H e_{R_3}  + y_1 \overline{L}_1 \left(\tilde{\eta}_1 N_{R_1} + \tilde{\eta}_2 N_{R_2} +\tilde{\eta}_3 N_{R_3}  \right) \nonumber \\ &+ y_2 \overline{L}_2 \left(\tilde{\eta}_1 N_{R_1} + \omega \tilde{\eta}_2 N_{R_2} + \omega^2 \tilde{\eta}_3 N_{R_3}  \right) + y_3 \overline{L}_3 \left(\tilde{\eta}_1 N_{R_1} + \omega^2 \tilde{\eta}_2 N_{R_2} + \omega \tilde{\eta}_3 N_{R_3}  \right) \nonumber \\ & + y'_1  \left( \overline{N}_{L_1} \xi_1 +   \overline{N}_{L_2} \xi_2 +  \overline{N}_{L_3} \xi_3  \right)\nu_{R_1} + y'_2 \left( \overline{N}_{L_1} \xi_1 + \omega  \overline{N}_{L_2} \xi_2 + \omega^2 \overline{N}_{L_3} \xi_3 \right)\nu_{R_2} \nonumber \\ & + M(\overline{N}_{L_1} N_{R_1} + \overline{N}_{L_2} N_{R_2} + \overline{N}_{L_3} N_{R_3}) +  h.c.\;
\end{align}
where $y_{ii}$, $y_i$ ($i = 1, 2, 3$) and $y'_j , (j=1, 2)$ are the  Yukawa couplings, $\omega$ is the cubic root of unity and  $M$ is the Dirac mass term of the BSM fermions. With this specific charge assignment, the mass matrix for charged leptons becomes diagonal. Consequently, the observed oscillation and mixing patterns of leptons originate solely from the neutrino sector as discussed next. 
\subsection{Neutrino Masses}
\begin{figure}[t!]
\centering
       \includegraphics[width=0.75\textwidth]{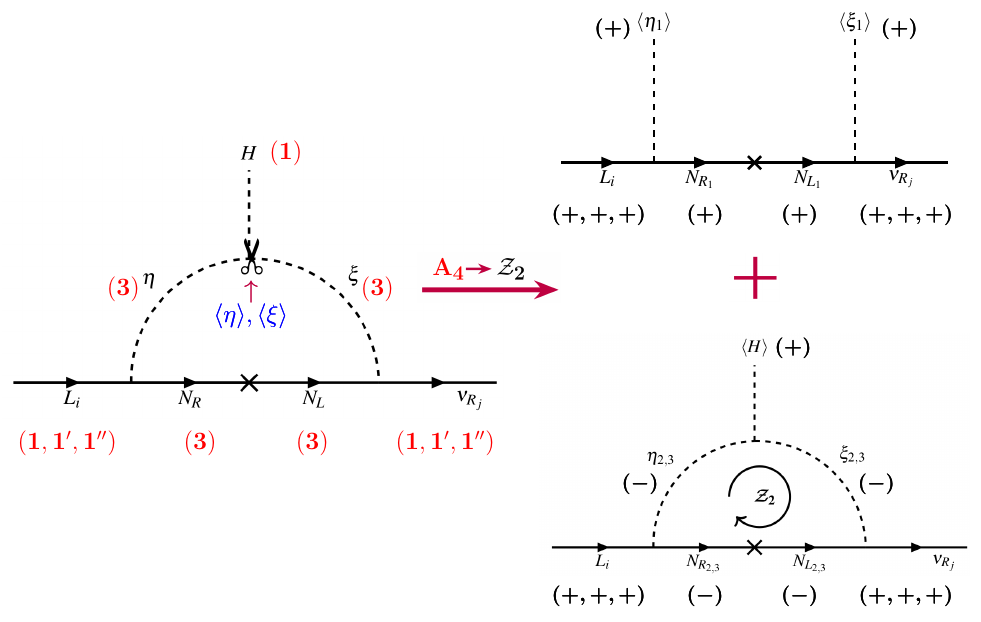}
    \caption{\footnotesize Breaking of $A_4 \to \mathcal{Z}_2$ leading to a scoto inverse-seesaw mass mechanism. The residual $\mathcal{Z}_2$ stabilizes the DM candidate running inside the loop. We have shown the \A4 ($\mathcal{Z}_2$) charges in the left (right) panel of the diagram.}
    \label{fig:breakA4}
\end{figure}

One of the primary objectives of the scoto-seesaw mechanism~\cite{Rojas:2018wym} is to account for the two observed neutrino mass-squared differences in oscillation experiments. To explain neutrino masses, we break the $A_4$ symmetry through the VEVs of $\eta_1$ and $\xi_1$. Consequently, this symmetry breaking divides the loop into two distinct segments: the type-I-like seesaw and the scoto-loop, as shown in Fig.~\ref{fig:breakA4}. The $A_4$ ($\mathcal{Z}_2$) charges before (after) symmetry breaking are shown in Fig.~\ref{fig:breakA4}. The breaking of the \A4 symmetry and the emergence of $\mathcal{Z}_2$ odd charges in the scoto-loop plays a very crucial role in generating neutrino masses and stabilizing the DM candidates.  
At the tree level, the $(6\times 6)$ mass matrix in the basis $(\overline{\nu}_{L_i}, \overline{N}_{L_j})$ and $(\nu_{R_i}, N_{R_j})^T$, (where $i\,, j =1,2,3$) is given by
\begin{align}
         \mathcal{M}_{\nu} &=  \begin{pmatrix}
                    0 & 0 & 0 & \frac{y_1v_2}{\sqrt{2}} & 0 & 0 \\
                    0 & 0 & 0 & \frac{y_2v_2}{\sqrt{2}} & 0 & 0 \\
                    0 & 0 & 0 & \frac{y_3v_2}{\sqrt{2}} & 0 & 0 \\
                    y^{\prime}_1u & y^{\prime}_2u & 0 & M & 0 & 0 \\
                    0 & 0 & 0 & 0 & M & 0 \\
                    0 & 0 & 0 & 0 & 0 & M \\
           \end{pmatrix}\,, \\
          & =  \begin{pmatrix}
          0  & \mathcal{M}_D \\
            \boldsymbol{\mu}
  & \mathcal{M}
             \end{pmatrix} \,,
           \end{align}
where,
  \begin{align}
  \boldsymbol{\mu}
 =  \begin{pmatrix}                   
                     y^{\prime}_1u & y^{\prime}_2u & 0 \\
                     0 & 0 & 0 \\
                     0 & 0 & 0 \\
           \end{pmatrix}  ,\,\,
         \mathcal{M}_D =  \frac{1}{\sqrt{2}}\begin{pmatrix}                   
                     y_1v_2 & 0 & 0 \\
                     y_2v_2 & 0 & 0 \\
                     y_3v_2 & 0 & 0 \\
           \end{pmatrix}  ,\,\,
\mathcal{M}=
\begin{pmatrix}
M & 0 & 0 \\
0 & M & 0 \\
0 & 0 & M 
\end{pmatrix}\;.       
  \end{align}  
Now, in the limit, $M \gg y'_j u, \ y_i v_2$  ($j = 1, 2$ and $i = 1, 2, 3$), the light Dirac neutrino mass matrix can be written as
\begin{equation}
 -m^{(1)}_{\nu} =  \mathcal{M}_D \mathcal{M}^{-1} \boldsymbol{\mu}
 = 
\frac{v_2u}{\sqrt{2}M} 
\begin{pmatrix}
y_1y'_1 & y_1y'_2 & 0 \\
y_2y'_1 & y_2y'_2 & 0 \\
y_3y'_1 & y_3y'_2 & 0\\
\end{pmatrix} \;.
\label{eqn:TreeNu}
\end{equation}
The resulting neutrino mass is proportional to the $\boldsymbol{\mu}$-matrix, where $u$ is the VEV of the singlet field $\xi$. Since $u$ can take very small values, this feature is characteristic of the inverse seesaw mechanism~\cite{CentellesChulia:2020dfh}, which will be discussed in the next section. It is noteworthy that the rank of $m^{(1)}_{\nu}$ is one, implying that only one of the neutrinos acquires mass at the tree level. 

Note that, even when we take the one-loop contribution associated with $N_1$ into account, the texture zeros of Eq.~\eqref{eqn:TreeNu} remain symmetry protected. This results in an additional mass matrix with a similar structure to Eq.~\eqref{eqn:TreeNu}. 
It is due to the fact that the residual $\mathcal{Z}_2$ symmetry remains unbroken, and as a consequence, mixing between the first generation singlet fermion $N_1$ (even under $\mathcal{Z}_2$) and the other generations $N_2$, $N_3$ (odd under $\mathcal{Z}_2$) is forbidden. Similarly, mixing between scalars $\eta_{1}$ or $\xi_{1}$ (even under $\mathcal{Z}_2$) and the dark scalars $\eta_{2,3}$ or $\xi_{2,3}$ (odd under $\mathcal{Z}_2$) are also not allowed by this residual symmetry. As a result, the loop correction do not induce any new non-zero entries in the neutrino mass matrix; the zero entries remain protected by symmetry. The loop induced corrections affect only the non-zero elements of the tree-level mass matrix in Eq.~\eqref{eqn:TreeNu}, 
which can be effectively absorbed into a redefinition of the Yukawa couplings.  We do not explicitly include these loop contribution in our analysis. Thus, the overall structure of the mass matrix remains unchanged and only one neutrino mass can be generated at this stage.

The additional neutrino masses can be generated at the loop level from the scoto-loop mechanism. The loop contribution is given by
\begin{align} \label{eq:nuloopmass}
(\mathcal{M}_{\nu})_{ij} =\sum_{k=1}^{3} Y_{ik}Y^{\prime}_{jk}(c_2+c_3) \,,
\end{align}
where, $Y_{ik}$ and $Y^{\prime}_{jk}$ are Yukawa couplings and
\begin{align} \label{eq:cexp}
c_2&=\frac{M}{32 \pi^2 }\sin (2 \theta_2)\left[ \frac{{m^{2}_{1R}}}{{m^{2}_{1R}}- M^{2}}\ln \left(\frac{{m^{2}_{1R}}}{M^2} \right) - \frac{{m^{2}_{2R}}}{{m^{2}_{2R}}- M^{2}}\ln \left(\frac{{m^{2}_{2R}}}{M^2} \right)\right]\,, \nonumber \\
c_3&=\frac{M}{32 \pi^2 }\sin (2 \theta_3)\left[ \frac{{m^{2}_{3R}}}{{m^{2}_{3R}}- M^{2}}\ln \left(\frac{{m^{2}_{3R}}}{M^2} \right) - \frac{{m^{2}_{4R}}}{{m^{2}_{4R}}- M^{2}}\ln \left(\frac{{m^{2}_{4R}}}{M^2} \right)\right] \,.
\end{align}
The Yukawa couplings $Y_{ik}$ and $Y^{\prime}_{jk}$ are as follows
\begin{align}
& Y_{11}= y_1,  \quad  Y_{12}= y_1,  \quad  Y_{13}= y_1,  \quad Y_{21}= y_2,  \quad  Y_{22}= \omega y_2, \quad Y_{23}= \omega^2 y_2\,,    \nonumber \\
&Y_{31}= y_3,  \quad  Y_{32}= \omega^2 y_3,  \quad      Y_{33}= \omega y_3\,,  \nonumber \\
&Y^{\prime}_{11}= y^{\prime}_1,  \quad  Y^{\prime}_{12}= y^{\prime}_1,  \quad  Y^{\prime}_{13}= y^{\prime}_1,  \quad Y^{\prime}_{21}= y^{\prime}_2,  \quad  Y^{\prime}_{22}= \omega y^{\prime}_2, \quad Y^{\prime}_{23}= \omega^2 y^{\prime}_2\,.
\end{align}
Hence, from the loop contribution, the neutrino mass matrix elements are given by
\begin{align}
&(\mathcal{M}_\nu)_{11}=  y_1 y'_1\left( c_2 + c_3 \right),  \quad   (\mathcal{M}_\nu)_{12}= y_1y'_2\left( \omega c_2 + \omega^2 c_3 \right),  \quad    (\mathcal{M}_\nu)_{13}= 0\,,
\nonumber \\
&(\mathcal{M}_\nu)_{21}= y_2y'_1\left(\omega c_2 + \omega^2 c_3 \right),  \quad  (\mathcal{M}_\nu)_{22}=  y_2y'_2\left(\omega^2 c_2 + \omega c_3 \right),  \quad  (\mathcal{M}_\nu)_{23}=  0\,, 
\nonumber \\
&(\mathcal{M}_\nu)_{31}= y_3y'_1\left(\omega^2 c_2 + \omega c_3 \right),  \quad  (\mathcal{M}_\nu)_{32}= y_3y'_2\left( c_2 + c_3 \right),  \quad   (\mathcal{M}_\nu)_{33}= 0\,.    
\end{align}
Thus, the light neutrino mass matrix arising from the scoto-loop level can be expressed as
\begin{equation}
m^{(2)}_{\nu}=
\begin{pmatrix}
y_1y'_1 d_1  & y_1y'_2d_2 & 0  \\
y_2y'_1d_2 & y_2 y'_2d_3 & 0 \\
y_3y'_1d_3 & y_3y'_2d_1 & 0 \\
\end{pmatrix}\,.
\label{eqn:loopnu}
\end{equation}
where, $ d_1  = c_2+c_3$,  $d_2 =  \omega c_2+\omega^2 c_3, $ and $d_3 =  \omega^2 c_2+\omega c_3$. Combining the contributions originating from both the tree-level seesaw and the scoto-loop,  known as the ``scoto-seesaw" scenario, yields the expression for the light neutrino mass matrix as follows:
\begin{equation} \label{eq:numat}
m^{(TOT)}_{\nu} = m^{(1)}_{\nu} +m^{(2)}_{\nu} \equiv 
\begin{pmatrix}
y_1y'_1 \left( d_1-\frac{v_2u}{\sqrt{2}M} \right)  & y_1y'_2\left( d_2-\frac{v_2u}{\sqrt{2}M} \right) & 0 \\
y_2y'_1\left( d_2-\frac{v_2u}{\sqrt{2}M} \right) & y_2 y'_2\left( d_3-\frac{v_2u}{\sqrt{2}M} \right) & 0 \\
y_3y'_1\left( d_3-\frac{v_2u}{\sqrt{2}M} \right) & y_3y'_2\left( d_1-\frac{v_2u}{\sqrt{2}M} \right) & 0 \\
\end{pmatrix} \;.
\end{equation}
In the matrix $m^{(TOT)}_{\nu}$, $d_1$ is a real parameter whereas $d_2$ and $d_3$ are complex and conjugate to each other due to the presence of $\omega$. Due to the $U(1)_{B-L}$ charge assignment (see Tab.~\ref{tab:FieldCharge}), $\nu_{R_3}$ has been decoupled.  Therefore, it is evident that within our model, only two neutrinos acquire mass, while one remains massless. This finding remains consistent with the neutrino oscillation data.
\section{Neutrino Sector Predictions} \label{sec:nusecres}
In this section, we discuss the numerical results concerning the neutrino sector. We start by analyzing the features of neutrino mass matrices in our model. The tree-level neutrino mass matrix given in Eq.~\eqref{eqn:TreeNu} is proportional to the factor $ u/M$. In this model, $\xi$ being a singlet of $SU(2)_L$, its VEV $\langle \xi \rangle =u$ can be chosen arbitrarily. Note that in the type-I seesaw mechanism, right-handed neutrino masses are typically around $\sim 10^{12}$ GeV for $\mathcal{O}(1)$ Yukawa couplings. In contrast, within this framework, the freedom to choose the value of $u$ allows us to reduce the fermion mass $M$ to a desired value while maintaining a fixed $u/M$ ratio.
This is similar to the inverse seesaw mass mechanism~\cite{Mohapatra:1986bd}, where a small fermion mass is achieved by introducing a small $\mu$ parameter, often referred to as the lepton number violating parameter. In our framework, this scenario is realized by inducing a VEV for $\xi$, leading us to refer to it as the ``\textit{inverse seesaw}'' mass mechanism. As discussed earlier, the two mass-squared differences, $\Delta m^2_{\rm sol}$ and $\Delta m^2_{\rm atm}$, are generated at the loop level (via the scotogenic mechanism) and tree level, respectively. Due to the $A_4$ symmetry in our model, both the loop and tree level contributions involve the same Yukawa couplings. For $\mathcal{O}(1)$ Yukawa couplings, the mass parameter $M$ is bounded from above (typically $\sim$ a few TeV), which also constrains the VEV $u$ (to $\sim$ 100 GeV) to simultaneously satisfy both the mass-squared differences. This upper bound on $u$ has direct implications for the scalar mass spectrum, as illustrated in Fig.~\ref{fig:masslim}, and significantly affects the ``dark sector" by restricting the parameter space available for viable DM candidates (see Sec.~\ref{sec:dm}).

We now turn to the numerical analysis of the neutrino sector. For our analysis, we consider two benchmark scenarios:
\begin{align}\label{eq:benchmark}
& {\rm Case-I:}\, y_1 = y_1^\prime\, {\rm and}\, y_2 = y_2^\prime\;,\nonumber \\
& {\rm Case-II:}\,  y_3 = \epsilon\, y_2, \epsilon = (0, 1)\,. 
 \end{align}
It is worth noting that our analysis is consistent with the latest global fit data~\cite{Capozzi:2021fjo, deSalas:2020pgw, Esteban:2020cvm}. In our model, one of the neutrinos is massless. Once the constraints from the two mass-squared differences are imposed, the neutrino mass spectrum becomes fully determined for both normal ordering (NO) and inverted ordering (IO). This prediction of a massless neutrino also leads to precise predictions for other neutrino sector observables in both NO and IO cases, as we discuss below.
\subsection{Normal Ordering}
For NO, $m_3>m_2>m_1$ and as the  $m_1 = 0$ in our scenario, we find $m_3$ and $m_2$ using global-fit data~\cite{Capozzi:2021fjo, deSalas:2020pgw, Esteban:2020cvm} as
\begin{align}
     &\Delta m^2_{31}=2.55 \times 10^{-3}\ \rm{eV^2}, \quad \Delta m^2_{21}=7.5 \times 10^{-5}\ \rm{eV^2}\,, \nonumber \\
     \implies& \hspace{1cm} m_3=5.05 \times 10^{-2}\ \rm{eV}, \quad  m_2= 8.66 \times 10^{-3}\ \rm{eV} \,.
\end{align}
Sum of neutrino masses confined to a narrow range, $\sum m_i \in \left[0.058, 0.061 \right] \rm{eV}$, which we obtain by varying $\Delta m^2_{i j}$ into their 3$\sigma$ ranges~\cite{Capozzi:2021fjo, deSalas:2020pgw, Esteban:2020cvm}. This value is consistent with the Planck data~\cite{Planck:2018vyg} as well as with the recent DESI  result~\cite{DESI:2024mwx}. 
\begin{figure}[!h]
\centering
       \includegraphics[width=0.5\textwidth]{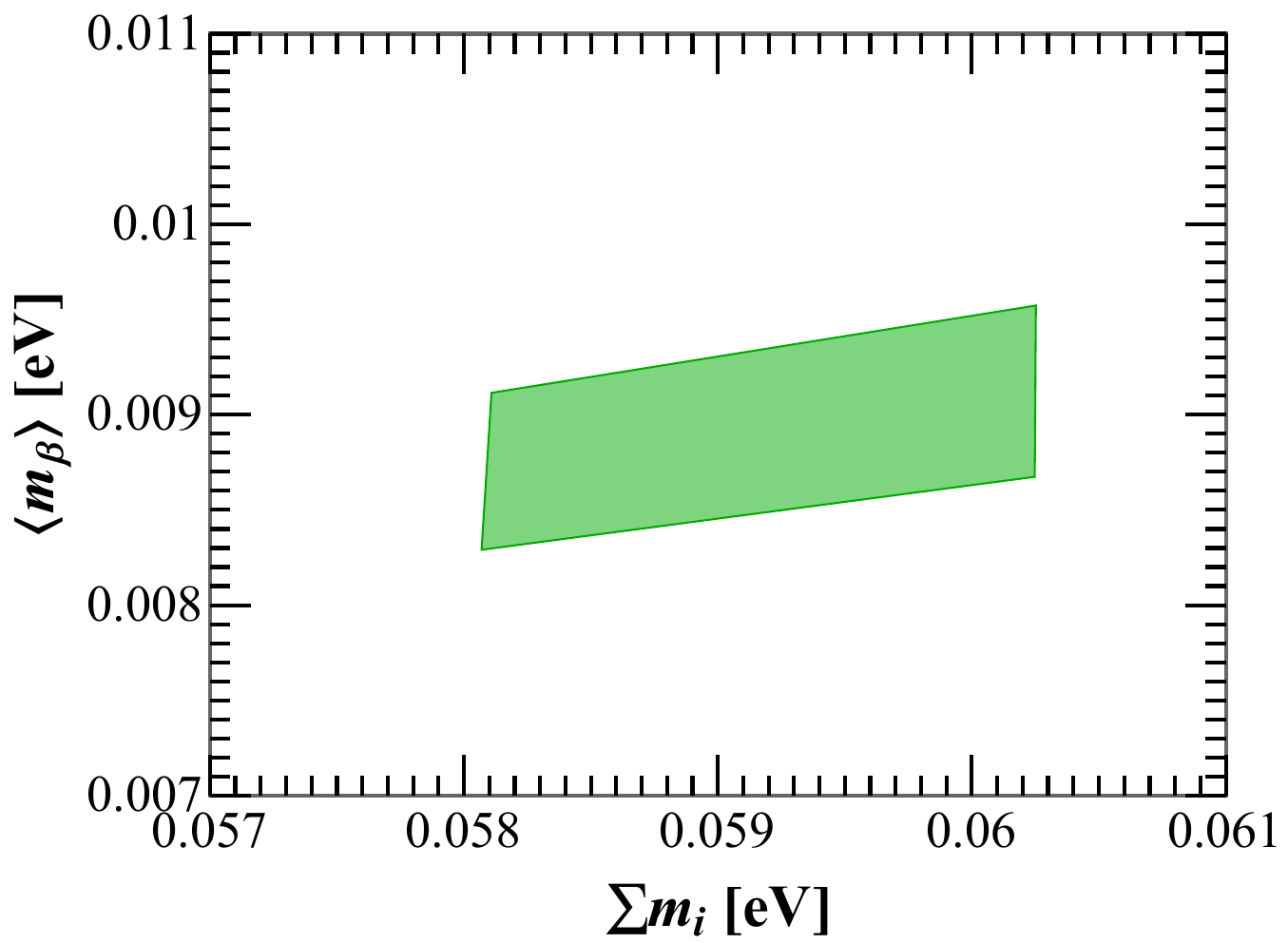}
    \caption{\footnotesize Correlation between sum of neutrino masses $\sum m_i$ and effective mass $\langle m_{\beta} \rangle$ of beta decay in the NO case.}
    \label{fig:nombetasum}
\end{figure}
Furthermore, the effective neutrino mass $\langle m_{\beta} \rangle$ is defined as:
\begin{align} \label{eq:me}
 \langle m_{\beta}\rangle \equiv \sqrt {\sum_{j}  |U_{ej} |^2 m^2_j}  = \sqrt{c^2_{12}c^2_{13}m^2_1 + s^2_{12}c^2_{13} m^2_2 +s^2_{13}m^2_3  }\;.
\end{align}
where $c_{ij}=\cos \theta_{ij}$ and $s_{ij}=\sin \theta_{ij}$ and that
falls in the range $\langle m_{\beta} \rangle \in \left[8.26, 9.57  \right] \rm{meV} $. This value is consistent with the expected sensitivity of KATRIN~\cite{KATRIN:2021uub}. In addition to the constrained values of both $\sum m_i$ and $\langle m_{\beta} \rangle$, they exhibit a correlation with each other as shown in Fig.~\ref{fig:nombetasum}. Predictions for the CP-violating phase \(\delta_{\rm{CP}}\) and its correlation with the atmospheric mixing angle \(\theta_{23}\) are displayed in the left panel of Fig.~\ref{fig:Neutrino1} for case-I, while the prediction for case-II is shown in the right panel. For comparison, the latest global analysis of neutrino oscillation data~\cite{deSalas:2020pgw} is included, represented by the cyan-colored contours in Fig.~\ref{fig:Neutrino1}, with the best-fit value indicated by a black dot.

\begin{figure}[!h]
\includegraphics[height=6.35cm]{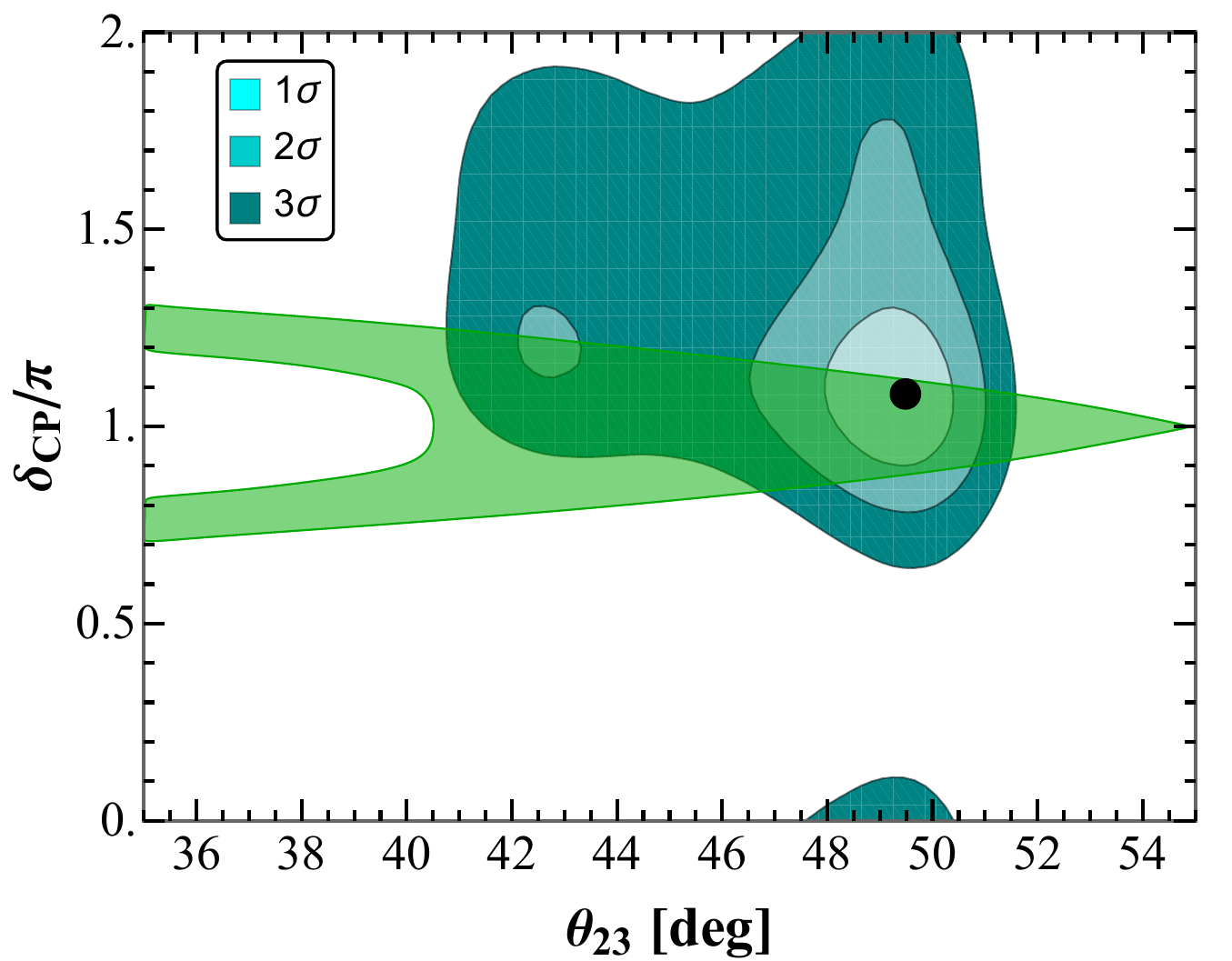}
        \includegraphics[height=6.35cm]{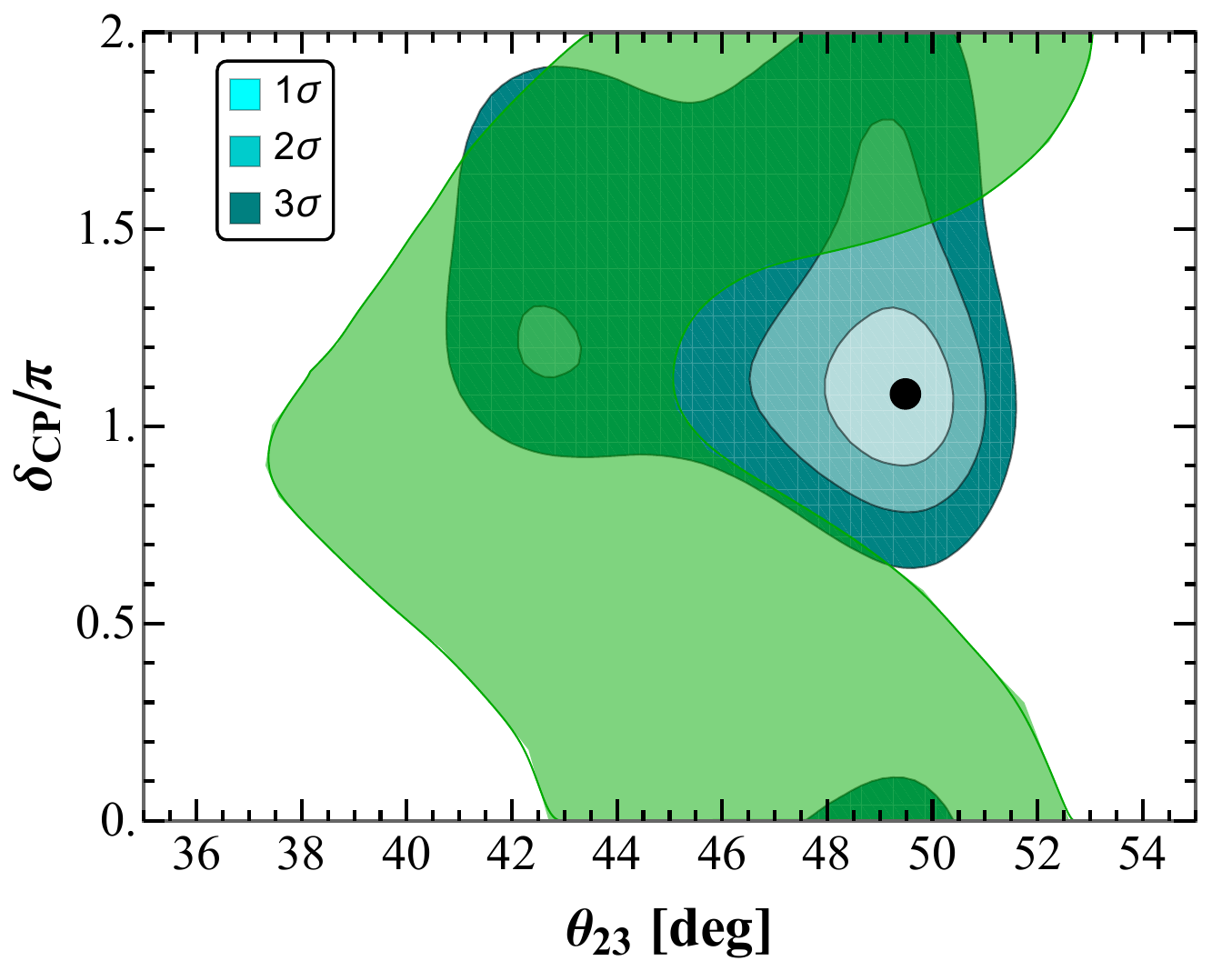}
        \caption{\centering \footnotesize Correlation between $\delta_{CP}$ and $\theta_{23}$ for NO, where left (right)-panel represents case-I (-II).}
                \label{fig:Neutrino1}
\end{figure}
From case-I, we observe a sharp and well-defined correlation between the CP-violating phase $\delta_{\rm{CP}}$ and the atmospheric mixing angle $\theta_{23}$, with predictions clustering closely around the global best-fit value, i.e., $\delta_{\rm{CP}} \sim \pi$. Interestingly, this correlation exhibits a bifurcated structure, with two distinct branches centered near $\delta_{\rm{CP}} = \pi$. Notably, this case excludes CP-conserving values ($\delta_{\rm{CP}} = \pi$) for a portion of the lower octant, specifically for $\theta_{23} \leq 42^\circ$. Furthermore, the latest global analysis of neutrino oscillation data disfavors the model predictions for $\theta_{23} \lesssim 41^\circ$ and $\theta_{23} \gtrsim 51^\circ$. In contrast, for case-II, the model predictions are found to be incompatible with the global-fit data at the $1\,\sigma$ confidence level. In particular, the CP-violating phase values in the range $0.1\pi \lesssim \delta_{\rm{CP}} \lesssim 0.7\pi$ are disfavored even at the $3\,\sigma$ confidence level. However, the model remains viable at the $2\,\sigma$ confidence level when the  CP values lie in the second or the third quadrants, offering a window for experimental verification in future neutrino oscillation studies.
\subsection{Inverted Ordering}
\begin{figure}[h!]
\centering
       \includegraphics[width=0.49\textwidth]{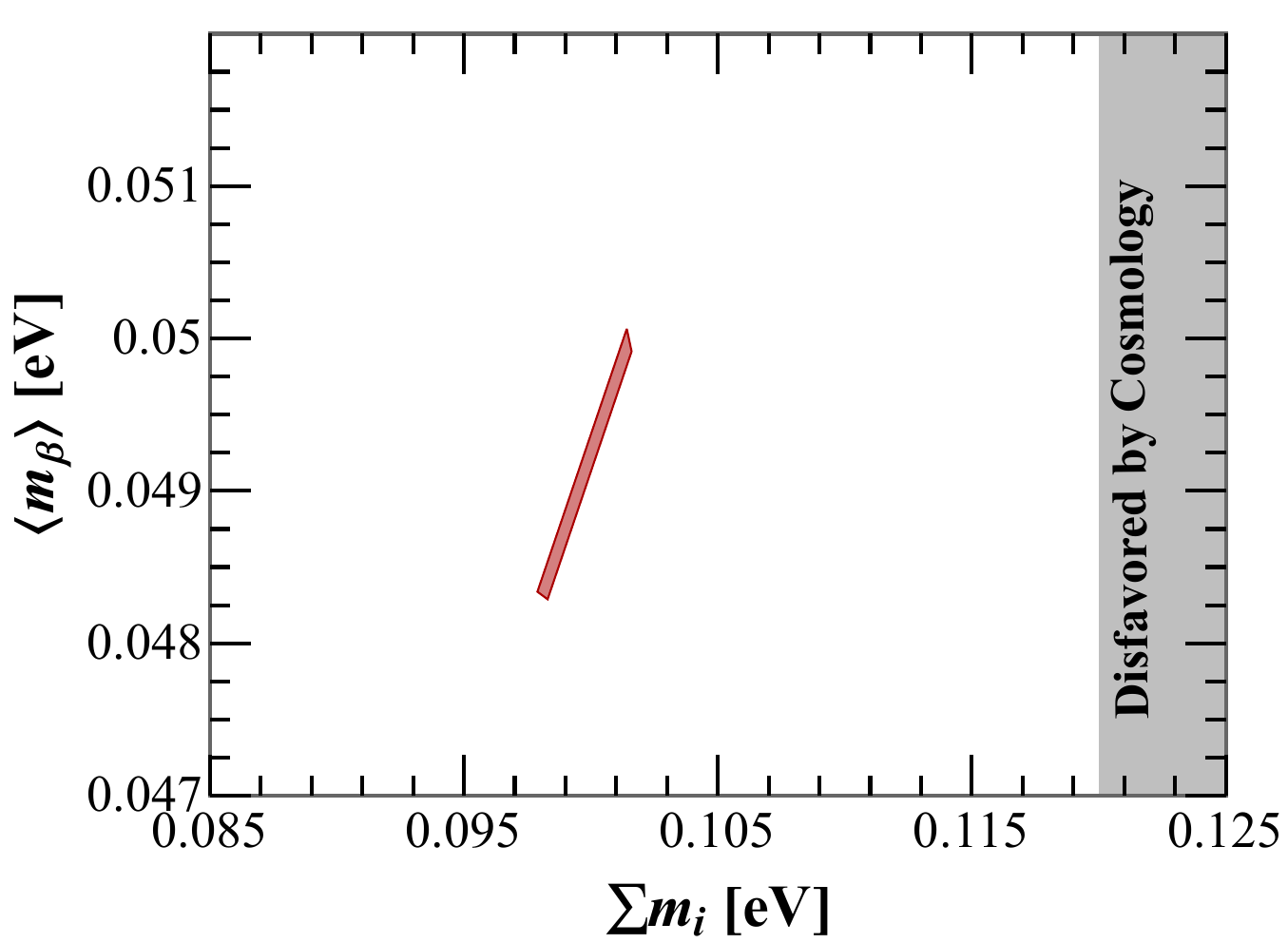}
       \includegraphics[width=0.49\textwidth]{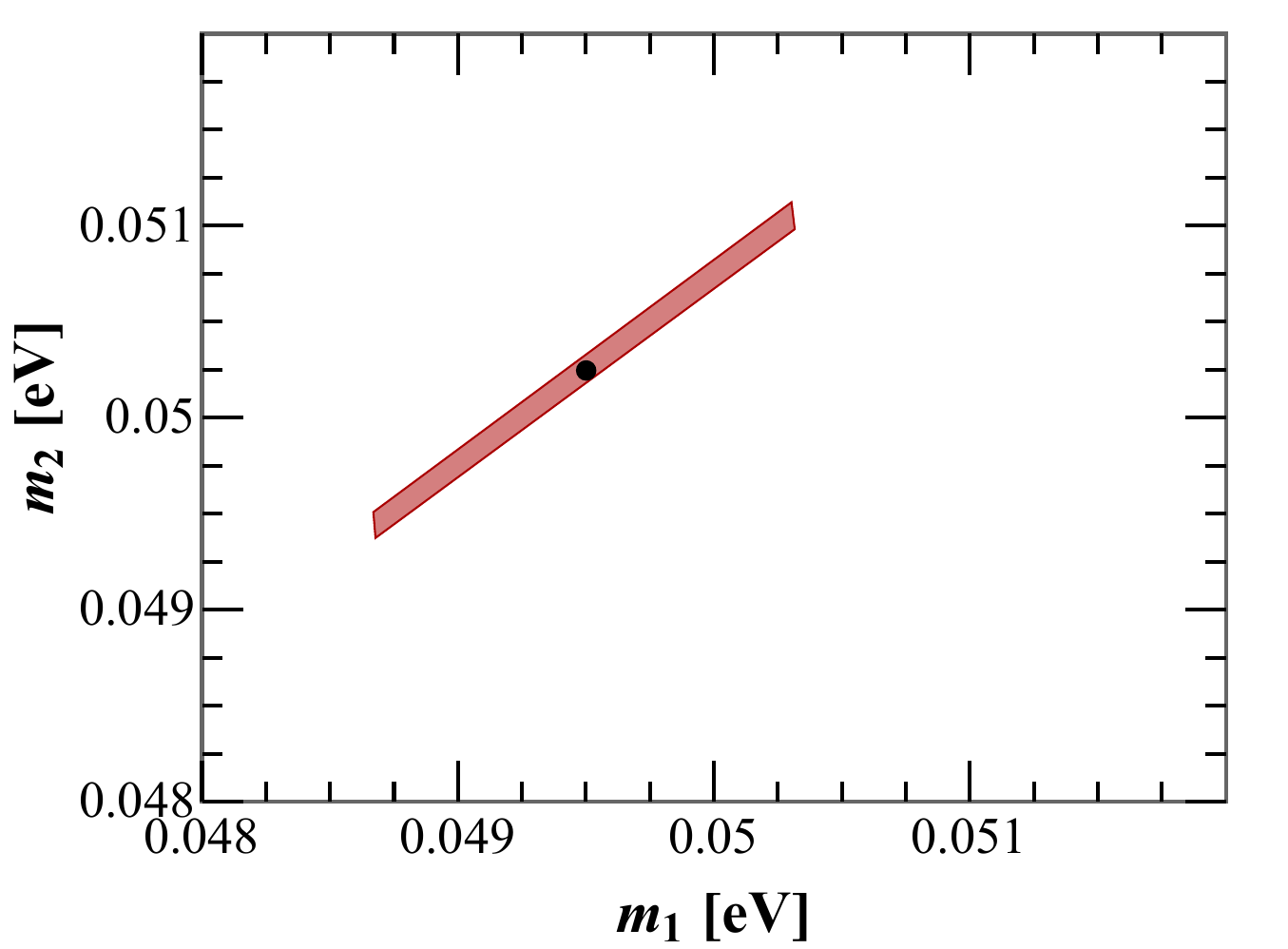}
    \caption{\footnotesize Model Predictions in the IO case. \textbf{Left panel}: Correlation between $\sum m_i$ and $\langle m_{\beta} \rangle$. The gray band depicts the cosmology limit. \textbf{Right panel:} Correlation between neutrino masses, $m_1$ and $m_2$. }
    \label{fig:io}
\end{figure}
For IO, it is noteworthy that, unlike the Majorana case~\cite{Kumar:2024zfb}, where IO was forbidden, in this Dirac neutrino formalism, IO is permitted and remains consistent with the observed neutrino oscillation data. As in IO, neutrino mass ordering follows $m_2>m_1>m_3$, hence the lightest one, $m_3=0$ for this case. Imposing $\Delta m^2_{i j}$ constraint, we find $m_1$ and $m_2$ are as follows: 
\begin{align}
     &\Delta m^2_{31}=-2.45 \times 10^{-3}\ \rm{eV^2}, \quad \Delta m^2_{21}=7.5 \times 10^{-5}\ \rm{eV^2}\,, \nonumber \\
     \implies& \hspace{1cm} m_1=4.950 \times 10^{-2} \ \rm{eV}, \quad  m_2= 5.025 \times 10^{-2}\ \rm{eV}\,.
\end{align}
In this case, we also find stringent range values for the sum of neutrino masses $\sum m_i$ and the effective mass of the beta decay $\langle m_{\beta} \rangle$ and are given as follows:
\begin{align*}
    \sum m_i \in \left[ 0.098, 0.101\right] \rm{eV}, \quad \langle m_{\beta} \rangle \in \left[0.048, 0.050  \right] \rm{eV}\;,
\end{align*}
which are consistent with the Planck results~\cite{Planck:2018vyg} and expected sensitivity of KATRIN~\cite{KATRIN:2021uub}, respectively. In the left panel of Fig.~\ref{fig:io}, we show the correlation of $\sum m_i$ and $\langle m_{\beta} \rangle$ with each other. In addition, there is a strong correlation between neutrino masses $m_1$ and $m_2$ as shown in the right panel of Fig.~\ref{fig:io}. Note that, unlike the NO, where $ m_1 = 0 \simeq m_2$, leading to no observable correlation, the IO has  $ m_3 = 0$  with  $ m_1$  and  $m_2$  being relatively large, resulting in a strong correlation between them.

\begin{figure}[h!]

        \includegraphics[height=6.35cm]{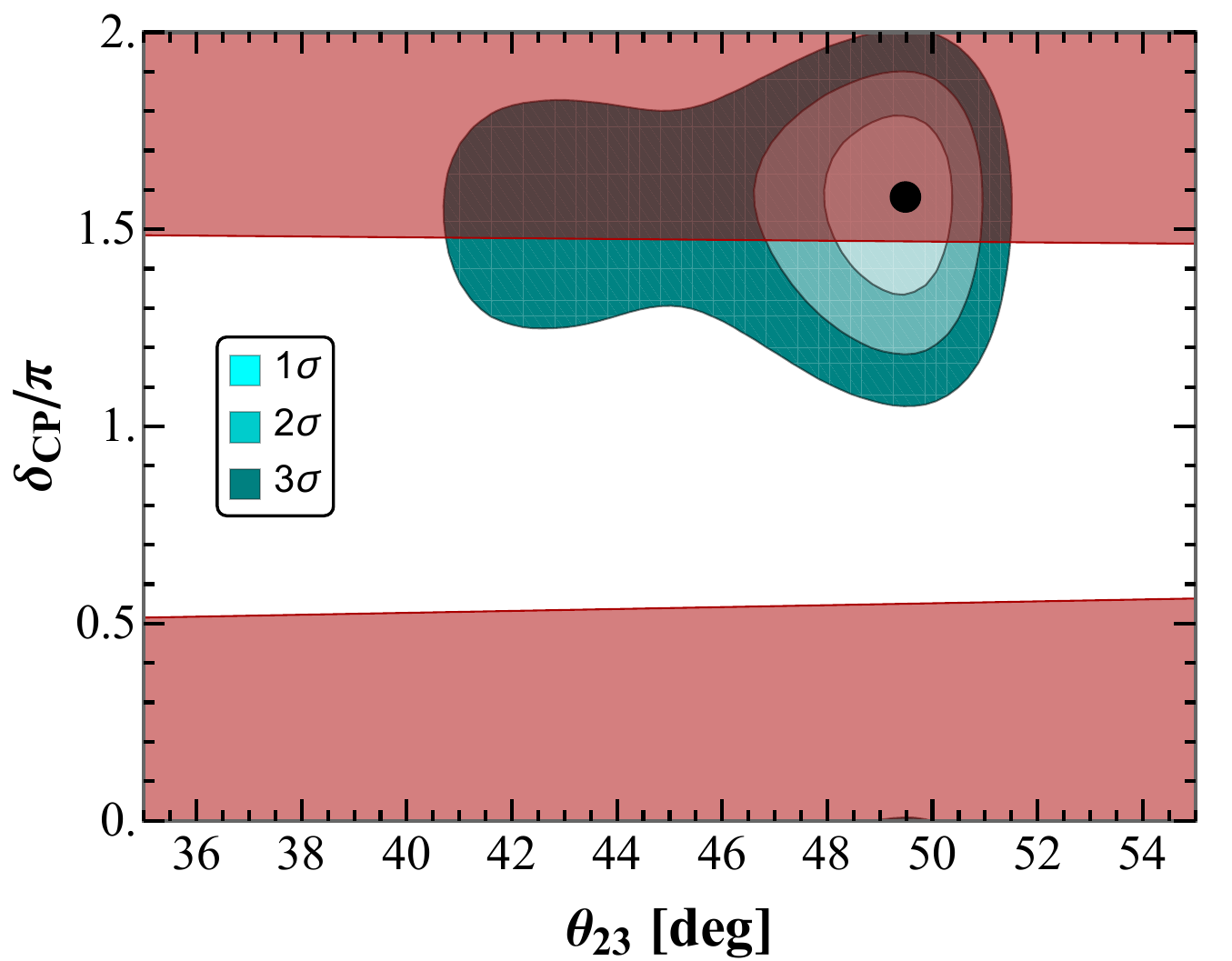}
        \includegraphics[height=6.35cm]{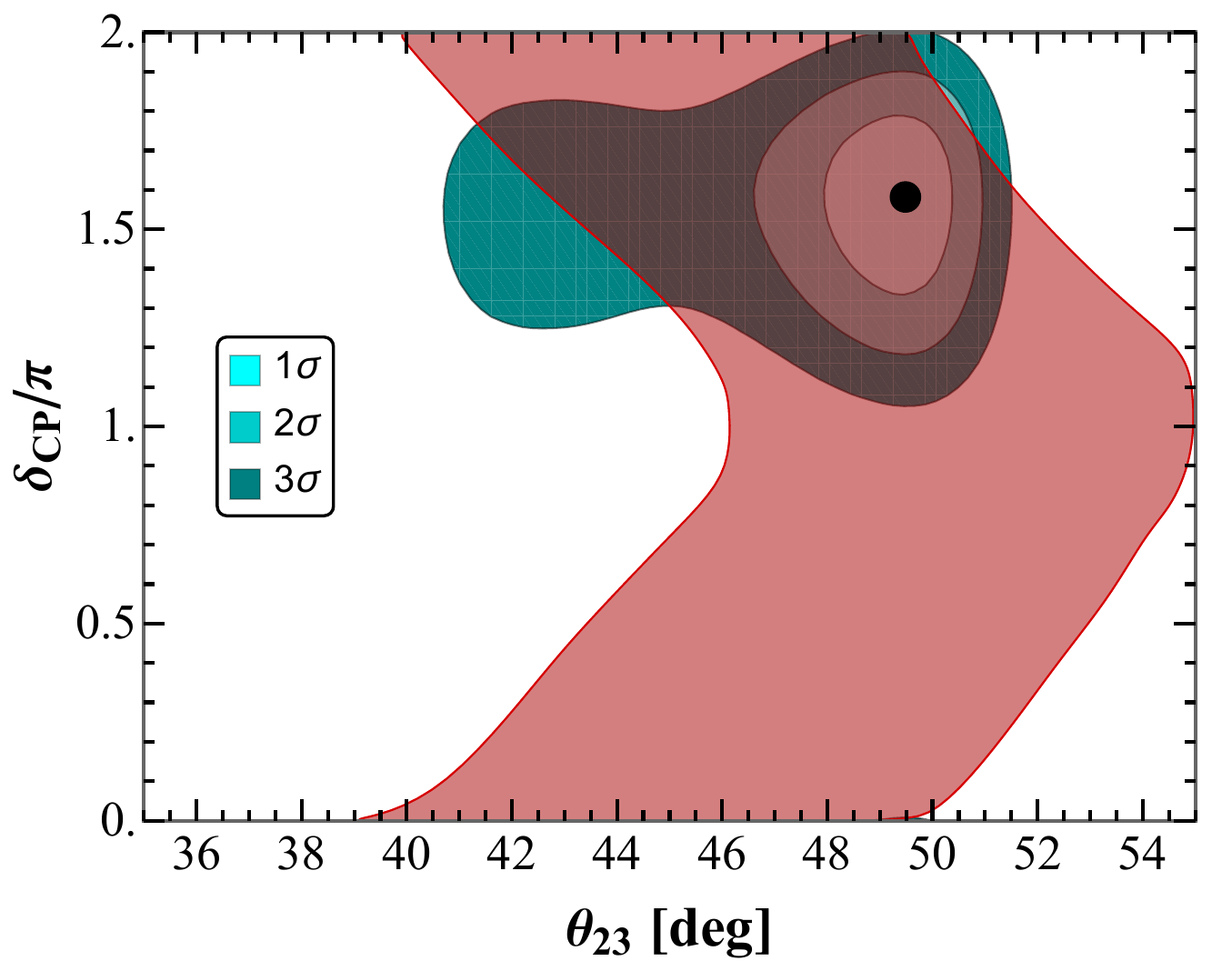}
        \caption{\centering \footnotesize Same as Fig.~\ref{fig:Neutrino1}, but for IO.}
                \label{fig:Neutrino2}
\end{figure}
In Fig.~\ref{fig:Neutrino2}, we present the correlation between $\delta_{\rm{CP}}$ and $\theta_{23}$ in the case of IO, for the two benchmark scenarios defined in Eq.~\eqref{eq:benchmark}. From the left panel, corresponding to case-I, we observe a pattern where the model predicts two distinct branches of solutions for $\delta_{\rm{CP}}$. However, one of these branches is excluded by the latest global analysis of neutrino oscillation data. The parameter space consistent with current experimental constraints can be summarized as $\theta_{23} \in [41^\circ, 51^\circ]$ and $\delta_{\rm{CP}} \in [1.5\pi, 2\pi]$. In the right panel, which shows case-II, we find that the model predicts a broader range of allowed values for $\delta_{\rm{CP}}$ compared to case-I. Specifically, the compatible region extends over $\delta_{\rm{CP}} \in [0.9\pi, 2\pi]$, while the allowed range for $\theta_{23}$ remains similar to that of case-I. This suggests that case-II is more flexible in accommodating current neutrino data within the IO scenario.
%
\section{Dark Sector Results} \label{sec:dm}
We now turn our attention to the DM phenomenology of our model. In the early universe, all the particles in the dark sector are in thermal equilibrium with the SM fields due to the production-annihilation processes depicted in App.~\ref{app:Feynman}. As the universe expands and cools, the temperature drops. For unstable particles, there is a threshold temperature below which the thermal bath lacks sufficient energy to produce them, while their annihilation and decay processes continue, leading to their eventual disappearance. However, the lightest particle of the dark sector remains stable. Once this stable particle decouples from the thermal bath, its relic density becomes `freezed out'. This relic density can be computed and compared with observations from the Planck satellite data \cite{Planck:2018vyg}. The current observed value for the DM relic density, as measured by Planck data, is $0.1126 \leq \Omega h^2 \leq 0.1246$ at the 3$\sigma$ confidence level \cite{Planck:2018vyg}.
This class of DM candidates could potentially be probed through nuclear recoil experiments such as XENONnT~\cite{XENON:2023cxc} and LZ~\cite{LZ:2022lsv}. In our model, we have both scalar and fermionic DM candidates. In the scalar sector, the viable DM candidates include the lightest neutral, predominantly doublet dark scalars ($\chi_{1,3}$), as well as the predominantly singlet dark scalars ($\chi_{2,4}$). On the other hand, the fermionic DM candidates are represented by the neutral fermions ($N_{2,3}$). The stability of the lightest dark sector particle is ensured by the residual $\mathcal{Z}_2$ symmetry.

As we have discussed in Sec.~\ref{sec:scsec}, the mass of the scalar is related to the VEV of the singlet scalar $(u)$ as shown in Fig.~\ref{fig:masslim}. From the neutrino sector analysis, we find that there is an upper bound for $u\sim 100$ GeV. This limit of VEV restricts the dark scalar mass to be around 700 GeV. Hence, in our model, the mass of the DM can not be greater than this value. Here, we would like to mention that in our numerical analysis of the dark sector as well as in LFV, we have presented the general scan and then shaded the region with the red color, which is not allowed by the neutrino oscillation constraints. 

Furthermore, we have imposed the collider constraint from LHC and LEP (LEP-I and LEP-II), which have a significant impact on the viable parameter space of the DM mass.
\textit{
\begin{itemize}
\item The most important LHC limit comes from the possibility of Higgs' ``invisible decay" to dark sector particles. The Higgs invisible decay comes through the channel $h_1 \to \chi_i \chi_j$ $(i,j=1,2,3,4)$. The current bound for the branching ratio of the Higgs invisible decay from LHC data has a limit~\cite{ATLAS:2022yvh}: 
\begin{align} \label{eq:LHC}
    BR(h_1 \to inv)<0.145.
\end{align}
\item The LEP-I measurements rule out SM gauge boson decays into dark sector particles~\cite{Cao:2007rm,Gustafsson:2007pc}. This condition leads to the following lower limit on the dark sector scalar masses:
\begin{align} \label{eq:LEP-I}
m_{\eta_{2/3}^{R/I}} + m_{\eta_{2/3}^{\pm}}> M_W, \quad m_{\eta_{2/3}^{R}} + m_{\eta_{2/3}^{I}}> M_Z, \quad m_{\eta_{2/3}^{\pm}} + m_{\eta_{2/3}^{\mp}}> M_Z \;.
\end{align}
\item Chargino searches in LEP-II in the context of singly-charged scalar production $e^+ e^- \rightarrow \eta_{2/3}^{\pm} \eta_{2/3}^{\mp}$ can be adapted to our analysis to set the limits~\cite{Pierce:2007ut} given by
\begin{align} \label{eq:LEP-II}
m_{\eta_{2/3}^{\pm}}> 70 \hspace{0.10cm} \text{\rm GeV} \; .
\end{align}
\end{itemize}}
We performed a detailed numerical scan with input parameters given in Tab.~\ref{tab:parameterrange}.

\begin{table}[!h]
\begin{center}
\begin{tabular}{| c | c | c | c |}
  \hline 
  Parameter    &   Range   &   Parameter    &   Range  \\
\hline
$\lambda_{1}$     &  	 $[10^{-4},\sqrt{4\pi}]$            &
$\lambda_{\eta_{l}}$   &  $[10^{-8},\sqrt{4\pi}]$            \\
$\lambda_{\xi_{k}}$   &   $[10^{-8},\sqrt{4\pi}]$           &
$|\lambda_{H \eta_{i} }|$   &   $[10^{-8},\sqrt{4\pi}]$   	   \\
$|\lambda_{H \xi_{m}}|$   &   $[10^{-8},\sqrt{4\pi}]$   	   &
$|\lambda_{\eta \xi_{n} }|$   &   $[10^{-8},\sqrt{4\pi}]$   	   \\
$|\kappa|$          &	     $[10^{-8},30]\text{ GeV}$       &
$u$          &	     $[10^{-4},10^{5}]\text{ GeV}$       \\
$M_{N_{i}}$ & $[10,10^{12}] \text{ GeV}$ 	    &	
$y_{i}$, $y^{\prime}_{j}$      &  	 $[10^{-6},10^{-1}]$                      \\
    \hline
  \end{tabular}
\end{center}
\caption{\footnotesize The ranges of values used in the numerical scan for the dark sector results correspond to the indices $i = 1, 2, 3$, $j = 1, 2$, $k = 1, 2, 3, 4, 5$, $l = 1, 2, 3, 4, 6, 7, 9, 10, 11, 12$, $m = 1, 2$, and $n = 1, 2, 4, 5, 8, 9, 11, 12, 14$, as defined in the scalar potential given in App.~\ref{app:pot}.}
 \label{tab:parameterrange} 
\end{table}

In App.~\ref{app:Feynman}, we present the relevant Feynman diagrams corresponding to the annihilation and co-annihilation channels that contribute to the relic density of the DM candidate. Additionally, we include diagrams associated with the direct detection (DD) prospects of the DM. For scalar DM, direct detection occurs at the tree level via Higgs and $Z$-boson exchange. In contrast, fermionic DM does not couple directly to the quarks at the tree level, and its DD interactions arise only at the loop level.
\subsection{Doublet scalar DM}
Results for the doublet scalar DM ($\chi_1$) are discussed in this section. Note that considering $\chi_3$ as a DM candidate will have the same results as the DM candidate $\chi_1$.
\begin{figure}[!h]
\centering
        \includegraphics[height=5.45cm]{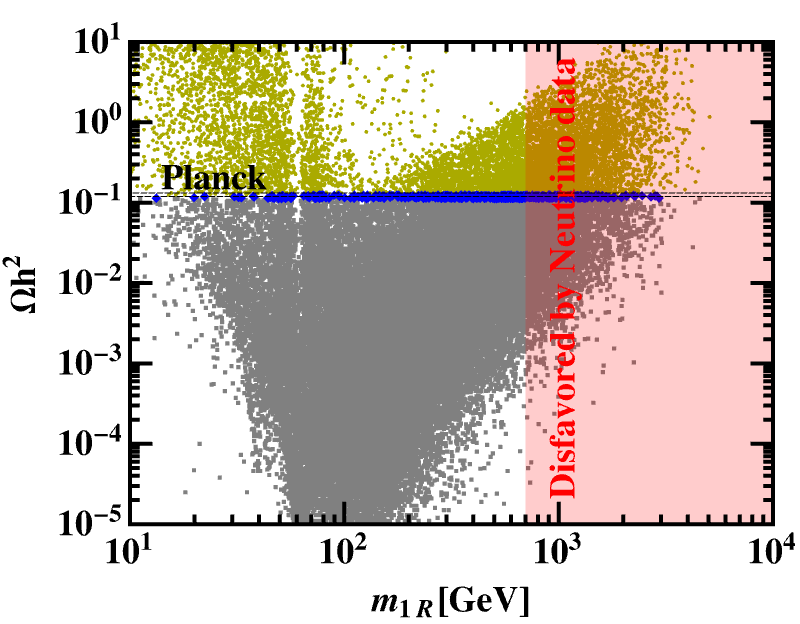}
        \includegraphics[height=5.45cm]{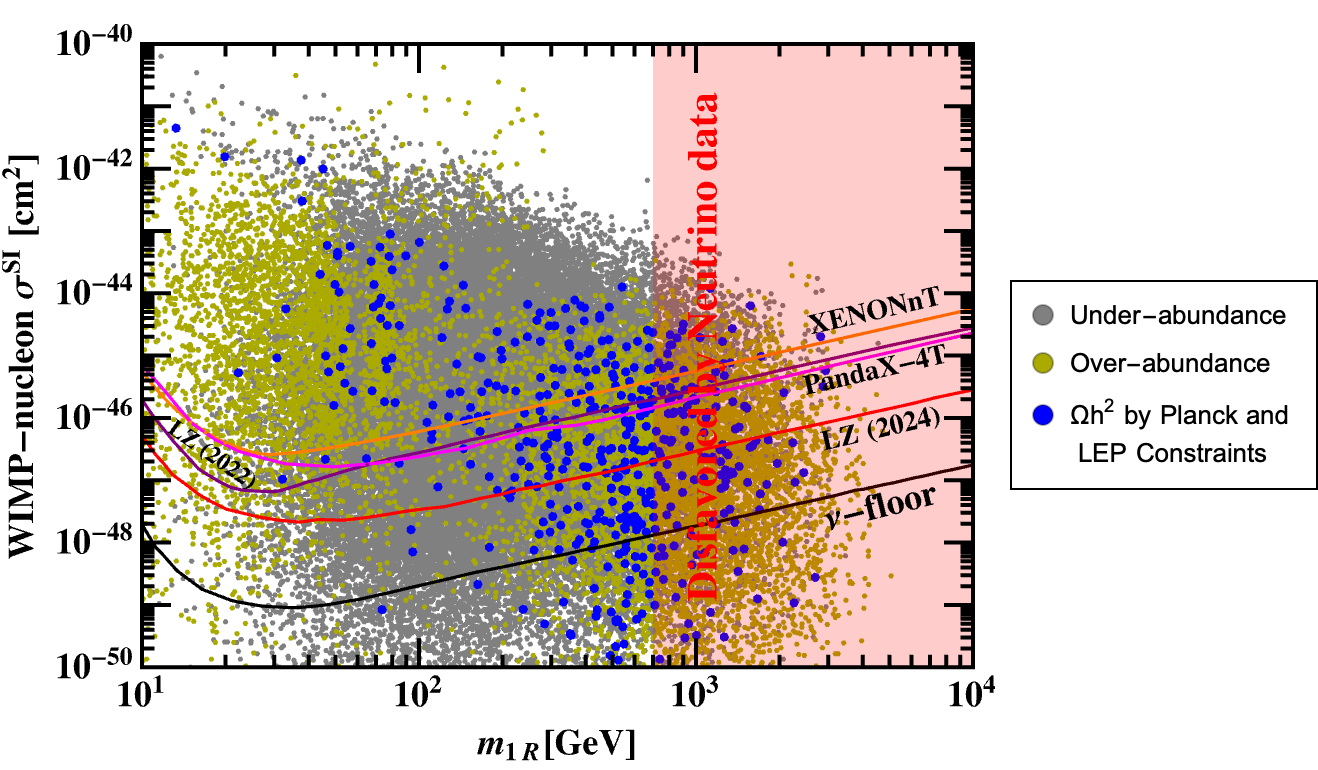}
        \caption{\footnotesize Predictions for the doublet DM candidate $\chi_1$. The left panel shows relic density and the right panel shows WIMP-nucleon cross section vs doublet scalar DM mass $m_{1R}$. The red shaded region is constrained by neutrino oscillation data.}
        \label{fig:doubletDM}
\end{figure}
In the left panel of Fig.~\ref{fig:doubletDM}, the dependence of the DM relic density on the mass of the doublet DM $m_{1R}$ has been shown. The blue points represent the $3 \sigma$ range of the relic density for doublet DM as measured by the Planck collaboration. The olive green and gray points represent the over-abundant and under-abundant relic density, respectively. Our analysis reveals that the blue points span a wide mass range from 10 GeV to approximately 3 TeV. The right side of Fig.~\ref{fig:doubletDM} illustrates the relationship between the spin-independent WIMP-nucleon cross-section and the mass of the DM particle. Various colored lines in Fig.~\ref{fig:doubletDM} denote the latest DD upper bounds from the XENONnT collaboration~\cite{XENON:2023cxc}, 
 the LZ collaboration~\cite{LZ:2022lsv,LZ:2024zvo}, and the PandaX-4T collaboration~\cite{PandaX:2024qfu}. The black line signifies the lower limit corresponding to the ``neutrino floor'' arising from coherent elastic neutrino scattering \cite{Billard:2013qya}. Observing the right panel of Fig.~\ref{fig:doubletDM}, one can deduce that the blue points positioned below the constraints of LZ  are considered permissible based on considerations of relic density, LHC, LEP, and DD constraints. Apart from these constraints, the DM mass has an upper bound $\sim$ 700 GeV (see Fig.~\ref{fig:masslim}). Note that we pointed out in Fig.~\ref{fig:masslim} that neutrino data could not be explained for DM mass greater than 700 GeV, which we show by the red shaded region. Therefore, the shaded region is ruled out for DM by neutrino oscillation constraints. Moreover, the mass region up to around 85 GeV is excluded by collider constraints from the LHC and LEP. Consequently, the doublet DM mass region is confined in the range $85 \leq m_{1R} \leq 700$ GeV, satisfying neutrino, DM, as well as collider constraints.
\subsection{Singlet scalar DM}
We now consider the scenario in which the DM is primarily composed of the $SU(2)_L$ singlet scalar, identified as $\chi_{2,4}$. In the following, we present our results for the DM candidate $\chi_2$, representing the singlet scalar case, noting that the analysis and results for $\chi_4$ are qualitatively similar.
\begin{figure}[th]
\centering
        \includegraphics[height=5.45cm]{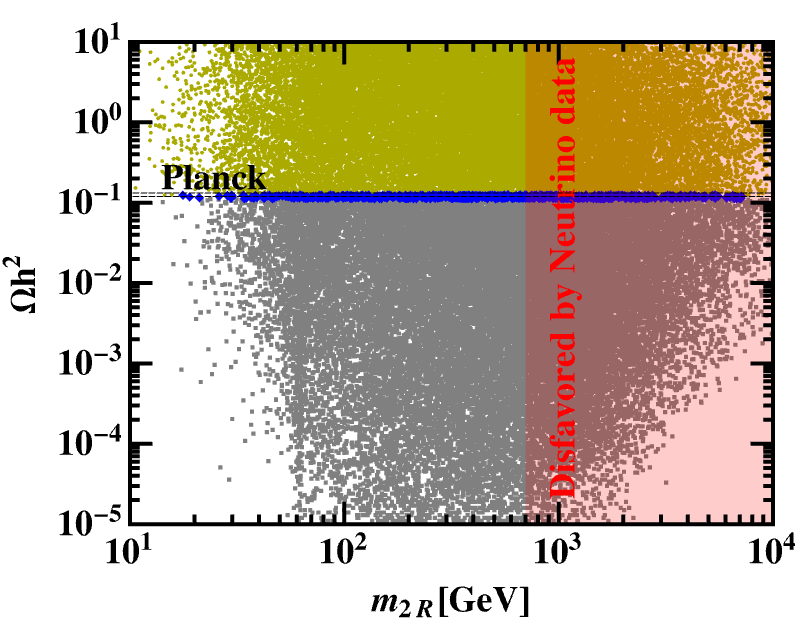}
        \includegraphics[height=5.45cm]{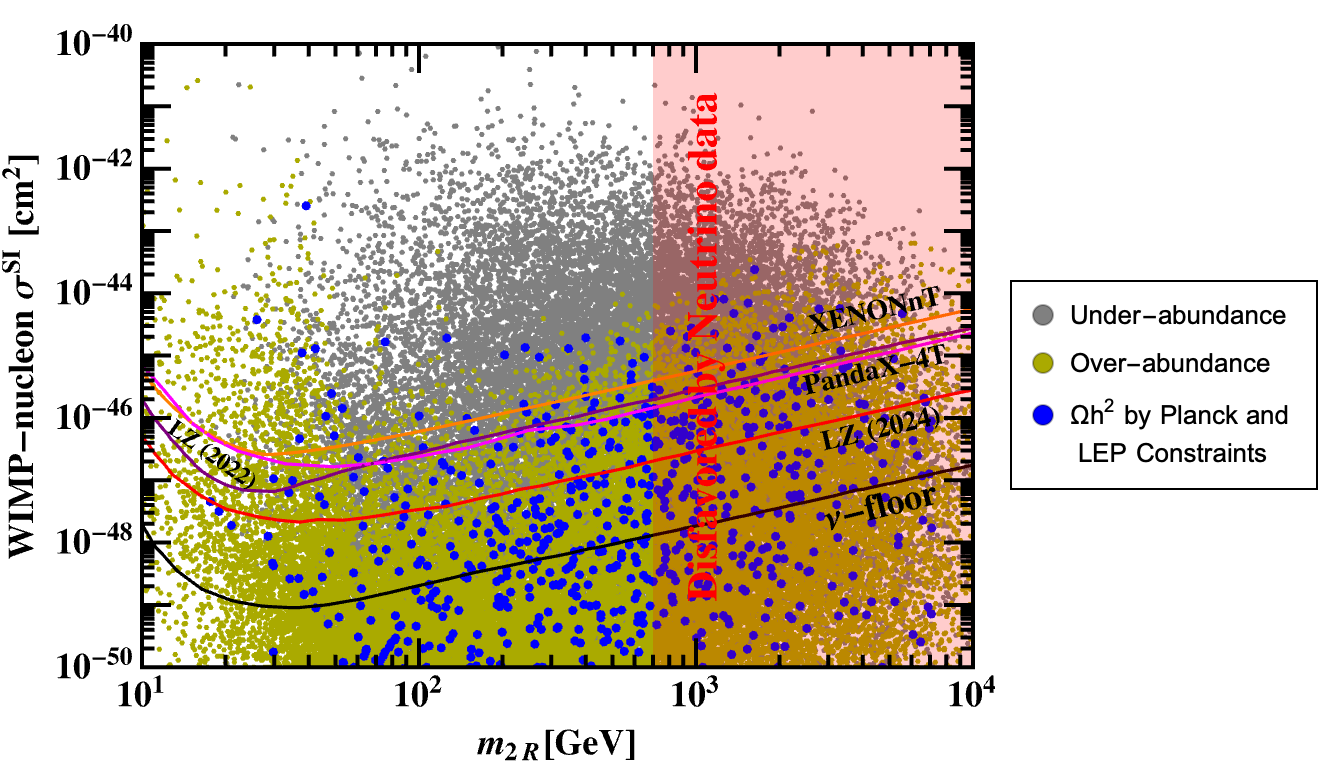}
        \caption{\footnotesize Predictions for the singlet scalar DM candidate $\chi_2$. The left panel shows relic density and the right panel shows WIMP-nucleon cross section vs singlet scalar DM mass $m_{2R}$.}
        \label{fig:singletDM}
\end{figure}
The left panel of Fig.~\ref{fig:singletDM} shows the dependence of the DM relic density on the DM mass $m_{2R}$. The color code remains the same as for the doublet scalar DM case. In our analysis, we observe that for the singlet DM candidate, the correct relic density points range from 20 GeV to 10 TeV of the DM mass. Hence, again constraining them from neutrino oscillation, DM constraints, and collider constraints, the allowed mass region for singlet DM is $20\leq m_{2R}\leq 700$ GeV.
\subsection{Fermionic DM}
\begin{figure}[h!]
\centering
        \includegraphics[height=7.5cm]{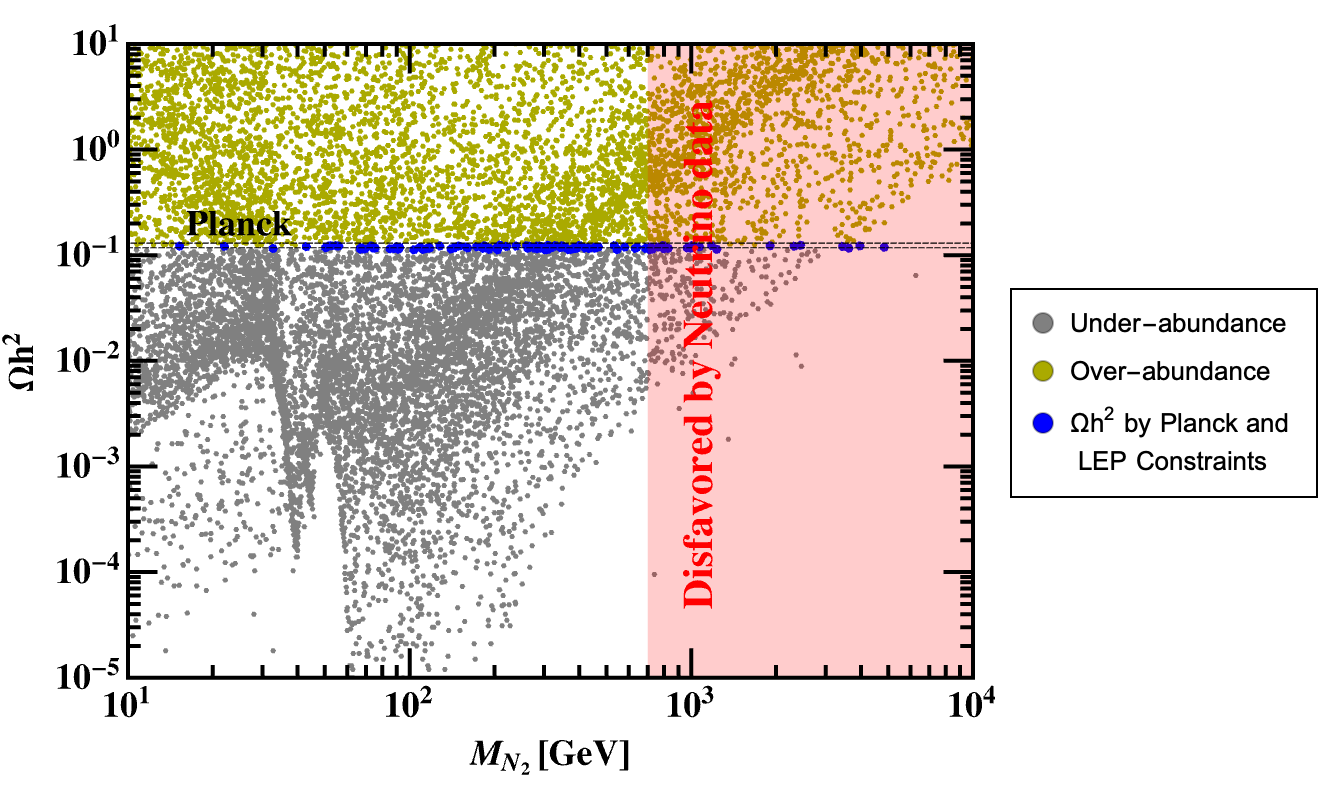}
        \caption{\centering \footnotesize Relic density as a function of fermionic DM mass $M_{N_2}$ has been presented.}
        \label{fig:fermionDM}
\end{figure}

The results for the fermionic DM are presented in Fig.~\ref{fig:fermionDM}, demonstrating that the fermionic DM is consistent with all current experimental constraints.
When the mass of the dark fermion is significantly smaller than that of the neutral doublet and singlet scalar, co-annihilation channels contribute negligibly to the computation of relic density. Consequently, the relic density is predominantly determined by fermion annihilation channels alone, leading to a higher relic density. This behavior can be understood by looking at Fig. 8 of the Ref.~\cite{Kumar:2024zfb}. To achieve the correct relic density, it is crucial to account for co-annihilation channels as well.

Importantly, it is observed that for the fermionic DM, the limitations from DD experiments such as XENONnT, LZ, and PandaX-4T are not significant. This is attributed to the fact that the fermionic DM candidate has no direct interaction with quarks at the tree level. It is noteworthy that, within the parameter space explored, the fermionic DM candidate $N_{2}$ with a mass range between 10 GeV to 5 TeV concurrently satisfies all the previously discussed constraints. Again, implementing the neutrino oscillation constraints, this region will be further narrowed down to the $10 \leq M_{N_2}\leq 700$ GeV. 

\section{Charged Lepton Flavor violation} \label{sec:lfv}
\begin{figure}[h!]
\includegraphics[width=8cm]{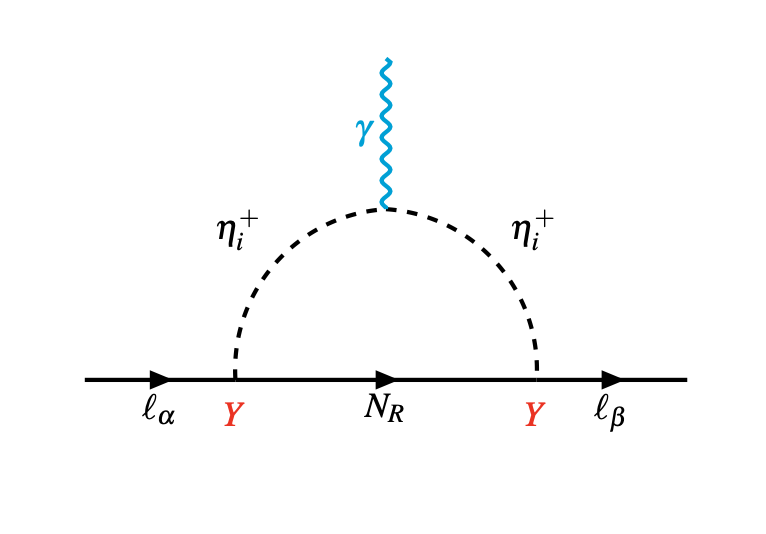}
\caption{\centering \footnotesize One loop feynman diagram for the process $\ell_{\alpha} \rightarrow \ell_{\beta} \gamma$. }
        \label{fig:LFV}
\end{figure}

The discovery of neutrino oscillations provides compelling evidence for the existence of charged lepton flavor violating interactions \cite{Gonzalez-Garcia:2002bkq}. In the SM, if neutrino masses originate similarly to those of other fermions (via Yukawa interactions with the SM Higgs), charged lepton flavor violating (CLFV) rates are expected to be extremely suppressed to practically unobservable levels. For instance, the branching fraction for $\mu \rightarrow e \gamma$ decay is estimated to be of the order of $10^{-54}$, significantly below the sensitivity threshold of any feasible experiment. However, alternative mechanisms for neutrino mass generation could yield to potentially significant CLFV effects. Extensive experimental efforts have been made to explore flavor-violating decays of muons and tau leptons. Yet, as of the present date, no conclusive positive signals have emerged from these experiments. Consequently, the absence of such signals has prompted the establishment of stringent bounds on the branching ratios associated with these flavor-violating decay processes. 
Presently, these limits stand at $BR(\mu \rightarrow e \gamma) < 3.1 \times 10^{-13}$ \cite{MEGII:2023ltw}, $BR(\tau \rightarrow e \gamma) < 3.3 \times 10^{-8}$ \cite{BaBar:2009hkt}, and $BR(\tau \rightarrow \mu \gamma) < 4.2 \times 10^{-8}$ \cite{Belle:2021ysv}, respectively. Projections indicate significant enhancements in the near future for detecting these decay processes, with anticipated limits of $BR(\mu \rightarrow e \gamma) < 6 \times 10^{-14}$ \cite{Baldini:2013ke}, $BR(\tau \rightarrow e \gamma) < 9 \times 10^{-9}$ \cite{Belle-II:2022cgf}, and $BR(\tau \rightarrow \mu \gamma) < 6.9 \times 10^{-9}$ \cite{Belle-II:2022cgf}, respectively. 
The leading contribution to $\ell_{\alpha} \rightarrow \ell_{\beta} \gamma$ arises at the one-loop level through the mediation of the charged scalar $\eta^+_{i}$ as shown in the diagram of Fig.~\ref{fig:LFV}. 
The branching ratio of this process is given by 
\begin{equation}   
\operatorname{Br}(\ell_{\alpha} \rightarrow \ell_{\beta} \gamma)=\operatorname{Br}\left(\ell_{\alpha} \rightarrow \ell_{\beta} \nu_{\alpha} \overline{\nu_{\beta}}\right) \times \frac{3 \alpha_{e m}}{16 \pi G_{F}^{2}}\left|\sum_{i} \frac{Y_{\beta i}Y^{*}_{\alpha i}}{m_{\eta^{+}_{i}}^{2}} j\left(\frac{M_{N_{i}}^{2}}{m_{\eta^{+}_{i}}^{2}}\right)\right|^{2} .
\label{eqn:cLFVeqn}
\end{equation}
We take the numerical  values of $\operatorname{Br}\left(\ell_{\alpha} \rightarrow \ell_{\beta} \nu_{\alpha} \overline{\nu_{\beta}}\right)$, $\alpha_{em}$ and $G_F$ from the PDG \cite{ParticleDataGroup:2022pth} and $j(x)$ is the loop function defined as
\begin{equation}  
j(x)=\frac{1-6x+3x^{2}+2x^{3}-6x^{2}\log(x)}{12(1-x)^{4}}\,.
\end{equation}

\begin{figure}[h!]
\centering
        \includegraphics[height=4.15cm]{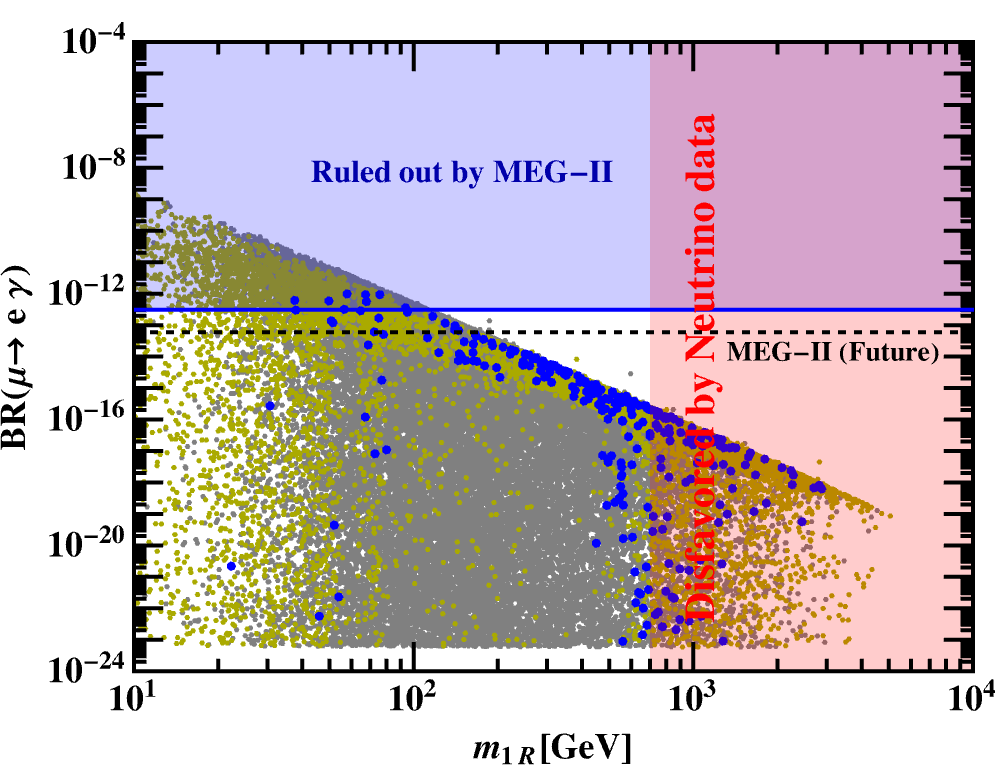}
    \includegraphics[height=4.15cm]{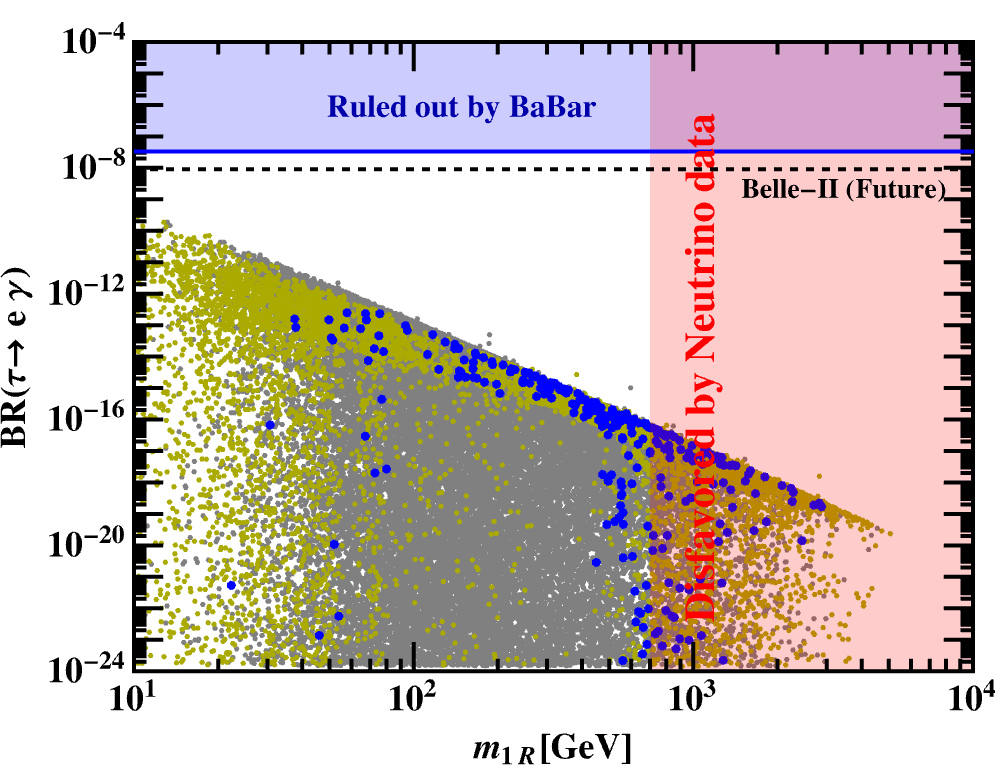}
      \includegraphics[height=4.15cm]{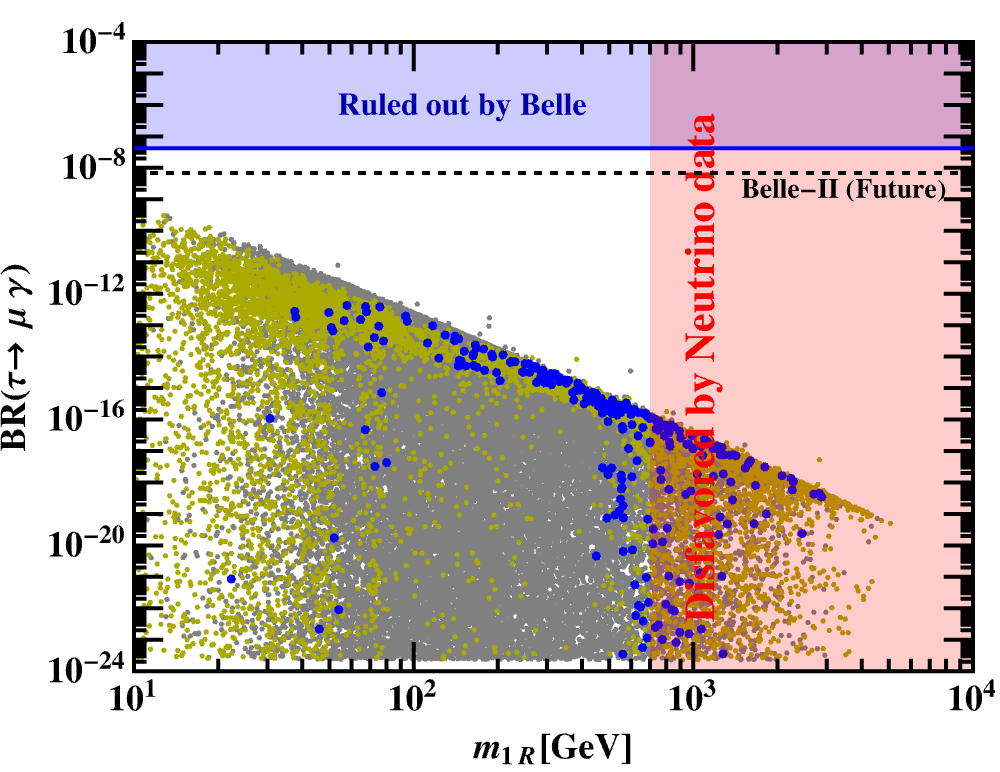}
\centering
        \includegraphics[height=4.15cm]{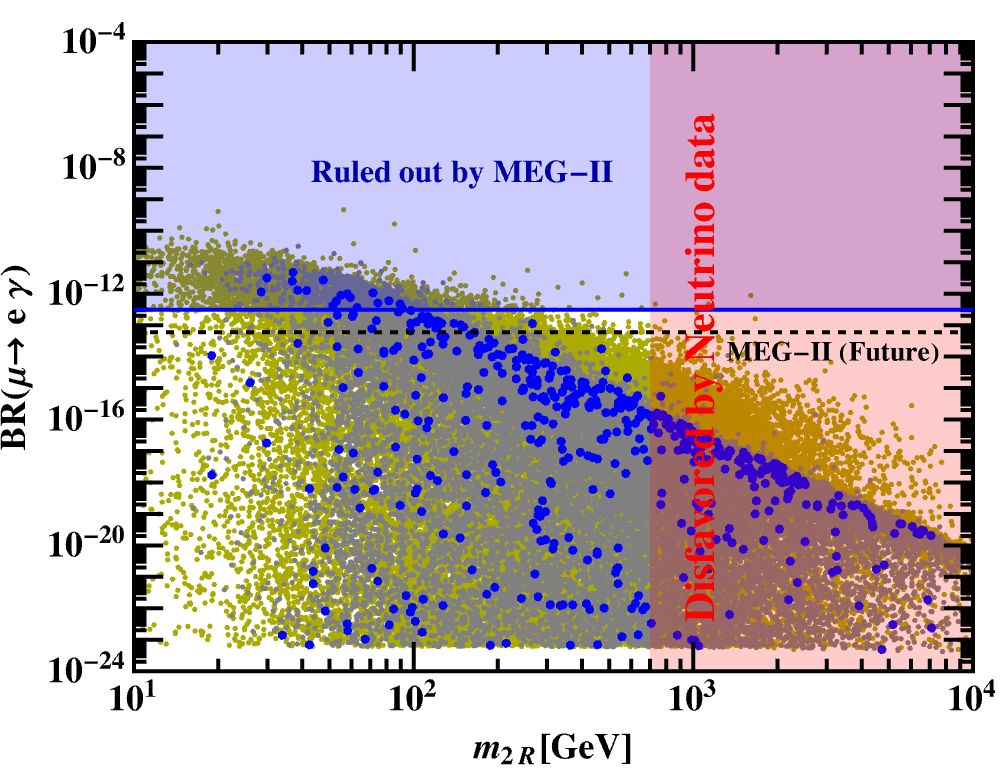}
    \includegraphics[height=4.15cm]{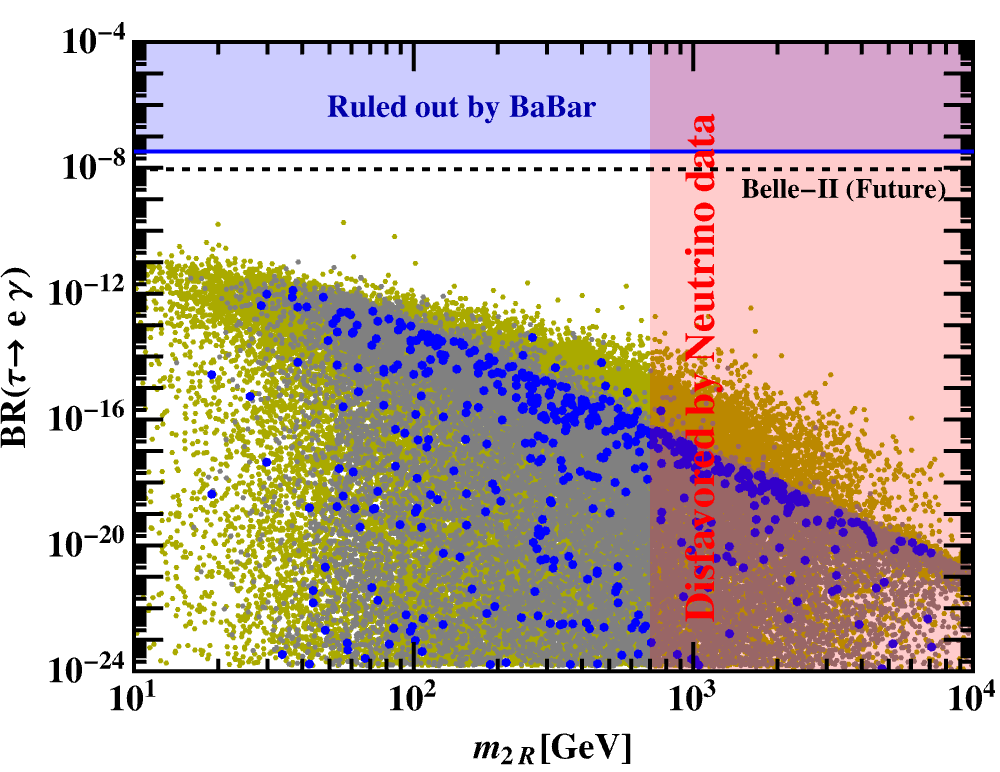}
      \includegraphics[height=4.15cm]{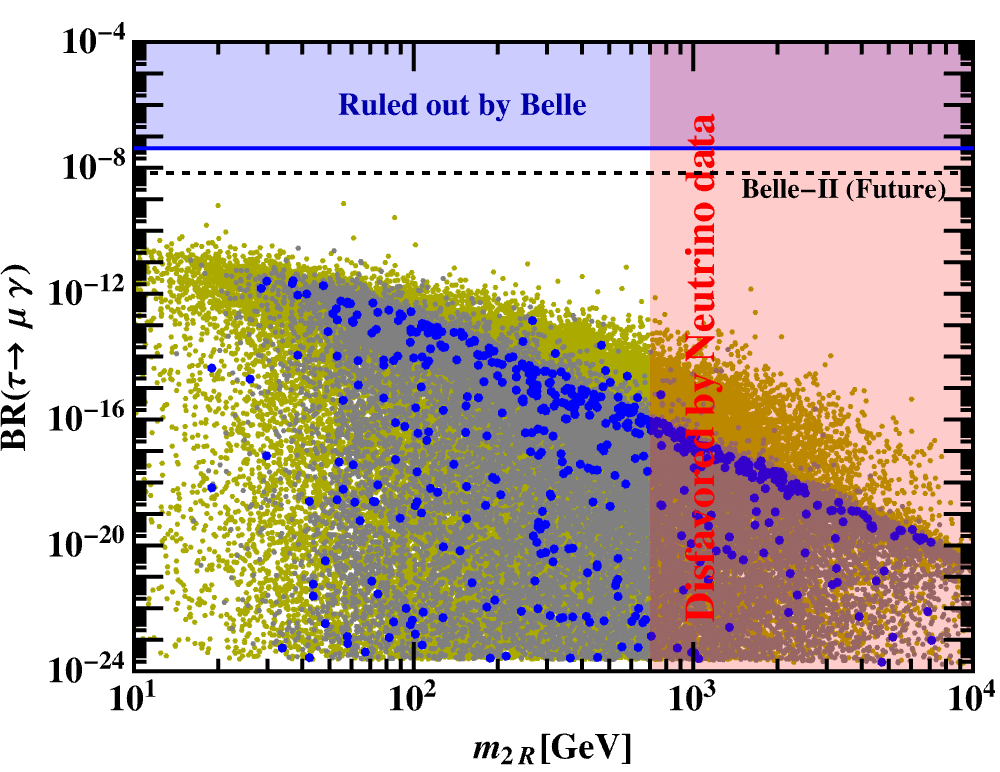}
\centering
        \includegraphics[height=4.15cm]{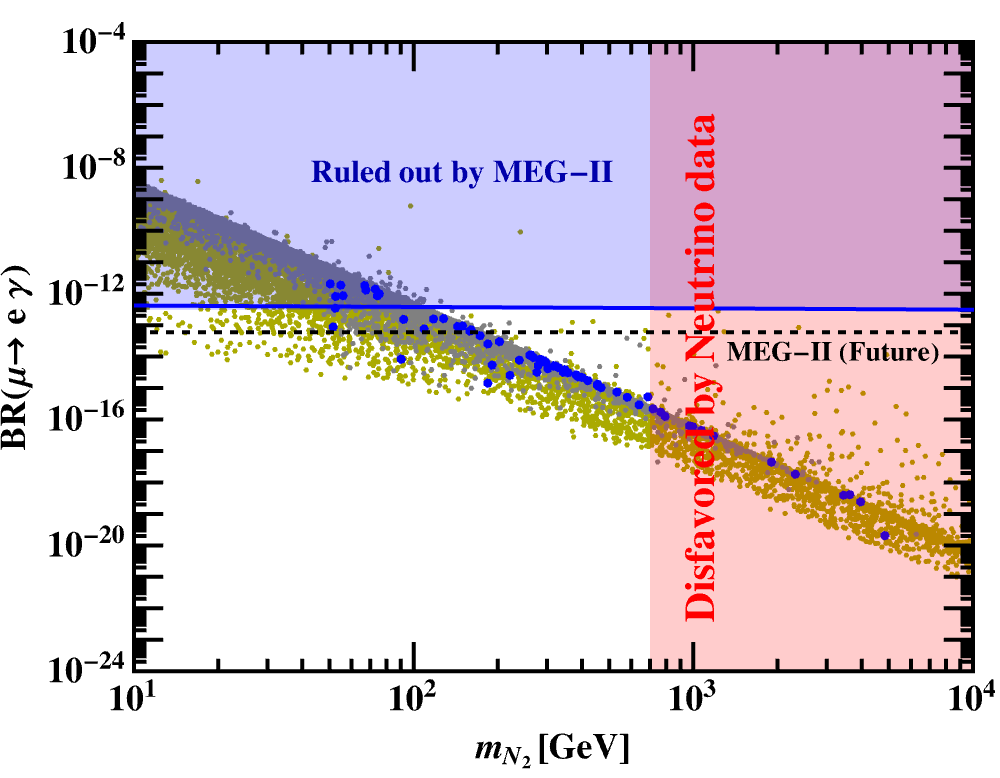}
    \includegraphics[height=4.15cm]{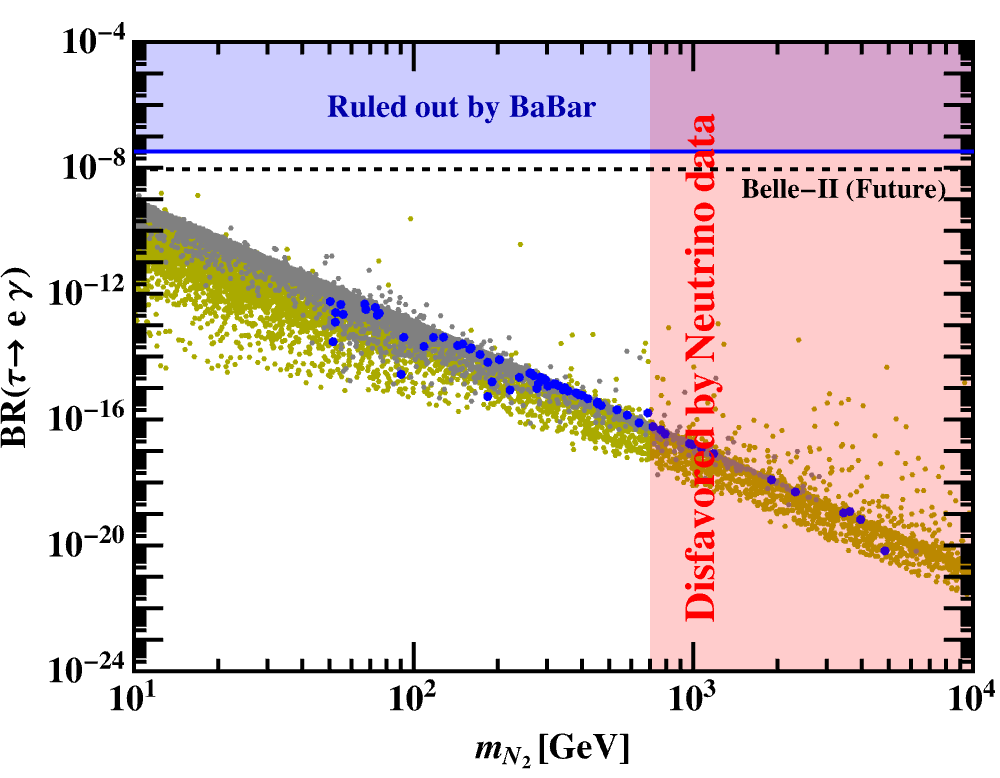}
      \includegraphics[height=4.15cm]{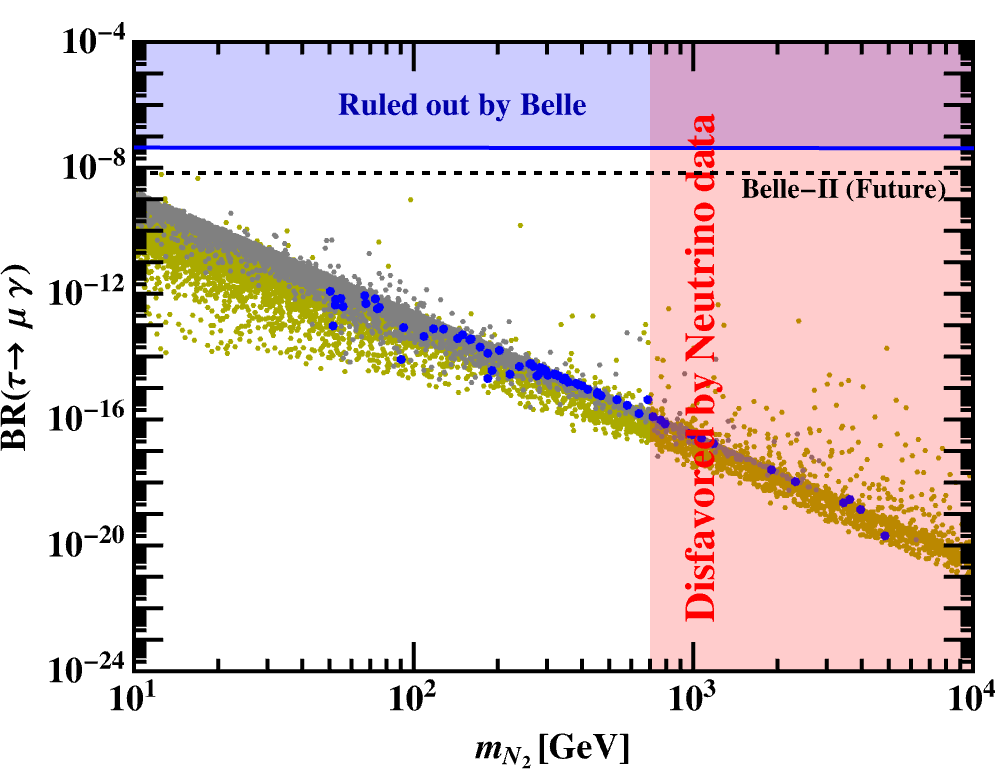}
        \caption{\footnotesize The CLFV rates vs DM mass have been shown. \textbf{1st column:} BR($\mu \rightarrow e \gamma$), \textbf{2nd column:} BR($\tau \rightarrow e \gamma$), \textbf{3rd column:} BR($\tau \rightarrow \mu \gamma$). \textbf{1st row:} Doublet scalar DM mass, \textbf{2nd row:} Singlet scalar  DM mass, \textbf{3rd row:} Fermionic DM mass. The gray and olive green points represent regions of under-abundance and over-abundance, respectively, while the blue points correspond to the correct relic density.}
        \label{fig:cLFV}
\end{figure}

We have analyzed various CLFV rates corresponding to different DM candidates. In Fig.~\ref{fig:cLFV}, we show the resulting CLFV decay rates as a function of the mass for each viable DM candidate in our model. 
The gray and olive green points represent regions of under-abundance and over-abundance, respectively, while the blue points correspond to the correct relic density. Note that all three processes show similar behavior: their maximum values drop as the DM mass increases. This behavior arises because DM is the lightest particle in the dark sector. As its mass increases, the other dark sector particles also become heavier, which lowers the CLFV rates, as can be seen from Eq.~\eqref{eqn:cLFVeqn}.
In our analysis, we identified viable data points in the mass ranges of (63–150) GeV for the doublet scalar, (48–265) GeV for the singlet scalar, and (50–160) GeV for the fermionic DM candidate, all of which satisfy the previously discussed experimental constraints. Notably, these points lie between the current experimental bounds and the projected future sensitivity for the decay $\mu \rightarrow e \gamma$. However, for the decay channels $\tau \rightarrow e \gamma$ and $\tau \rightarrow \mu \gamma$, no points within our scanned parameter space fall within the current or expected experimental limits. This is primarily due to the comparatively weaker experimental bounds for these decay modes, which our model does not reach within the explored parameter range. Thus, CLFV points that are both within experimental reach and consistent with DM constraints appear only in a few distinct regions, making CLFV measurements a valuable probe for testing our model.
\section{Conclusions}  \label{sec:conc}
We propose a Dirac scotogenic-like model where DM stability arises from the breaking of an $A_4$ flavor symmetry. The same breaking also induces a ``scoto inverse-seesaw'' mechanism, which combines a tree-level inverse-seesaw with a one-loop scotogenic contribution to generate neutrino masses. The inverse seesaw mechanism at the tree level is enabled by the smallness of the singlet VEV $u$. This framework naturally explains the observed neutrino mass-squared differences, $\Delta m^2_{\rm{atm}}$ and $\Delta m^2_{\rm{sol}}$, from oscillation data. The model also addresses leptonic flavor patterns via the $A_4$ symmetry. It includes right-handed neutrinos $\nu_R$, $SU(2)_L$ singlet fermions $N_L$, $N_R$, a singlet scalar $\xi$, and a doublet scalar $\eta$. Here, $\nu_R$'s are $A_4$ singlets, while all other BSM particles transform as $A_4$ triplets. To forbid tree-level neutrino couplings before symmetry breaking, we assign specific $B-L$ charges $(-4, -4, 5)$ to $\nu_R$. The breaking $U(1)_{B-L} \to \mathcal{Z}_3$ then ensures neutrinos remain Dirac in nature and results in one massless neutrino.

\setlength{\parskip}{0pt}
Our model is compatible with both normal (NO) and inverted (IO) neutrino mass orderings. In both cases, we find a strong correlation between the sum of neutrino masses $\sum m_i$ and the effective beta decay mass $\langle m_{\beta} \rangle$, with both quantities confined within narrow ranges. For the IO scenario, neutrino masses exhibit particularly sharp correlations (see right panel of Fig.~\ref{fig:io}). We also examine the correlation between the CP-violating phase $\delta_{CP}$ and the atmospheric mixing angle $\theta_{23}$ for two benchmark scenarios. For NO, case-I aligns more closely with the global best-fit value, whereas case-II remains consistent within the 2$\sigma$ range. For IO, both scenarios allow a broader range of $\delta_{CP}$ values, still compatible with the best-fit value. These predictions provide testable signatures that can be probed in upcoming neutrino oscillation experiments.

\setlength{\parskip}{0pt}
Constraints from neutrino oscillation data impose an upper bound on the VEV of the singlet scalar $u$, which in turn limits scalar masses due to the requirement of perturbative Yukawa couplings. These theoretical bounds significantly influence dark sector predictions, restricting the DM mass to an upper limit of $\sim 700~\rm{GeV}$. Incorporating collider (LHC and LEP) and DD constraints further narrows the viable DM parameter space. For doublet scalar DM, the allowed mass range is $85 \leq m_{1R} \leq 700~\rm{GeV}$, though the lower region is nearly ruled out by DD bounds. Singlet scalar DM remains viable in the $20 \leq m_{2R} \leq 700~\rm{GeV}$ range. Fermionic DM remains viable in the mass range $10 \leq M_{N_2} \leq 700~\rm{GeV}$. Thus, the model predicts DM candidates with masses confined to narrow, experimentally and theoretically motivated windows.

\setlength{\parskip}{0pt}
We have also analyzed lepton LFV decays, such as $\mu \to e\gamma$, for the three DM candidates in our model. By comparing the predicted branching ratios with current limits from MEG, BaBar, and Belle, as well as projected future sensitivities, we find that viable regions of parameter space remain consistent with all LFV constraints. Overall, the flavor symmetry-breaking framework not only ensures the stability of DM but also leads to a novel \emph{scoto inverse-seesaw} mechanism. This yields a unified and predictive framework for both neutrino mass generation and DM phenomenology. As a result, the model offers tightly constrained and testable predictions in both the neutrino and dark sectors, potentially verifiable in upcoming experimental efforts.

\section{Acknowledgments}
Authors would like to acknowledge the SARAH-4.14.5 \cite{Staub:2015kfa} and SPheno-4.0.5 \cite{Porod:2011nf} to compute all vertex diagrams, mass matrices, and tadpole equations. Additionally, the thermal contribution to the DM relic density, as well as the cross sections for DM-nucleon scattering, are evaluated using micrOMEGAs-5.3.41 \cite{Belanger:2014vza}. RK and SY would like to acknowledge the funding support by the CSIR SRF-NET fellowship.
\enlargethispage{15cm}
\appendix  
\section{Scalar Potential and Mass spectrum} \label{app:pot}
In the scalar sector, the $SU(2)_L$ doublet $H$ is a singlet under the $A_4$ symmetry, while the doublet $\eta$ and the singlet $\xi$ transform as triplets. The potential is formulated based on the specified $\mathrm{SU(2)_L \otimes U(1)_Y}$ and $A_{4}$ charge assignment for the scalar fields
\begin{align}\label{eq:scalarpot1} 
V&=  V_H +V_{\eta} +V_{\xi} +V_{H \eta} + V_{H\xi} + V_{\eta \xi} + V_{H\eta \xi} + h.c. \ .
\end{align}
\newpage
The individual terms in the potential are explicitly written as:
\begin{align} \label{eq:pot1}
&V_H=\mu_H^2 H^\dagger H +\lambda_1 (H^\dagger H)^2 ,\nonumber \\
&V_{\eta}= \mu_{\eta}^2 (\eta^\dagger\eta)_\mathbf{1} + \lambda_{\eta_{1}}(\eta^\dagger\eta)_{\mathbf{1}}(\eta^\dagger\eta)_{\mathbf{1}}+\lambda_{\eta_{2}}(\eta^\dagger\eta)_{\mathbf{1'}}(\eta^\dagger\eta)_{\mathbf{1''}}+\lambda_{\eta_{3}}(\eta^\dagger \eta^{\dagger})_{\mathbf{1}}(\eta \eta)_{\mathbf{1}}+\lambda_{\eta_{4}}(\eta^\dagger \eta^{\dagger})_{\mathbf{1'}}(\eta \eta)_{\mathbf{1''}} \nonumber \\ &+\lambda_{\eta_{5}}(\eta^\dagger \eta^{\dagger})_{\mathbf{1''}}(\eta \eta)_{\mathbf{1'}}+ \lambda_{\eta_{6}}(\eta^\dagger\eta)_{\mathbf{3_1}}(\eta^\dagger\eta)_{\mathbf{3_1}} + \lambda_{\eta_{7}}(\eta^\dagger\eta)_{\mathbf{3_1}}(\eta^\dagger\eta)_{\mathbf{3_2}} + \lambda_{\eta_{8}}(\eta^\dagger\eta)_{\mathbf{3_2}}(\eta^\dagger\eta)_{\mathbf{3_2}} \nonumber \\ 
&+ \lambda_{\eta_{9}}(\eta^\dagger \eta^\dagger)_{\mathbf{3_1}}(\eta \eta)_{\mathbf{3_1}} + \lambda_{\eta_{10}}(\eta^\dagger \eta^\dagger)_{\mathbf{3_1}}(\eta \eta)_{\mathbf{3_2}}+ \lambda_{\eta_{11}}(\eta^\dagger \eta^\dagger)_{\mathbf{3_2}}(\eta \eta)_{\mathbf{3_1}} + \lambda_{\eta_{12}}(\eta^\dagger \eta^\dagger)_{\mathbf{3_2}}(\eta \eta)_{\mathbf{3_2}} ,\nonumber \\
&V_{\xi}= \mu_{\xi}^2 (\xi \xi)_\mathbf{1} + \lambda_{\xi_{1}}(\xi \xi)_{\mathbf{1}}(\xi \xi)_{\mathbf{1}} + \lambda_{\xi_{2}}(\xi \xi)_{\mathbf{1'}}(\xi \xi)_{\mathbf{1''}} +{\lambda_{\xi_{3}}(\xi \xi)_{\mathbf{3_1}}(\xi \xi)_{\mathbf{3_1}}}+{\lambda_{\xi_{4}}(\xi \xi)_{\mathbf{3_1}}(\xi \xi)_{\mathbf{3_2}}} \nonumber \\ &+ {\lambda_{\xi_{5}}(\xi \xi)_{\mathbf{3_2}}(\xi \xi)_{\mathbf{3_2}}} , \nonumber \\
&V_{H\eta}= \lambda_{H\eta_{1}} (H^\dagger H)(\eta^\dagger\eta)_\mathbf{1} 
+\lambda_{H\eta_{2}}(H^\dagger\eta)_\mathbf{3}(\eta^\dagger H)_\mathbf{3} +\lambda_{H\eta_{3}}(H^\dagger\eta^\dagger)_\mathbf{3}(\eta H)_\mathbf{3} ,\nonumber \\
&V_{H\xi}=\lambda_{H\xi_{1}} (H^\dagger H)(\xi \xi)_\mathbf{1} + \lambda_{H\xi_{2}}(H^\dagger\xi)_\mathbf{3}(\xi H)_\mathbf{3} ,\nonumber \\
&V_{\eta \xi}=[{\lambda_{\eta \xi_{1}}(\eta^\dagger\eta)_{\mathbf{1}}(\xi \xi)_{\mathbf{1}}} + {\lambda_{\eta \xi_{2}}(\eta^\dagger\eta)_{\mathbf{1'}}(\xi \xi)_{\mathbf{1''}}}+ {\lambda_{\eta \xi_{3}}(\eta^\dagger\eta)_{\mathbf{1''}}(\xi \xi)_{\mathbf{1'}}}] \nonumber \\ 
&+[\lambda_{\eta \xi_{4}}(\eta^\dagger\eta)_{\mathbf{3_1}}(\xi \xi)_{\mathbf{3_1}} + \lambda_{\eta \xi_{5}}(\eta^\dagger\eta)_{\mathbf{3_1}}(\xi \xi)_{\mathbf{3_2}} + \lambda_{\eta \xi 6}(\eta^\dagger\eta)_{\mathbf{3_2}}(\xi \xi)_{\mathbf{3_1}} + \lambda_{\eta \xi 7}(\eta^\dagger\eta)_{\mathbf{3_2}}(\xi \xi)_{\mathbf{3_2}}]    \nonumber \\ &+
[{\lambda_{\eta \xi_{8}}(\eta^\dagger\xi)_{\mathbf{1}}(\eta \xi)_{\mathbf{1}}} + {\lambda_{\eta \xi_{9}}(\eta^\dagger\xi)_{\mathbf{1'}}(\eta \xi)_{\mathbf{1''}}}+ {\lambda_{\eta \xi_{10}}(\eta^\dagger\xi)_{\mathbf{1''}}(\eta \xi)_{\mathbf{1'}}}] \nonumber \\ &+ 
[\lambda_{\eta \xi_{11}}(\eta^\dagger\xi)_{\mathbf{3_1}}(\eta \xi)_{\mathbf{3_1}} + \lambda_{\eta \xi_{12}}(\eta^\dagger\xi)_{\mathbf{3_1}}(\eta \xi)_{\mathbf{3_2}} + \lambda_{\eta \xi_{13}}(\eta^\dagger\xi)_{\mathbf{3_2}}(\eta \xi)_{\mathbf{3_1}} + \lambda_{\eta \xi_{14}}(\eta^\dagger\xi)_{\mathbf{3_2}}(\eta \xi)_{\mathbf{3_2}}]  ,\nonumber \\
&V_{H\eta \xi}= \kappa (H^\dagger)_{\mathbf{1}} (\eta \xi)_{\mathbf{1}} + h.c. 
\end{align}
Now, following the  $A_4$ multiplication rule as given in Eqs.~\eqref{eq:a4mrule} and \eqref{eq:pr}, the expressions given in Eq.~\eqref{eq:pot1} can be simplified in the component form as
\begin{align} \label{eq:pot2}
&V_H=\mu_H^2 H^\dagger H +\lambda_1 (H^\dagger H)^2, \nonumber \\
&V_{\eta}= \mu_{\eta}^2 (\eta^\dagger_{1}\eta_{1} + \eta^\dagger_{2}\eta_{2} + \eta^\dagger_{3}\eta_{3}) + (\lambda_{\eta_{1}} + \lambda_{\eta_{2}} + \lambda_{\eta_{3}}+ 2 \lambda_{\eta_{4}})(\eta^\dagger_{1}\eta_{1} \eta^\dagger_{1}\eta_{1} + \eta^\dagger_{2}\eta_{2} \eta^\dagger_{2}\eta_{2}+ \eta^\dagger_{3}\eta_{3} \eta^\dagger_{3}\eta_{3}) \nonumber \\
&+ (2 \lambda_{\eta_{1}} - \lambda_{\eta_{2}} + \lambda_{\eta_{10}} + \lambda_{\eta_{11}})(\eta^\dagger_{1}\eta_{1} \eta^\dagger_{2}\eta_{2} + \eta^\dagger_{1}\eta_{1} \eta^\dagger_{3}\eta_{3} + \eta^\dagger_{2}\eta_{2} \eta^\dagger_{3}\eta_{3}) \nonumber \\
&+ (\lambda_{\eta_{3}} - \lambda_{\eta_{4}} + \lambda_{\eta_{6}}) (\eta^\dagger_{2}\eta_{3} \eta^\dagger_{2}\eta_{3} + \eta^\dagger_{3}\eta_{1} \eta^\dagger_{3}\eta_{1} + \eta^\dagger_{1}\eta_{2} \eta^\dagger_{1}\eta_{2} +h.c.) \nonumber \\
&+ (\lambda_{\eta_{7}} + \lambda_{\eta_{9}} + \lambda_{\eta_{12}}) (\eta^\dagger_{2}\eta_{3} \eta^\dagger_{3}\eta_{2} + \eta^\dagger_{3}\eta_{1} \eta^\dagger_{1}\eta_{3} + \eta^\dagger_{1}\eta_{2} \eta^\dagger_{2}\eta_{1}), \nonumber \\
&V_{\xi}= \mu_{\xi}^2 (\xi_{1} \xi_{1} + \xi_{2} \xi_{2} + \xi_{3} \xi_{3}) + (\lambda_{\xi_{1}} + \lambda_{\xi_{2}})(\xi_{1} \xi_{1} \xi_{1} \xi_{1} + \xi_{2} \xi_{2} \xi_{2} \xi_{2} + \xi_{3} \xi_{3} \xi_{3} \xi_{3}) \nonumber \\
&+ (2 \lambda_{\xi_{1}}-\lambda_{\xi_{2}} + \lambda_{\xi_{3}} + \lambda_{\xi_{4}} + \lambda_{\xi_{5}})( \xi_{1} \xi_{1} \xi_{2} \xi_{2} + \xi_{1} \xi_{1} \xi_{3} \xi_{3} + \xi_{2} \xi_{2} \xi_{3} \xi_{3}), \nonumber \\
&V_{H\eta}= (\lambda_{H\eta_{1}} + \lambda_{H\eta_{3}}) (H^\dagger H \eta^\dagger_{1}\eta_{1} + H^\dagger H \eta^\dagger_{2}\eta_{2} + H^\dagger H \eta^\dagger_{3}\eta_{3}) + \lambda_{H\eta_{2}}(H^\dagger\eta_1 \eta^\dagger_1 H + H^\dagger\eta_2 \eta^\dagger_2 H + H^\dagger\eta_3 \eta^\dagger_3 H), \nonumber \\
&V_{H\xi}= (\lambda_{H\xi_{1}} + \lambda_{H\xi_{2}}) \left(H^\dagger H \xi_{1}\xi_{1} + H^\dagger H \xi_{2}\xi_{2} + H^\dagger H \xi_{3}\xi_{3}\right), \nonumber \\
&V_{\eta\xi}= (\lambda_{\eta \xi_{1}} +2 \lambda_{\eta \xi_{2}} + \lambda_{\eta \xi_{8}} + 2 \lambda_{\eta \xi_{9}}) (\eta^\dagger_1 \eta_1 \xi_1 \xi_1 + \eta^\dagger_2 \eta_2 \xi_2 \xi_2 + \eta^\dagger_3 \eta_3 \xi_3 \xi_3) \nonumber \\
&+ (\lambda_{\eta \xi_{4}} + \lambda_{\eta \xi_{5}} + \lambda_{\eta \xi_{12}}+\lambda_{\eta \xi_{8}}-\lambda_{\eta \xi_{9}}) (\eta^\dagger_2 \eta_3 \xi_2 \xi_3 + \eta^\dagger_3 \eta_1 \xi_3 \xi_1 + \eta^\dagger_1 \eta_2 \xi_1 \xi_2 +h.c.) \nonumber \\
&+ (\lambda_{\eta \xi_{1}} - \lambda_{\eta \xi_{2}} + \lambda_{\eta \xi_{11}}) (\eta^\dagger_2 \xi_3 \eta_2 \xi_3 + \eta^\dagger_3 \xi_1 \eta_3 \xi_1 + \eta^\dagger_1 \xi_2 \eta_1 \xi_2) \nonumber \\
&+ (\lambda_{\eta \xi_{1}} - \lambda_{\eta \xi_{2}} + \lambda_{\eta \xi_{14}}) (\eta^\dagger_3 \xi_2 \eta_3 \xi_2 + \eta^\dagger_1 \xi_3 \eta_1 \xi_3 + \eta^\dagger_2 \xi_1 \eta_2 \xi_1), \nonumber \\
&V_{H\eta \xi}= \kappa (H^\dagger \eta_1 \xi_1 + H^\dagger \eta_2 \xi_2 + H^\dagger \eta_3 \xi_3) + h.c. \ .
\end{align}

We can now compute the scalar mass spectrum of the model. The particles in the model are classified as $\mathcal{Z}_2$ even or $\mathcal{Z}_2$ odd, on their transformation properties under $A_4$. Since the $\mathcal{Z}_2$ symmetry remains unbroken, only $\mathcal{Z}_2$ even particles mix among themselves, and similarly, $\mathcal{Z}_2$ odd particles mix only with each other. The $\mathcal{Z}_2$ even particles $(\phi_1, \phi_2, \phi_3)$ mix with each other giving $\mathcal{M}^2_{H}$ matrix. Similarly, the $\mathcal{Z}_2$ even particles $(\sigma_1, \sigma_2)$ mix to give the mass matrix $\mathcal{M}^2_{A}$, while $(H^{\pm}, \eta^{\pm}_1 )$ mix to form the mass matrix $\mathcal{M}^2_{H^{\pm}}$. The $\mathcal{Z}_2$ odd particles $(\eta_{2}^R, \xi_{2}^R)$ and $(\eta_{3}^R, \xi_{3}^R)$  do mix as well as $(\eta_2^{\pm}, \eta_3^{\pm})$ among themselves giving $\mathcal{M}^2_{2}$, $\mathcal{M}^2_{3}$ and $\mathcal{M}^2_{\eta^{\pm}_{2/3}}$ mass matrices, respectively.

These matrices are as follows:
\begin{align}
  &\mathcal{M}^2_{H} =  \begin{bmatrix}                   
 2 \lambda_1 v_1^2-\frac{\kappa v_2 u}{v_1} & \Lambda_{12}v_1 v_2 + \kappa u & \kappa v_2 + 2 \lambda_{H\xi_{1}} + \lambda_{H\xi_{2}}v_1 u \\
 \Lambda_{12}v_1 v_2 + \kappa u & 2 \Lambda_{\eta}v_2^2-\frac{\kappa v_1 u}{v_2} & 2 \Lambda_{23}v_2 u +\kappa v_1 \\
                     \kappa v_2 + 2 \lambda_{H\xi_{1}} + \lambda_{H\xi_{2}}v_1 u & 2 \Lambda_{23}v_2 u +\kappa v_1 & 8 \Lambda_{\xi}u^2-\frac{\kappa v_1 v_2}{u} \\
           \end{bmatrix}  ,\,\, \nonumber \\
         &\mathcal{M}^2_A =  \begin{bmatrix}                   
-\frac{\kappa v_2 u}{v_1}&  \kappa u  \\
 \kappa u & -\frac{\kappa v_1 u}{v_2}   \\
           \end{bmatrix}  ,\,\, 
\mathcal{M}^2_{H^{\pm}}=
\begin{bmatrix}
-\left(  \frac{\lambda_{H\eta_{2}}}{2}v_2^2 + \frac{\kappa v_2u}{v_1}  \right)  &  \frac{\lambda_{H\eta_{2}}}{2} v_1 v_2 + \kappa u   \\
\frac{\lambda_{H\eta_{2}}}{2} v_1 v_2 + \kappa u  & -\left(  \frac{\lambda_{H\eta_{2}}}{2}v_2^2 + \frac{\kappa v_1u}{v_2}  \right)  \\
\end{bmatrix} ,\,\, \nonumber \\ 
&\mathcal{M}^2_{2R}= \begin{bmatrix}
\Lambda_{\eta_{2}} v_2^2-\alpha_{2} u^2 & \alpha_1 v_2 u \\
\alpha_1 v_2 u &\Lambda_{\xi_{2}} u^2-\alpha_{3} v_2^2 \\
\end{bmatrix}, \, 
\mathcal{M}^2_{3R}= \begin{bmatrix}
\Lambda_{\eta_{2}} v_2^2-\alpha_{3} u^2 & \alpha_1 v_2 u \\
\alpha_1 v_2 u &\Lambda_{\xi_{2}} u^2-\alpha_{2} v_2^2 \\
\end{bmatrix} ,\,\, \nonumber \\ 
%
&
\mathcal{M}^2_{\eta^{\pm}_{2/3}}= \begin{bmatrix}
m_2^{\pm} & 0 \\
0 & m_3^{\pm}\\
\end{bmatrix}, \,  \quad
\mathcal{M}^2_{\eta^{I}_{2/3}}= \begin{bmatrix}
m_2^{I} & 0 \\
0 & m_3^{I}\\
\end{bmatrix}
\;.
\end{align}
where 
\begin{align*}
m_{2}^{\pm}=-\frac{\lambda_{H \eta_{2}}}{2} v_1^2-\left(\frac{3}{2}\lambda_{\eta_{2}}+\lambda_{\eta_{3}}+2\lambda_{\eta_{4}}-\frac{1}{2}\lambda_{\eta_{10}} + \lambda_{\eta_{11}}\right)v_2^2-\alpha_{2} u^2, 
\end{align*}
\begin{align*}
&m_{3}^{\pm}=-\frac{\lambda_{H \eta_{2}}}{2} v_1^2-\left(\frac{3}{2}\lambda_{\eta_{2}}+\lambda_{\eta_{3}}+2\lambda_{\eta_{4}}-\frac{1}{2}\lambda_{\eta_{10}} + \lambda_{\eta_{11}}\right)v_2^2-\alpha_{3} u^2, 
\nonumber \\ 
&m_{2}^{I}=\left(-3\lambda_{\eta_{2}}+\lambda_{\eta_{10}} + \lambda_{\eta_{11}}+\lambda_{\eta_{7}} + \lambda_{\eta_{9}} + \lambda_{\eta_{12}}-4\lambda_{\eta_{3}}-2\lambda_{\eta_{4}}-2\lambda_{\eta_{6}}\right)\frac{v_2^2}{2} \nonumber \\ &-\kappa \frac{v_1 u}{v_2}+\left(\lambda_{\eta \xi_{14}}-3\lambda_{\eta \xi_{2}}-\lambda_{\eta \xi_{8}}-2\lambda_{\eta \xi_{9}}\right)v_3^2,
\nonumber \\
&m_{3}^{I}=\left(-3\lambda_{\eta_{2}}+\lambda_{\eta_{10}} + \lambda_{\eta_{11}}+\lambda_{\eta_{7}} + \lambda_{\eta_{9}} + \lambda_{\eta_{12}}-4\lambda_{\eta_{3}}-2\lambda_{\eta_{4}}-2\lambda_{\eta_{6}}\right)\frac{v_2^2}{2} \nonumber \\& -\kappa \frac{v_1 u}{v_2}+\left(\lambda_{\eta \xi_{11}}-3\lambda_{\eta \xi_{2}}-\lambda_{\eta \xi_{8}}-2\lambda_{\eta \xi_{9}}\right)v_3^2.
\end{align*}
The following combinations of couplings have been used in above mass matrices
\begin{align} 
& \Lambda_{\eta} \equiv \lambda_{\eta_{1}}+ \lambda_{\eta_{2}} + \lambda_{\eta_{3}}+ 2 \lambda_{\eta_{4}}, \quad \quad \Lambda_{\xi} \equiv \lambda_{\xi_{1}}+ \lambda_{\xi_{2}}  , \nonumber \\
& \alpha_2 \equiv \frac{\kappa v_1}{v_2 u}+ (3 \lambda_{\eta \xi_{2}}+\lambda_{\eta \xi_{8}}+2\lambda_{\eta \xi_{9}}-\lambda_{\eta \xi_{14}}), \quad \quad \alpha_1 \equiv \frac{\kappa v_1}{ v_2u}+ \lambda_{\eta \xi_{4}} + \lambda_{\eta \xi_{5}} + \lambda_{\eta \xi_{12}}+ \lambda_{\eta \xi_{8}}- \lambda_{\eta \xi_{9}}, \nonumber \\
& \Lambda_{12} \equiv \lambda_{H \eta_{2}} + \lambda_{H \eta_{4}}, \quad \quad \Lambda_{23} \equiv \lambda_{\eta \xi_{1}} + 2 \lambda_{\eta \xi_{2}} + \lambda_{\eta \xi_{8}} + 2 \lambda_{\eta \xi_{9}}, \nonumber \\
& \Lambda_{\eta_{2}} \equiv \lambda_{\eta_{6}} - \frac{3}{2}\lambda_{\eta_{2}} -3 \lambda_{\eta_{4}}+\frac{1}{2}\lambda_{\eta_{10}} + \lambda_{\eta_{11}}+\frac{1}{2}\lambda_{\eta_{7}} + \lambda_{\eta_{9}} + \lambda_{\eta_{12}}, \quad \quad  \quad \Lambda_3 \equiv \lambda_{\xi_{3}} + \frac{1}{2}\lambda_{\xi_{4}} -\frac{3}{2} \lambda_{\xi_{2}}. \nonumber \\
& \Lambda_{\xi_{2}} \equiv 2 \lambda_{\xi_{3}} + \lambda_{\xi_{4}} + \lambda_{\xi_{5}} - 6 \lambda_{\xi_{2}}, \quad \alpha_3 \equiv \frac{\kappa v_1}{v_2 u}+ (3 \lambda_{\eta \xi_{2}}+\lambda_{\eta \xi_{8}}+2\lambda_{\eta \xi_{9}}-\lambda_{\eta \xi_{11}}).
\end{align}
%

\section{$A_4$ symmetry and its multiplication rule} \label{app:A4}
The $A_4$ symmetry is a non-abelian discrete flavor group, corresponding to the even permutation group of four objects. This symmetry is also the group of symmetries for a regular tetrahedron. With its 12 elements, $A_4$ can be generated by two generators, denoted as $S$ and $T$, which satisfy the following relations
\begin{equation} \label{eq:a4genrel}
S^2=T^3=(ST)^3=\mathcal{I}.
\end{equation}
\A4 symmetry group features four distinct irreducible representations: one triplet, denoted as 3, and three singlets, labeled as 1, $1'$, and $1''$. Their multiplication rules are as follows:
\begin{eqnarray} \label{eq:a4mrule}
&1\times1&=1=1' \times 1'', \quad 1'\times 1'=1'', \quad 1''\times 1''=1',  \nonumber \\
&1 \times 3&=3, \quad 3\times 3= 1+ 1' + 1'' + 3_1 + 3_2 \quad .
\end{eqnarray}
In the basis where generators $S$ and $T$ are real matrices, they are given by 
\begin{equation} \label{eq:a4genmat}
S=\left(
\begin{array}{ccc}
1&0&0\\
0&-1&0\\
0&0&-1\\
\end{array}
\right)\,, \quad
T=\left(
\begin{array}{ccc}
0&1&0\\
0&0&1\\
1&0&0\\
\end{array}
\right)\;.
\end{equation}
If $a=\left(a_1,a_2,a_3\right)$ and $b=\left(b_1,b_2,b_3\right)$ are two triplets of \A4, then they obey the  following multiplication rules~\cite{Boucenna:2011tj,Ishimori:2010au}
\begin{equation}\label{eq:pr}
\begin{array}{lll}
\left(ab\right)_1&=&a_1b_1+a_2b_2+a_3b_3\, ,\\
\left(ab\right)_{1'}&=&a_1b_1+\omega a_2b_2+\omega^2a_3b_3\, ,\\
\left(ab\right)_{1''}&=&a_1b_1+\omega^2 a_2b_2+\omega a_3b_3\, ,\\
\left(ab\right)_{3_1}&=&\left(a_2b_3,a_3b_1,a_1b_2\right)\, ,\\
\left(ab\right)_{3_2}&=&\left(a_3b_2,a_1b_3,a_2b_1\right) .
\end{array}
\end{equation}
Here, $\omega^3=1$, denotes a cube root of unity.  
The generator $S$ remains invariant under VEV alignments $\langle \varphi \rangle \sim (v,0,0)$, which leads to the breaking of the $A_4$ symmetry into its subgroup $\mathcal{Z}_2$.
\begin{equation} \label{eq:a4genvev}
S \langle \varphi \rangle=\langle \varphi \rangle .
\end{equation}
For a \A4 triplet, any field $\psi \equiv \left(a_1,a_2,a_3 \right)^T$transforms under $S$ as follows
\begin{align} \label{eq:a4z2con}
S \psi =\left(
\begin{array}{ccc}
1&0&0\\
0&-1&0\\
0&0&-1\\
\end{array}
\right) \left(
\begin{array}{ccc}
a_1\\
a_2\\
a_3 \\
\end{array}
\right)= \left(
\begin{array}{ccc}
a_1\\
-a_2\\
-a_3 \\
\end{array}
\right)
\end{align}
Once \A4 symmetry is broken, the $\mathcal{Z}_2$ residual symmetry for triplets $N$, $\eta$ and $\xi$ is given as
\begin{align} \label{eq:a4fielddec}
N_1 &\to +N_1\,,\quad  \eta_1 \to +\eta_1\,,\quad \xi_1 \to +\xi_1 \, , \nonumber \\  
N_{2,3} &\to -N_{2,3}\,,\quad \eta_{2,3} \to -\eta_{2,3}\,,\quad \eta_{2,3} \to -\eta_{2,3} \;.
\end{align}
The remaining fields of our model are $\mathcal{Z}_2$ even, because they are singlets of \A4 and transform trivially under $S$.
\section{Annihilation, production and detection of DM}
\label{app:Feynman}
In this section, we present the possible Feynman diagrams that play an important role in determining the relic density of different DM candidates, doublet and singlet scalar, as well as fermion. In addition, we include Feynman diagrams for DD, for scalars at tree level, and for fermionic DM at loop level. These Feynman diagrams serve to assist our understanding of the dark sector results presented in Sec.~\ref{sec:dm}.

\begin{figure}[th]
        \includegraphics[width=14cm]{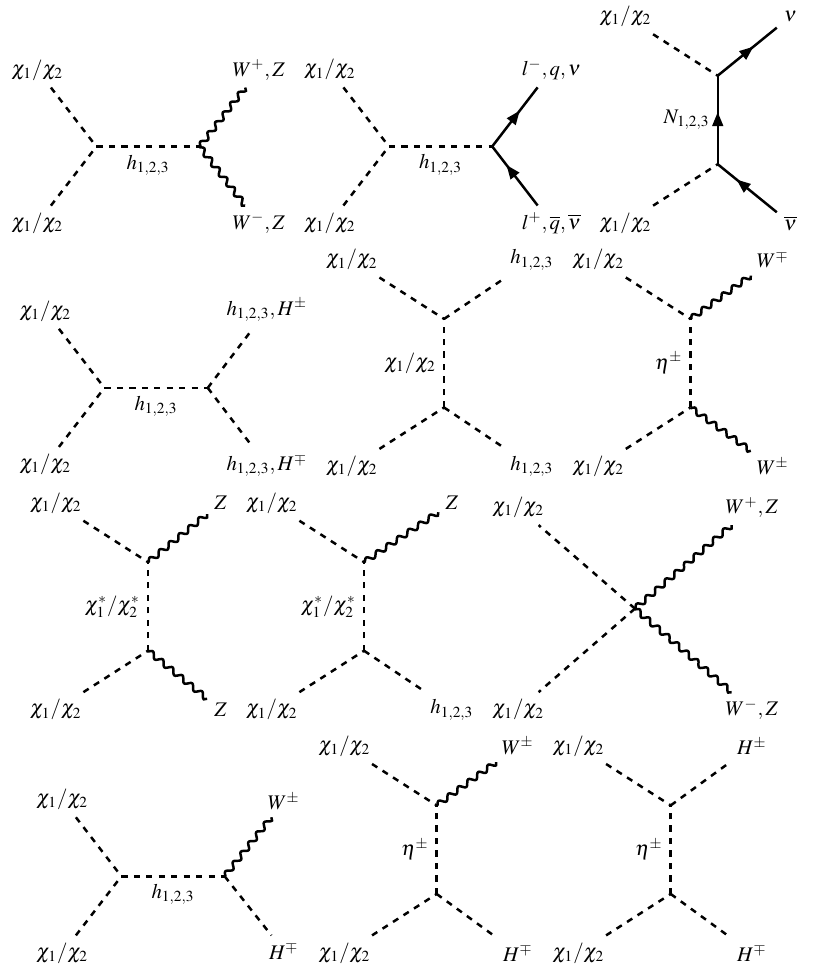}
                \label{fig:anihixi1}
\end{figure}

\begin{figure}[th]
        \includegraphics[width=14cm]{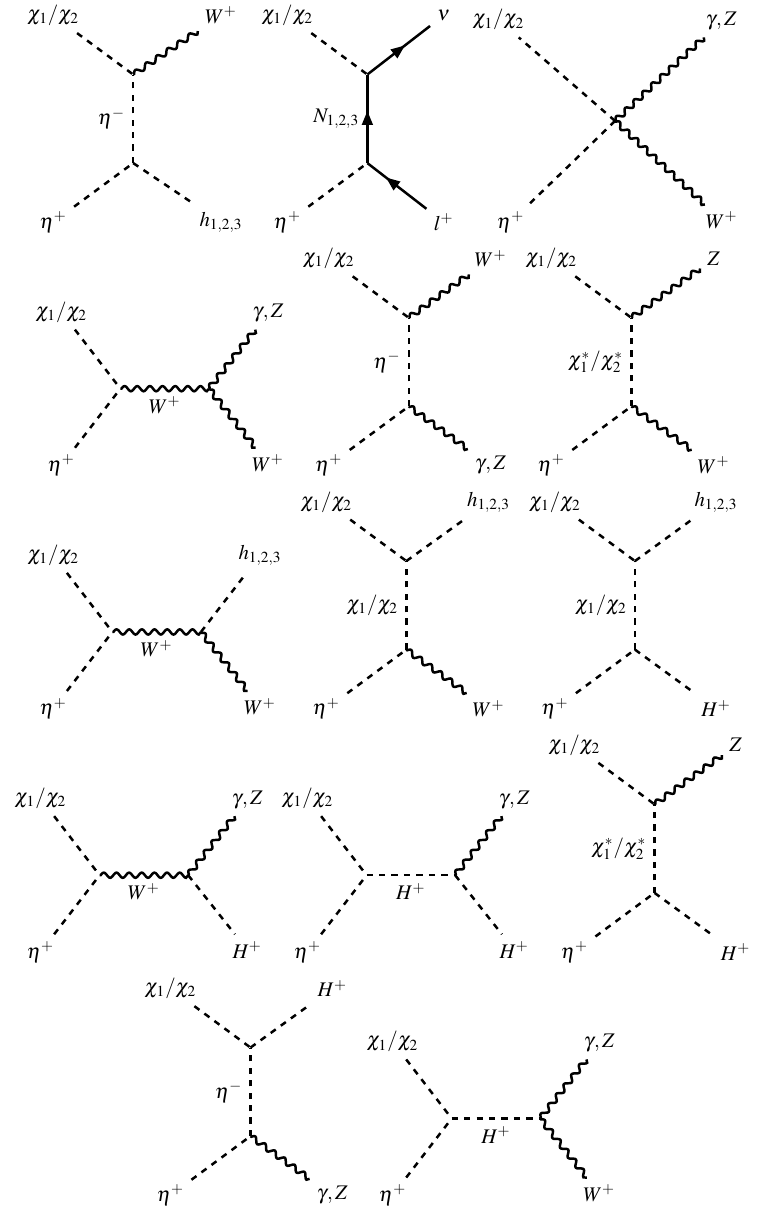}
                \label{fig:coanihixi1}
\end{figure}

\begin{figure}[th]
        \includegraphics[width=14cm]{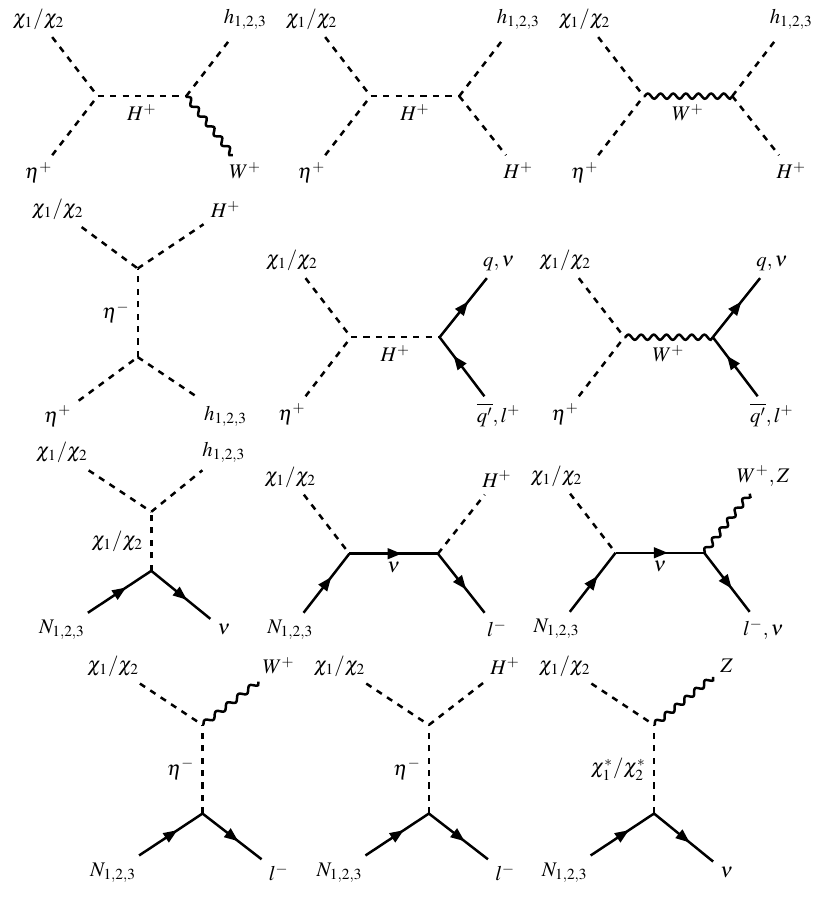}
        \caption{\centering \footnotesize Relevant diagrams for the computation of relic density of the scalar DM candidate.}
                \label{fig:coanihixi2}
\end{figure}

\begin{figure}[th]
        \includegraphics[width=9cm]{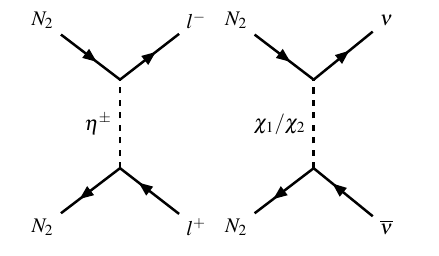}
        \caption{\centering \footnotesize Relevant diagrams for the computation of relic density of the fermion DM candidate.}
                \label{fig:anihiN2}
\end{figure}

\begin{figure}[th]
        \includegraphics[width=9cm]{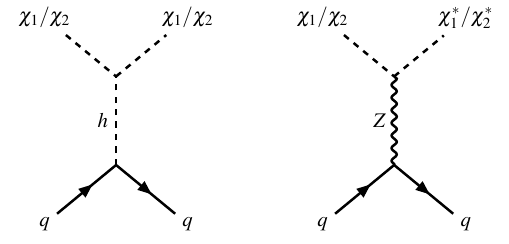}
        \caption{\centering \footnotesize Relevant diagrams for the direct detection of the scalar DM candidate.}
                \label{fig:DDscalar}
\end{figure}
\begin{figure}[th]
        \includegraphics[width=12cm]{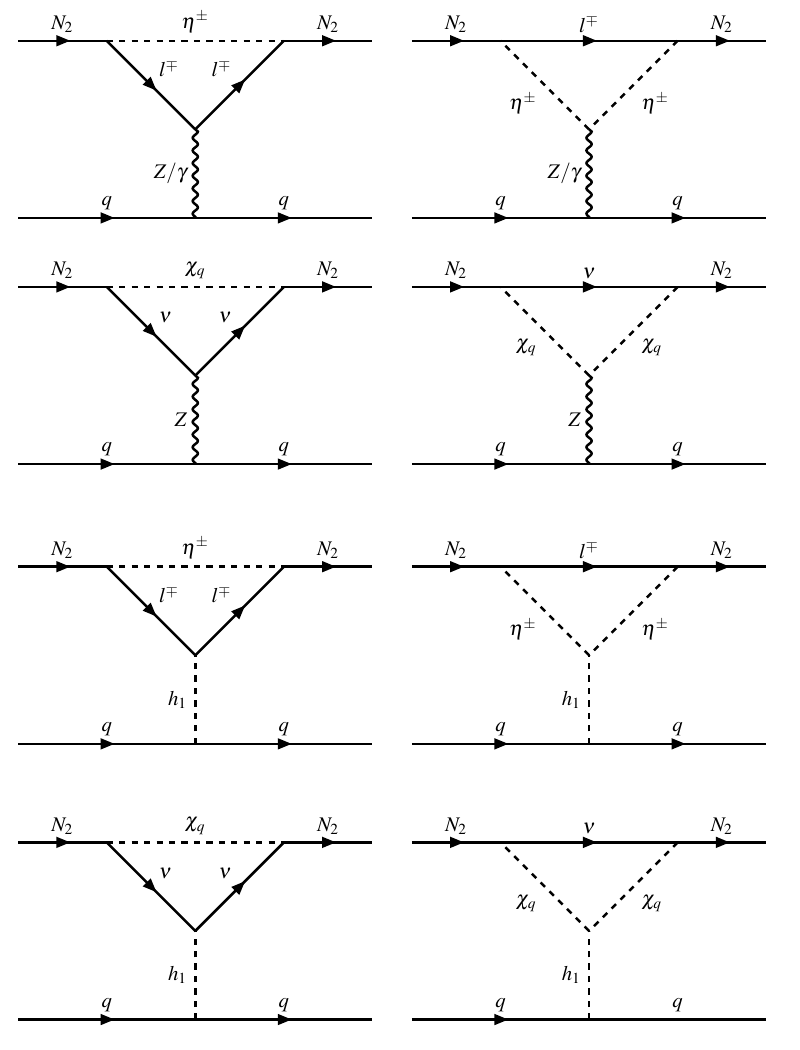}
        \caption{\centering \footnotesize Relevant diagrams for the direct detection of the fermion DM candidate.}
                \label{fig:DDfermion}
\end{figure}
\FloatBarrier

\bibliographystyle{utphys}
\bibliography{bibliography} 

\providecommand{\href}[2]{#2}\begingroup\raggedright\begin{thebibliography}{10}

\bibitem{SNO:2001kpb}
{\bfseries SNO} Collaboration, Q.~R. Ahmad {\em et~al.}, ``{Measurement of the
  rate of $\nu_e+d \to p+p+e^-$ interactions produced by $^8$B solar neutrinos
  at the Sudbury Neutrino Observatory},''
  \href{http://dx.doi.org/10.1103/PhysRevLett.87.071301}{{\em Phys. Rev. Lett.}
  {\bfseries 87} (2001) 071301},
  \href{http://arxiv.org/abs/nucl-ex/0106015}{{\ttfamily
  arXiv:nucl-ex/0106015}}.

\bibitem{Super-Kamiokande:1998kpq}
{\bfseries Super-Kamiokande} Collaboration, Y.~Fukuda {\em et~al.}, ``{Evidence
  for oscillation of atmospheric neutrinos},''
  \href{http://dx.doi.org/10.1103/PhysRevLett.81.1562}{{\em Phys. Rev. Lett.}
  {\bfseries 81} (1998) 1562--1567},
  \href{http://arxiv.org/abs/hep-ex/9807003}{{\ttfamily arXiv:hep-ex/9807003}}.

\bibitem{Planck:2018vyg}
{\bfseries Planck} Collaboration, N.~Aghanim {\em et~al.}, ``{Planck 2018
  results. VI. Cosmological parameters},''
  \href{http://dx.doi.org/10.1051/0004-6361/201833910}{{\em Astron. Astrophys.}
  {\bfseries 641} (2020) A6}, \href{http://arxiv.org/abs/1807.06209}{{\ttfamily
  arXiv:1807.06209 [astro-ph.CO]}}. [Erratum: Astron.Astrophys. 652, C4
  (2021)].

\bibitem{ma2006verifiable}
E.~Ma, ``Verifiable radiative seesaw mechanism of neutrino mass and dark
  matter,'' {\em Physical Review D} {\bfseries 73} no.~7, (2006) 077301.

\bibitem{Capozzi:2021fjo}
F.~Capozzi, E.~Di~Valentino, E.~Lisi, A.~Marrone, A.~Melchiorri, and
  A.~Palazzo, ``{Unfinished fabric of the three neutrino paradigm},''
  \href{http://dx.doi.org/10.1103/PhysRevD.104.083031}{{\em Phys. Rev. D}
  {\bfseries 104} no.~8, (2021) 083031},
  \href{http://arxiv.org/abs/2107.00532}{{\ttfamily arXiv:2107.00532
  [hep-ph]}}.

\bibitem{deSalas:2020pgw}
P.~de~Salas {\em et~al.}, ``{2020 Global reassessment of the neutrino
  oscillation picture},'' \href{http://arxiv.org/abs/2006.11237}{{\ttfamily
  arXiv:2006.11237 [hep-ph]}}.

\bibitem{Esteban:2020cvm}
I.~Esteban, M.~C. Gonzalez-Garcia, M.~Maltoni, T.~Schwetz, and A.~Zhou, ``{The
  fate of hints: updated global analysis of three-flavor neutrino
  oscillations},'' \href{http://dx.doi.org/10.1007/JHEP09(2020)178}{{\em JHEP}
  {\bfseries 09} (2020) 178}, \href{http://arxiv.org/abs/2007.14792}{{\ttfamily
  arXiv:2007.14792 [hep-ph]}}.

\bibitem{Rojas:2018wym}
N.~Rojas, R.~Srivastava, and J.~W. Valle, ``{Simplest Scoto-Seesaw
  Mechanism},'' \href{http://dx.doi.org/10.1016/j.physletb.2018.12.014}{{\em
  Phys.Lett.} {\bfseries B789} (2019) 132--136},
  \href{http://arxiv.org/abs/1807.11447}{{\ttfamily arXiv:1807.11447
  [hep-ph]}}.

\bibitem{Barreiros:2020gxu}
D.~M. Barreiros, F.~R. Joaquim, R.~Srivastava, and J.~W.~F. Valle, ``{Minimal
  scoto-seesaw mechanism with spontaneous CP violation},''
  \href{http://dx.doi.org/10.1007/JHEP04(2021)249}{{\em JHEP} {\bfseries 04}
  (2021) 249}, \href{http://arxiv.org/abs/2012.05189}{{\ttfamily
  arXiv:2012.05189 [hep-ph]}}.

\bibitem{Mandal:2021yph}
S.~Mandal, R.~Srivastava, and J.~W.~F. Valle, ``{The simplest scoto-seesaw
  model: WIMP dark matter phenomenology and Higgs vacuum stability},''
  \href{http://dx.doi.org/10.1016/j.physletb.2021.136458}{{\em Phys. Lett. B}
  {\bfseries 819} (2021) 136458},
  \href{http://arxiv.org/abs/2104.13401}{{\ttfamily arXiv:2104.13401
  [hep-ph]}}.

\bibitem{Barreiros:2022aqu}
D.~M. Barreiros, H.~B. Camara, and F.~R. Joaquim, ``{Flavour and dark matter in
  a scoto/type-II seesaw model},''
  \href{http://dx.doi.org/10.1007/JHEP08(2022)030}{{\em JHEP} {\bfseries 08}
  (2022) 030}, \href{http://arxiv.org/abs/2204.13605}{{\ttfamily
  arXiv:2204.13605 [hep-ph]}}.

\bibitem{Ganguly:2023jml}
J.~Ganguly, J.~Gluza, B.~Karmakar, and S.~Mahapatra, ``{Phenomenology of the
  flavor symmetric scoto-seesaw model with dark matter and TM$_1$ mixing},''
  \href{http://arxiv.org/abs/2311.15997}{{\ttfamily arXiv:2311.15997
  [hep-ph]}}.

\bibitem{Ganguly:2022qxj}
J.~Ganguly, J.~Gluza, and B.~Karmakar, ``{Common origin of
  \ensuremath{\theta}$_{13}$ and dark matter within the flavor symmetric
  scoto-seesaw framework},''
  \href{http://dx.doi.org/10.1007/JHEP11(2022)074}{{\em JHEP} {\bfseries 11}
  (2022) 074}, \href{http://arxiv.org/abs/2209.08610}{{\ttfamily
  arXiv:2209.08610 [hep-ph]}}.

\bibitem{VanDong:2023xbd}
P.~Van~Dong and D.~Van~Loi, ``{Scotoseesaw model implied by dark right-handed
  neutrinos},'' \href{http://arxiv.org/abs/2311.09795}{{\ttfamily
  arXiv:2311.09795 [hep-ph]}}.

\bibitem{Leite:2023gzl}
J.~Leite, S.~Sadhukhan, and J.~W.~F. Valle, ``{Dynamical scoto-seesaw mechanism
  with gauged $B-L$},'' \href{http://arxiv.org/abs/2307.04840}{{\ttfamily
  arXiv:2307.04840 [hep-ph]}}.

\bibitem{Kumar:2023moh}
R.~Kumar, P.~Mishra, M.~K. Behera, R.~Mohanta, and R.~Srivastava,
  ``{Predictions from scoto-seesaw with $A_4$ modular symmetry},''
  \href{http://arxiv.org/abs/2310.02363}{{\ttfamily arXiv:2310.02363
  [hep-ph]}}.

\bibitem{VanDong:2024lry}
P.~Van~Dong, D.~Van~Loi, D.~T. Huong, N.~T. Duy, and D.~Van~Soa, ``{Dark
  symmetry implication for right-handed neutrinos},''
  \href{http://arxiv.org/abs/2407.02324}{{\ttfamily arXiv:2407.02324
  [hep-ph]}}.

\bibitem{Schechter:1981bd}
J.~Schechter and J.~W.~F. Valle, ``{Neutrinoless Double beta Decay in SU(2) x
  U(1) Theories},'' \href{http://dx.doi.org/10.1103/PhysRevD.25.2951}{{\em
  Phys. Rev. D} {\bfseries 25} (1982) 2951}.

\bibitem{Gu:2007ug}
P.-H. Gu and U.~Sarkar, ``{Radiative Neutrino Mass, Dark Matter and
  Leptogenesis},'' \href{http://dx.doi.org/10.1103/PhysRevD.77.105031}{{\em
  Phys. Rev. D} {\bfseries 77} (2008) 105031},
  \href{http://arxiv.org/abs/0712.2933}{{\ttfamily arXiv:0712.2933 [hep-ph]}}.

\bibitem{Farzan:2012sa}
Y.~Farzan and E.~Ma, ``{Dirac neutrino mass generation from dark matter},''
  \href{http://dx.doi.org/10.1103/PhysRevD.86.033007}{{\em Phys. Rev. D}
  {\bfseries 86} (2012) 033007},
  \href{http://arxiv.org/abs/1204.4890}{{\ttfamily arXiv:1204.4890 [hep-ph]}}.

\bibitem{Ma:2016mwh}
E.~Ma and O.~Popov, ``{Pathways to Naturally Small Dirac Neutrino Masses},''
  \href{http://dx.doi.org/10.1016/j.physletb.2016.11.027}{{\em Phys. Lett. B}
  {\bfseries 764} (2017) 142--144},
  \href{http://arxiv.org/abs/1609.02538}{{\ttfamily arXiv:1609.02538
  [hep-ph]}}.

\bibitem{Bonilla:2018ynb}
C.~Bonilla, S.~Centelles-Chuli\'a, R.~Cepedello, E.~Peinado, and R.~Srivastava,
  ``{Dark matter stability and Dirac neutrinos using only Standard Model
  symmetries},'' \href{http://dx.doi.org/10.1103/PhysRevD.101.033011}{{\em
  Phys. Rev. D} {\bfseries 101} no.~3, (2020) 033011},
  \href{http://arxiv.org/abs/1812.01599}{{\ttfamily arXiv:1812.01599
  [hep-ph]}}.

\bibitem{Dasgupta:2019rmf}
A.~Dasgupta, S.~K. Kang, and O.~Popov, ``{Radiative Dirac neutrino mass,
  neutrinoless quadruple beta decay, and dark matter in B-L extension of the
  standard model},'' \href{http://dx.doi.org/10.1103/PhysRevD.100.075030}{{\em
  Phys. Rev. D} {\bfseries 100} no.~7, (2019) 075030},
  \href{http://arxiv.org/abs/1903.12558}{{\ttfamily arXiv:1903.12558
  [hep-ph]}}.

\bibitem{Guo:2020qin}
S.-Y. Guo and Z.-L. Han, ``{Observable Signatures of Scotogenic Dirac Model},''
  \href{http://dx.doi.org/10.1007/JHEP12(2020)062}{{\em JHEP} {\bfseries 12}
  (2020) 062}, \href{http://arxiv.org/abs/2005.08287}{{\ttfamily
  arXiv:2005.08287 [hep-ph]}}.

\bibitem{CentellesChulia:2024iom}
S.~Centelles~Chuli\'a, R.~Srivastava, and S.~Yadav, ``{Comprehensive
  phenomenology of the Dirac Scotogenic Model: Novel low-mass dark matter},''
  \href{http://dx.doi.org/10.1007/JHEP04(2025)038}{{\em JHEP} {\bfseries 04}
  (2025) 038}, \href{http://arxiv.org/abs/2409.18513}{{\ttfamily
  arXiv:2409.18513 [hep-ph]}}.

\bibitem{Kumar:2024zfb}
R.~Kumar, N.~Nath, and R.~Srivastava, ``{Cutting the scotogenic loop: adding
  flavor to dark matter},''
  \href{http://dx.doi.org/10.1007/JHEP12(2024)036}{{\em JHEP} {\bfseries 12}
  (2024) 036}, \href{http://arxiv.org/abs/2406.00188}{{\ttfamily
  arXiv:2406.00188 [hep-ph]}}.

\bibitem{Borah:2024gql}
D.~Borah, P.~Das, B.~Karmakar, and S.~Mahapatra, ``{Discrete dark matter with
  light Dirac neutrinos},'' \href{http://arxiv.org/abs/2406.17861}{{\ttfamily
  arXiv:2406.17861 [hep-ph]}}.

\bibitem{Kumar:2024jot}
R.~Kumar, N.~Nath, and R.~Srivastava, ``{Cutting the~Scotogenic Loop: $A_4$
  Flavor Symmetry to~$Z_2$ Dark Symmetry},''
  \href{http://dx.doi.org/10.1007/978-981-97-0289-3_331}{{\em Springer Proc.
  Phys.} {\bfseries 304} (2024) 1183--1185}.

\bibitem{Ma:2001dn}
E.~Ma and G.~Rajasekaran, ``{Softly broken A(4) symmetry for nearly degenerate
  neutrino masses},'' \href{http://dx.doi.org/10.1103/PhysRevD.64.113012}{{\em
  Phys. Rev. D} {\bfseries 64} (2001) 113012},
  \href{http://arxiv.org/abs/hep-ph/0106291}{{\ttfamily arXiv:hep-ph/0106291}}.

\bibitem{Babu:2002dz}
K.~S. Babu, E.~Ma, and J.~W.~F. Valle, ``{Underlying A(4) symmetry for the
  neutrino mass matrix and the quark mixing matrix},''
  \href{http://dx.doi.org/10.1016/S0370-2693(02)03153-2}{{\em Phys. Lett. B}
  {\bfseries 552} (2003) 207--213},
  \href{http://arxiv.org/abs/hep-ph/0206292}{{\ttfamily arXiv:hep-ph/0206292}}.

\bibitem{Altarelli:2005yx}
G.~Altarelli and F.~Feruglio, ``{Tri-bimaximal neutrino mixing, A(4) and the
  modular symmetry},''
  \href{http://dx.doi.org/10.1016/j.nuclphysb.2006.02.015}{{\em Nucl. Phys. B}
  {\bfseries 741} (2006) 215--235},
  \href{http://arxiv.org/abs/hep-ph/0512103}{{\ttfamily arXiv:hep-ph/0512103}}.

\bibitem{Hirsch:2010ru}
M.~Hirsch, S.~Morisi, E.~Peinado, and J.~W.~F. Valle, ``{Discrete dark
  matter},'' \href{http://dx.doi.org/10.1103/PhysRevD.82.116003}{{\em Phys.
  Rev. D} {\bfseries 82} (2010) 116003},
  \href{http://arxiv.org/abs/1007.0871}{{\ttfamily arXiv:1007.0871 [hep-ph]}}.

\bibitem{Boucenna:2011tj}
M.~S. Boucenna, M.~Hirsch, S.~Morisi, E.~Peinado, M.~Taoso, and J.~W.~F. Valle,
  ``{Phenomenology of Dark Matter from $A_4$ Flavor Symmetry},''
  \href{http://dx.doi.org/10.1007/JHEP05(2011)037}{{\em JHEP} {\bfseries 05}
  (2011) 037}, \href{http://arxiv.org/abs/1101.2874}{{\ttfamily arXiv:1101.2874
  [hep-ph]}}.

\bibitem{DeLaVega:2018bkp}
L.~M.~G. De~La~Vega, R.~Ferro-Hernandez, and E.~Peinado, ``{Simple $A_4$ models
  for dark matter stability with texture zeros},''
  \href{http://dx.doi.org/10.1103/PhysRevD.99.055044}{{\em Phys. Rev. D}
  {\bfseries 99} no.~5, (2019) 055044},
  \href{http://arxiv.org/abs/1811.10619}{{\ttfamily arXiv:1811.10619
  [hep-ph]}}.

\bibitem{Mohapatra:1986bd}
R.~N. Mohapatra and J.~W.~F. Valle, ``{Neutrino Mass and Baryon Number
  Nonconservation in Superstring Models},''
  \href{http://dx.doi.org/10.1103/PhysRevD.34.1642}{{\em Phys. Rev. D}
  {\bfseries 34} (1986) 1642}.

\bibitem{CentellesChulia:2020dfh}
S.~Centelles~Chuli\'a, R.~Srivastava, and A.~Vicente, ``{The inverse seesaw
  family: Dirac and Majorana},''
  \href{http://dx.doi.org/10.1007/JHEP03(2021)248}{{\em JHEP} {\bfseries 03}
  (2021) 248}, \href{http://arxiv.org/abs/2011.06609}{{\ttfamily
  arXiv:2011.06609 [hep-ph]}}.

\bibitem{Ma:2014qra}
E.~Ma and R.~Srivastava, ``{Dirac or inverse seesaw neutrino masses with $B-L$
  gauge symmetry and $S_3$ flavor symmetry},''
  \href{http://dx.doi.org/10.1016/j.physletb.2014.12.049}{{\em Phys. Lett. B}
  {\bfseries 741} (2015) 217--222},
  \href{http://arxiv.org/abs/1411.5042}{{\ttfamily arXiv:1411.5042 [hep-ph]}}.

\bibitem{Ma:2015raa}
E.~Ma and R.~Srivastava, ``{Dirac or inverse seesaw neutrino masses from gauged
  $B–L$ symmetry},'' \href{http://dx.doi.org/10.1142/S0217732315300207}{{\em
  Mod. Phys. Lett. A} {\bfseries 30} no.~26, (2015) 1530020},
  \href{http://arxiv.org/abs/1504.00111}{{\ttfamily arXiv:1504.00111
  [hep-ph]}}.

\bibitem{Ma:2015mjd}
E.~Ma, N.~Pollard, R.~Srivastava, and M.~Zakeri, ``{Gauge $B-L$ Model with
  Residual $Z_3$ Symmetry},''
  \href{http://dx.doi.org/10.1016/j.physletb.2015.09.010}{{\em Phys. Lett. B}
  {\bfseries 750} (2015) 135--138},
  \href{http://arxiv.org/abs/1507.03943}{{\ttfamily arXiv:1507.03943
  [hep-ph]}}.

\bibitem{DESI:2024mwx}
{\bfseries DESI} Collaboration, A.~G. Adame {\em et~al.}, ``{DESI 2024 VI:
  Cosmological Constraints from the Measurements of Baryon Acoustic
  Oscillations},'' \href{http://arxiv.org/abs/2404.03002}{{\ttfamily
  arXiv:2404.03002 [astro-ph.CO]}}.

\bibitem{KATRIN:2021uub}
{\bfseries KATRIN} Collaboration, M.~Aker {\em et~al.}, ``{Direct neutrino-mass
  measurement with sub-electronvolt sensitivity},''
  \href{http://dx.doi.org/10.1038/s41567-021-01463-1}{{\em Nature Phys.}
  {\bfseries 18} no.~2, (2022) 160--166},
  \href{http://arxiv.org/abs/2105.08533}{{\ttfamily arXiv:2105.08533
  [hep-ex]}}.

\bibitem{XENON:2023cxc}
{\bfseries XENON} Collaboration, E.~Aprile {\em et~al.}, ``{First Dark Matter
  Search with Nuclear Recoils from the XENONnT Experiment},''
  \href{http://dx.doi.org/10.1103/PhysRevLett.131.041003}{{\em Phys. Rev.
  Lett.} {\bfseries 131} no.~4, (2023) 041003},
  \href{http://arxiv.org/abs/2303.14729}{{\ttfamily arXiv:2303.14729
  [hep-ex]}}.

\bibitem{LZ:2022lsv}
{\bfseries LZ} Collaboration, J.~Aalbers {\em et~al.}, ``{First Dark Matter
  Search Results from the LUX-ZEPLIN (LZ) Experiment},''
  \href{http://dx.doi.org/10.1103/PhysRevLett.131.041002}{{\em Phys. Rev.
  Lett.} {\bfseries 131} no.~4, (2023) 041002},
  \href{http://arxiv.org/abs/2207.03764}{{\ttfamily arXiv:2207.03764
  [hep-ex]}}.

\bibitem{ATLAS:2022yvh}
{\bfseries ATLAS} Collaboration, G.~Aad {\em et~al.}, ``{Search for invisible
  Higgs-boson decays in events with vector-boson fusion signatures using 139
  fb$^{-1}$ of proton-proton data recorded by the ATLAS experiment},''
  \href{http://dx.doi.org/10.1007/JHEP08(2022)104}{{\em JHEP} {\bfseries 08}
  (2022) 104}, \href{http://arxiv.org/abs/2202.07953}{{\ttfamily
  arXiv:2202.07953 [hep-ex]}}.

\bibitem{Cao:2007rm}
Q.-H. Cao, E.~Ma, and G.~Rajasekaran, ``{Observing the Dark Scalar Doublet and
  its Impact on the Standard-Model Higgs Boson at Colliders},''
  \href{http://dx.doi.org/10.1103/PhysRevD.76.095011}{{\em Phys. Rev. D}
  {\bfseries 76} (2007) 095011},
  \href{http://arxiv.org/abs/0708.2939}{{\ttfamily arXiv:0708.2939 [hep-ph]}}.

\bibitem{Gustafsson:2007pc}
M.~Gustafsson, E.~Lundstrom, L.~Bergstrom, and J.~Edsjo, ``{Significant Gamma
  Lines from Inert Higgs Dark Matter},''
  \href{http://dx.doi.org/10.1103/PhysRevLett.99.041301}{{\em Phys. Rev. Lett.}
  {\bfseries 99} (2007) 041301},
  \href{http://arxiv.org/abs/astro-ph/0703512}{{\ttfamily
  arXiv:astro-ph/0703512}}.

\bibitem{Pierce:2007ut}
A.~Pierce and J.~Thaler, ``{Natural Dark Matter from an Unnatural Higgs Boson
  and New Colored Particles at the TeV Scale},''
  \href{http://dx.doi.org/10.1088/1126-6708/2007/08/026}{{\em JHEP} {\bfseries
  08} (2007) 026}, \href{http://arxiv.org/abs/hep-ph/0703056}{{\ttfamily
  arXiv:hep-ph/0703056}}.

\bibitem{LZ:2024zvo}
{\bfseries LZ} Collaboration, J.~Aalbers {\em et~al.}, ``{Dark Matter Search
  Results from 4.2 Tonne-Years of Exposure of the LUX-ZEPLIN (LZ)
  Experiment},'' \href{http://arxiv.org/abs/2410.17036}{{\ttfamily
  arXiv:2410.17036 [hep-ex]}}.

\bibitem{PandaX:2024qfu}
{\bfseries PandaX} Collaboration, Z.~Bo {\em et~al.}, ``{Dark Matter Search
  Results from 1.54 Tonne$\cdot$Year Exposure of PandaX-4T},''
  \href{http://arxiv.org/abs/2408.00664}{{\ttfamily arXiv:2408.00664
  [hep-ex]}}.

\bibitem{Billard:2013qya}
J.~Billard, L.~Strigari, and E.~Figueroa-Feliciano, ``{Implication of neutrino
  backgrounds on the reach of next generation dark matter direct detection
  experiments},'' \href{http://dx.doi.org/10.1103/PhysRevD.89.023524}{{\em
  Phys. Rev. D} {\bfseries 89} no.~2, (2014) 023524},
  \href{http://arxiv.org/abs/1307.5458}{{\ttfamily arXiv:1307.5458 [hep-ph]}}.

\bibitem{Gonzalez-Garcia:2002bkq}
M.~C. Gonzalez-Garcia and Y.~Nir, ``{Neutrino Masses and Mixing: Evidence and
  Implications},'' \href{http://dx.doi.org/10.1103/RevModPhys.75.345}{{\em Rev.
  Mod. Phys.} {\bfseries 75} (2003) 345--402},
  \href{http://arxiv.org/abs/hep-ph/0202058}{{\ttfamily arXiv:hep-ph/0202058}}.

\bibitem{MEGII:2023ltw}
{\bfseries MEG II} Collaboration, K.~Afanaciev {\em et~al.}, ``{A search for
  $\mu^+\to e^+\gamma$ with the first dataset of the MEG II experiment},''
  \href{http://arxiv.org/abs/2310.12614}{{\ttfamily arXiv:2310.12614
  [hep-ex]}}.

\bibitem{BaBar:2009hkt}
{\bfseries BaBar} Collaboration, B.~Aubert {\em et~al.}, ``{Searches for Lepton
  Flavor Violation in the Decays tau+- ---\ensuremath{>} e+- gamma and tau+-
  ---\ensuremath{>} mu+- gamma},''
  \href{http://dx.doi.org/10.1103/PhysRevLett.104.021802}{{\em Phys. Rev.
  Lett.} {\bfseries 104} (2010) 021802},
  \href{http://arxiv.org/abs/0908.2381}{{\ttfamily arXiv:0908.2381 [hep-ex]}}.

\bibitem{Belle:2021ysv}
{\bfseries Belle} Collaboration, A.~Abdesselam {\em et~al.}, ``{Search for
  lepton-flavor-violating tau-lepton decays to $\ell\gamma$ at Belle},''
  \href{http://dx.doi.org/10.1007/JHEP10(2021)019}{{\em JHEP} {\bfseries 10}
  (2021) 19}, \href{http://arxiv.org/abs/2103.12994}{{\ttfamily
  arXiv:2103.12994 [hep-ex]}}.

\bibitem{Baldini:2013ke}
A.~M. Baldini {\em et~al.}, ``{MEG Upgrade Proposal},''
  \href{http://arxiv.org/abs/1301.7225}{{\ttfamily arXiv:1301.7225
  [physics.ins-det]}}.

\bibitem{Belle-II:2022cgf}
{\bfseries Belle-II} Collaboration, L.~Aggarwal {\em et~al.}, ``{Snowmass White
  Paper: Belle II physics reach and plans for the next decade and beyond},''
  \href{http://arxiv.org/abs/2207.06307}{{\ttfamily arXiv:2207.06307
  [hep-ex]}}.

\bibitem{ParticleDataGroup:2022pth}
{\bfseries Particle Data Group} Collaboration, R.~L. Workman {\em et~al.},
  ``{Review of Particle Physics},''
  \href{http://dx.doi.org/10.1093/ptep/ptac097}{{\em PTEP} {\bfseries 2022}
  (2022) 083C01}.

\bibitem{Staub:2015kfa}
F.~Staub, ``{Exploring new models in all detail with SARAH},''
  \href{http://dx.doi.org/10.1155/2015/840780}{{\em Adv. High Energy Phys.}
  {\bfseries 2015} (2015) 840780},
\href{http://arxiv.org/abs/1503.04200}{{\ttfamily arXiv:1503.04200 [hep-ph]}}.

\bibitem{Porod:2011nf}
W.~Porod and F.~Staub, ``{SPheno 3.1: Extensions including flavour, CP-phases
  and models beyond the MSSM},''
  \href{http://dx.doi.org/10.1016/j.cpc.2012.05.021}{{\em Comput. Phys.
  Commun.} {\bfseries 183} (2012) 2458--2469},
  \href{http://arxiv.org/abs/1104.1573}{{\ttfamily arXiv:1104.1573 [hep-ph]}}.

\bibitem{Belanger:2014vza}
G.~B\'elanger, F.~Boudjema, A.~Pukhov, and A.~Semenov, ``{micrOMEGAs4.1: two
  dark matter candidates},''
  \href{http://dx.doi.org/10.1016/j.cpc.2015.03.003}{{\em Comput. Phys.
  Commun.} {\bfseries 192} (2015) 322--329},
  \href{http://arxiv.org/abs/1407.6129}{{\ttfamily arXiv:1407.6129 [hep-ph]}}.

\bibitem{Ishimori:2010au}
H.~Ishimori, T.~Kobayashi, H.~Ohki, Y.~Shimizu, H.~Okada, and M.~Tanimoto,
  ``{Non-Abelian Discrete Symmetries in Particle Physics},''
  \href{http://dx.doi.org/10.1143/PTPS.183.1}{{\em Prog. Theor. Phys. Suppl.}
  {\bfseries 183} (2010) 1--163},
  \href{http://arxiv.org/abs/1003.3552}{{\ttfamily arXiv:1003.3552 [hep-th]}}.

\end{thebibliography}\endgroup
\end{document}